\documentclass[iop]{emulateapj}

\usepackage{amsmath}

\usepackage{threeparttable}
\usepackage{graphicx, subfigure}
\usepackage{times}
\usepackage[english]{babel}
\usepackage{listings}
\usepackage{color}

\usepackage{epstopdf}
\usepackage{url}

\usepackage{xcolor}
\definecolor{green}{HTML}{33CC33}
\definecolor{red}{HTML}{FF3300}
\definecolor{blue}{HTML}{3333FF}

\usepackage[colorlinks=true,citecolor=blue,urlcolor=green]{hyperref}

\usepackage{mathrsfs}

\usepackage{natbib,twoopt}
\usepackage[varg]{txfonts}

\usepackage[normalem]{ulem}

\renewcommand{\eqref}[1]{Equation~(\ref{#1})}
\renewcommand{\S}[1]{Section~}
\newcommand{\fref}[1]{Figure~\ref{#1}}
 
\newcommand{\ie}{i.\,e.}

\newcommand{\eg}{e.\,g.}
 
\def\note #1]{{\bf #1]}}

\newcommand{\numax}{$\nu_{\rm max}$}

\numberwithin{equation}{section}

\makeatletter
\def\maketag@@@#1{\hbox{\m@th\normalfont\normalsize#1}}
\makeatother

\bibpunct{(}{)}{;}{a}{}{,} 

\shorttitle{Differential rotation and asteroseismology}
\shortauthors{Lund et al.}

\begin{document}

\submitted{Received 4 February 2014 -- Accepted 11 June 2014 -- Published 11 July 2014}

\title{\large Differential rotation in main-sequence solar-like stars: \\  Qualitative inference from asteroseismic data \vspace*{0.3cm}} 
\author{Mikkel~N.~Lund$^{1,2\star}$}
\author{Mark~S.~Miesch$^3$}
\author{J\o rgen Christensen-Dalsgaard$^1$\\ \vspace*{0.4cm}}

\affil{$^1$Stellar Astrophysics Centre (SAC), Department of Physics and Astronomy, Aarhus University,\\ Ny Munkegade 120, DK-8000 Aarhus C, Denmark}
\affil{$^2$Sydney Institute for Astronomy (SIfA), School of Physics, University of Sydney, NSW 2006, Australia}
\affil{$^3$High Altitude Observatory (HAO), National Center for Atmospheric Research, Boulder, CO 80307-3000, USA}

\email{$^{\star}$mikkelnl@phys.au.dk}


\begin{abstract}
Understanding differential rotation of Sun-like stars is of great importance for insight into the angular momentum transport in these stars. One means of gaining such information is that of asteroseismology. By a forward modeling approach we analyze in a qualitative manner the impact of different differential rotation profiles on the splittings of \emph{p}-mode oscillation frequencies. The optimum modes for inference on differential rotation are identified along with the best value of the stellar inclination angle. We find that in general it is not likely that asteroseismology can be used to make an unambiguous distinction between a rotation profile such as, \eg, a conical Sun-like profile and a cylindrical profile. In addition, it seems unlikely that asteroseismology of Sun-like stars will result in inferences on the radial profile of the differential rotation, such as can be done for, \eg, red giants. At best one could possibly obtain the sign of the radial differential rotation gradient. Measurements of the extent of the latitudinal differential from frequency splitting are, however, more promising. One very interesting aspect that could likely be tested from frequency splittings is whether the differential rotation is solar-like or anti-solar-like in nature, in the sense that a solar-like profile has an equator rotating faster than the poles.
\end{abstract}



\section{Introduction}
\label{sec:intro}

A better understanding of stellar rotation is of great importance for our understanding of stellar dynamics in general, and especially the evolution and transport of angular momentum. These aspects are intimately linked to the differential rotation of the star, \ie, the notion that different parts of the star rotate at different rates \citep[][]{2005LRSP....2....1M,2009AnRFM..41..317M,2013IAUS..294..399K}. Moreover, differential rotation is thought to be a major player in the driving of the solar dynamo responsible for the solar 11 yr cycle\footnote{The 11-yr cycle (a.k.a. the Schwabe cycle) is half of the 22-yr cycle (a.k.a. the Hale cycle) for the reversal of the solar magnetic field.}.

A means of obtaining information on the rotation in stars comes with asteroseismology. Using asteroseismology on time-series data from, \eg, the \emph{Kepler} \citep[][]{2010PASP..122..131G} or \emph{CoRoT}\footnote{Convection Rotation and planetary Transits.} \citep[][]{2009A&A...506..411A} missions we are in a position to extract information on the rotational characteristics of the stars. One major strength of asteroseismology is, furthermore, that not only can the aspects of differential rotation possibly be inferred, but from the pulsation frequencies observed in the star a realistic model can be found \citep[see][for a recent review]{2013ARA&A..51..353C}. This enables tests of models for differential rotation, be it mean field models \citep[see, \eg,][]{2011MNRAS.411.1059K,2011A&A...530A..48K} or more elaborate 3D simulations \citep[see, \eg,][]{2004ApJ...614.1073B,2008ApJ...676.1262B,2012ApJ...755L..22K}.

With \emph{Kepler} data much has already been done for, \eg, red giants \citep[][]{2012Natur.481...55B,2012ApJ...756...19D,2012A&A...548A..10M,2014A&A...564A..27D}. Here, for instance, it has been possible to estimate the ratio between the rotation rate in the small helium core and the large envelope of these stars. 
Similarly, much effort is being invested in analysis of more active stars using their modulations of the light curve from stellar surface spots \citep[see, \eg,][]{2012A&A...543A.146F,2012A&A...547A..37B,2013MNRAS.432.1203M,2014A&A...564A..50L}.

However, in the context of differential rotation, little has been done with asteroseismology for slowly rotating main-sequence solar-like oscillators -- this is where we have focused our attention in the present work. One reason for this is presumably that the analysis is much more difficult than, for instance, for red giants. This comes from the fact that only pressure-dominated \emph{p}-modes are available in the analysis and these propagate to a large extent in the outer parts of the star, leaving little information on the radial profile of the rotation. In evolved solar-like oscillators one often has the advantage of having, in addition to \emph{p}-modes, so-called mixed modes that are sensitive to the deep interior while still being readily detectable. Also, the fact that stars like the Sun more often than not are rotating quite slowly ($P_{\rm rot}>10 \, \rm days$), compared to the mean lifetime of their oscillation modes, further complicates the asteroseismic analysis. This is a complications because the line width of the modes in the power spectrum used in the asteroseismic analysis is inversely proportional to the mode lifetime, while the rotational splitting of modes is inversely proportional to the rotation period. 
For the Sun many studies have been carried out using helioseismology, and the rotation profile has here been pinned down in high detail \citep[see, \eg,][]{1998ApJ...505..390S,2003ARA&A..41..599T}. The study of stars like the Sun is naturally important in order to better understand the observations of our own Sun and to put our knowledge here into a greater context.

We have analyzed the impact on the splittings of stellar pulsation frequencies in a main-sequence solar-like model star for a number of differential rotation profiles. This is done in order to see how a distinction between these profiles might be made in asteroseismic data and to get a better feel for what one should expect in the analysis of such stars. In our analysis of the different profiles for differential rotation we will be solving the so-called \emph{forward problem}, in which one assumes to know the rotation profile and the structure of the star and from this calculates the effect from the rotation on the pulsation frequencies. This is opposite to the \emph{inverse problem}, where the structure of the star and the perturbed pulsation frequencies are assumed known and from these the underlying rotation profile is inferred.

The structure of the paper is as follows: In \S~\ref{sec:split} we briefly describe the formalism for rotational frequency splittings in the context of asteroseismology, with emphasis on the so-called \emph{rotation kernels} and the regions of the star upon which the frequency splittings are sensitive in \S~\ref{sec:ker}, and in \S~\ref{sec:optinc} we take a look at which values for the stellar inclination angle will give the best results in an observational context. The stellar model used throughout the paper is described in \S~\ref{sec:model}. In \S~\ref{sec:rates} we give a short presentation of the theory of differential rotation, especially how the different rotational profiles are set up, and we further give an account for the rotation rates used in our analysis. The differential rotation profiles tested are presented in \S~\ref{sec:profiles}, and here the obtained splittings for the profiles are also given. In \S~\ref{sec:depen} we test some dependencies of the splittings for a solar-like differential rotation profile, for instance, the effect of changing the radial or latitudinal gradient of rotation. A rough estimate on the feasibility of detecting differential rotation from splitting is given in \S~\ref{sec:esti}.  
In \S~\ref{sec:con} we finally discuss which main conclusions can be drawn from our results, and which effects other than rotation can affect the observed splittings for a real star.


\section{Rotational splitting}
\label{sec:split}

The splitting of frequencies can come from any effect that distorts the spherical symmetry of the star, \eg\ magnetic fields, large-scale flows, and rotation.
A pulsation mode is characterized by three quantum numbers, viz., the angular degree $l$, giving the number of surface nodal lines, the radial order $n$, giving the number of concentric nodal shells along the radius of the star, and lastly the azimuthal order $m$, giving the number of nodal surface lines that run along constant longitude. The azimuthal order can take on values from $m=-l$ to $m=l$, such that for a mode of a given $l$ there will be $2l+1$ $m$-components. For a non-perturbed star the values of the different $m$-components will be degenerate, but if the star is perturbed from its spherical symmetry, \eg\ by rotation, these $m$-components will split up in frequency. 
In considering only the effects of rotation the splitting of modes can to first order be given as
\begin{align}
\nu_{nl m} = \nu_{nl} &+ m\, \delta\nu_{nlm}  \, ,
\label{eq:split2}
\end{align}
with $\nu_{nl}$ being the central frequency of a given multiplet having azimuthal order $m=0$ (the \emph{zonal} component).
By considering how the pulsation modes interact with the rotation, the perturbation to the frequencies of the pulsation modes, \ie, the rotational splitting $\delta\nu_{nlm}$ can be found. 
This perturbation is generally given by a double integral over radius, $r$, and co-latitude\footnote{The co-latitude goes from $\theta=0^{\circ}$ at the pole to $\theta=90^{\circ}$ at the equator.}, $\theta$ \citep[][]{1977ApJ...217..151H, 1981MNRAS.196..731G}:
\begin{equation}
\delta\nu_{nl m} = \int_0^{\pi}\int_0^R \mathcal{K}_{nl m}(r, \theta)\mathcal{N}(r, \theta) r dr  d\theta \, .
\label{eq:split}
\end{equation}
In this equation $\mathcal{N}(r, \theta)$ is the 2D rotation profile of the star, which we throughout will give in units of cyclic frequency instead of angular frequency, \ie,
\begin{equation}
\mathcal{N}(r, \theta) = (2\pi)^{-1} \Omega(r, \theta) \, .
\end{equation}
$\mathcal{K}_{nlm}(r, \theta)$ in \eqref{eq:split} is the so-called \textit{rotation kernel}. The rotation kernel, or kernel in short, is essentially a weighting function that describes where in the star, \ie, at which $(r,\theta)$ positions, the splittings of the pulsation frequencies are sensitive to the rotation. The rotation kernel is symmetric around the equator, which also means that only the north-south symmetric parts of the rotation profile $\mathcal{N}(r, \theta)$ will contribute to the calculated splittings. 
It follows that in order to actually obtain the largest possible rotational splitting the maximum regions of the kernel must coincide with the maximum regions of the rotation.
The kernel is the condensed result of the perturbation analysis where multiple terms have been collected into one single quantity. It depends largely on (1) the displacement eigenfunctions of the specific modes, which are found in the adiabatic calculation of the mode frequencies; (2) the mean (spherically symmetric) stellar density profile; and (3) a co-latitude dependence from the associated Legendre polynomials entering the spherical harmonic description of the pulsation modes.       
We refer the reader to Appendix~\ref{app:ker} for a detailed account of the construction of the rotation kernels. We will work with frequency splittings via the generalized symmetric splitting given by \citep[see, \eg,][]{2004ESASP.538..133G}
\begin{align}
S_{nl m} &= \frac{1}{2|m|}\left(\nu_{nl m} - \nu_{nl -m} \right)=\delta\nu_{nlm}\, .
\end{align}
The reason for using this construct is that \eqref{eq:split} only accounts for first-order effects of rotation. Determining $\delta\nu_{nlm}$ simply from $\nu_{nl m} - \nu_{nl0}$ can be biased by terms to second order in $m$, as these will offset the $m=0$ component of a multiplet from the midpoint between the $\pm m$ components. This asymmetry can be given as $A_{nlm} = \nu_{nl-m} + \nu_{nlm} - 2\nu_{nl0}$ -- see \citet[][]{2010AN....331.1073S} for details on how this asymmetry in fact can be used to probe the stellar rotation profile. This effect, in addition to the $m$ dependence of the splitting, is taken out by using $S_{nl m}$, enabling a more reliable comparison between different $(n,l,m)$ combinations.


\subsection{Rotation Kernels}
\label{sec:ker}
\begin{figure*}[ht]
\centering
\subfigure[$l,n,m$ = (1,20,1)]{
   \includegraphics[width=0.3\textwidth, trim = 16mm 0mm 16mm 0mm, clip] {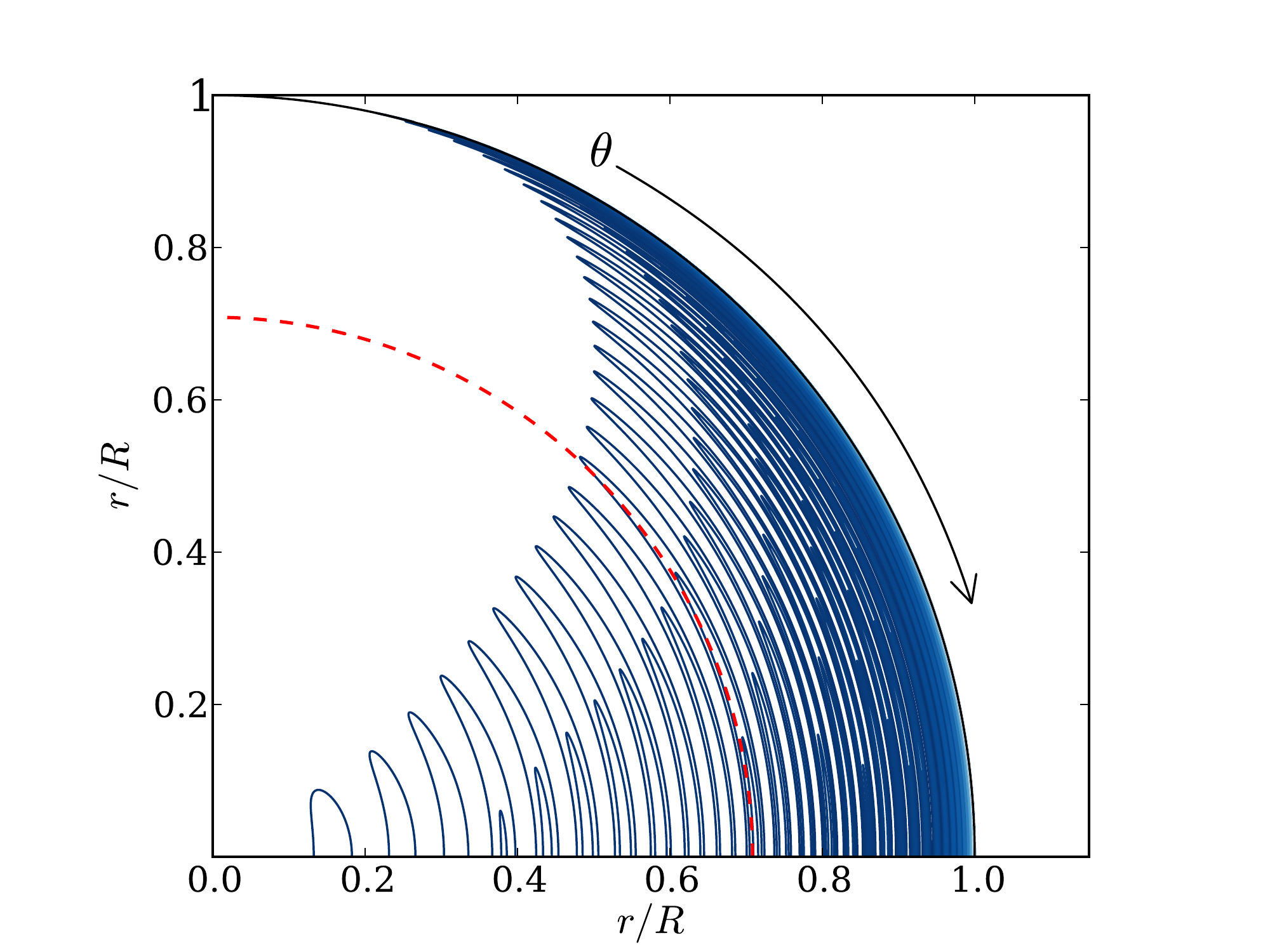}
   \label{fig:l1m1n20}
 }
 \subfigure[$l,n,m$ = (2,20,1)]{
   \includegraphics[width=0.3\textwidth, trim = 16mm 0mm 16mm 0mm, clip] {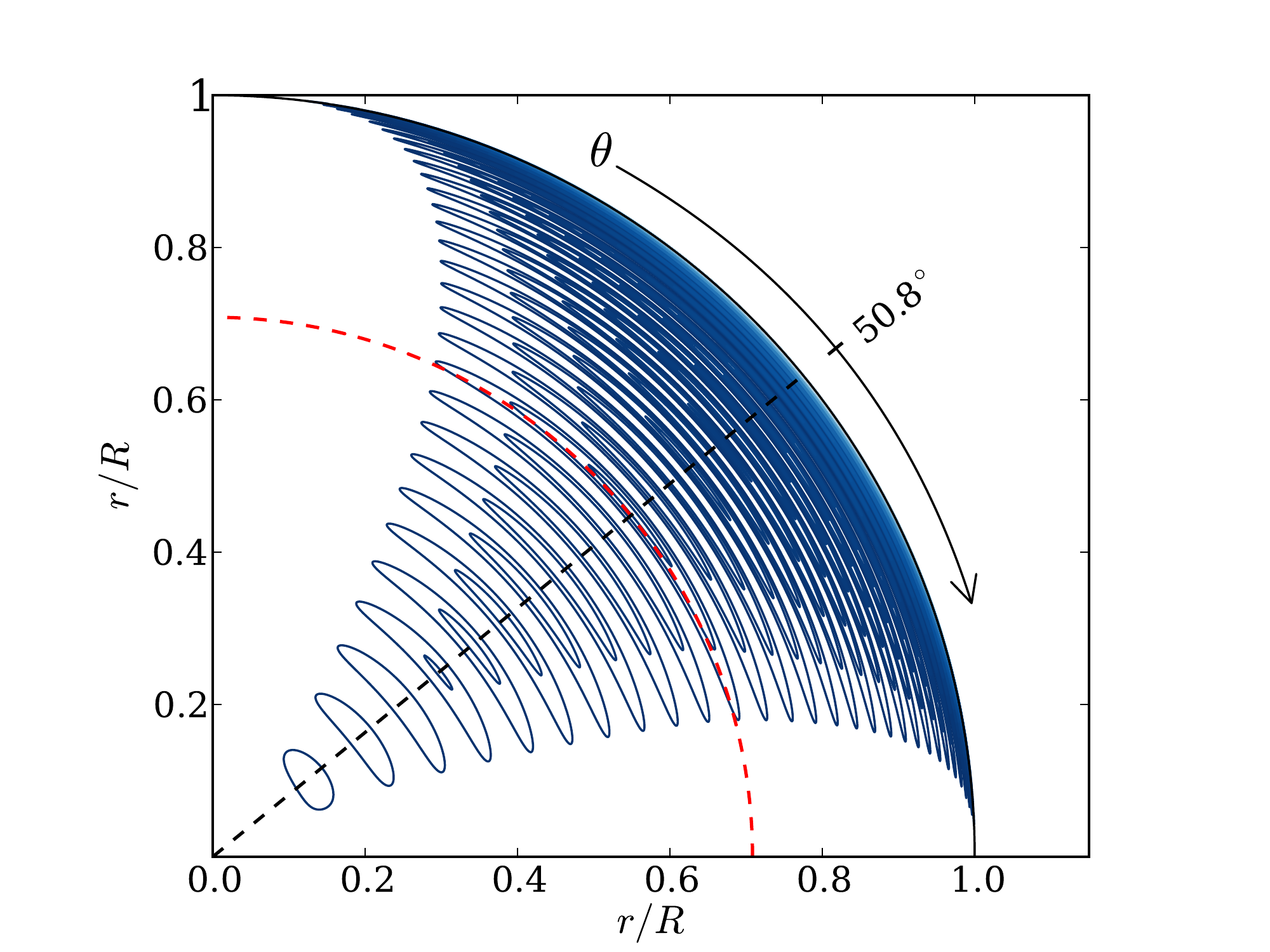}
   \label{fig:l2m1n20}
 }
 \subfigure[$l,n,m$ = (2,20,2)]{
   \includegraphics[width=0.3\textwidth, trim = 16mm 0mm 16mm 0mm, clip] {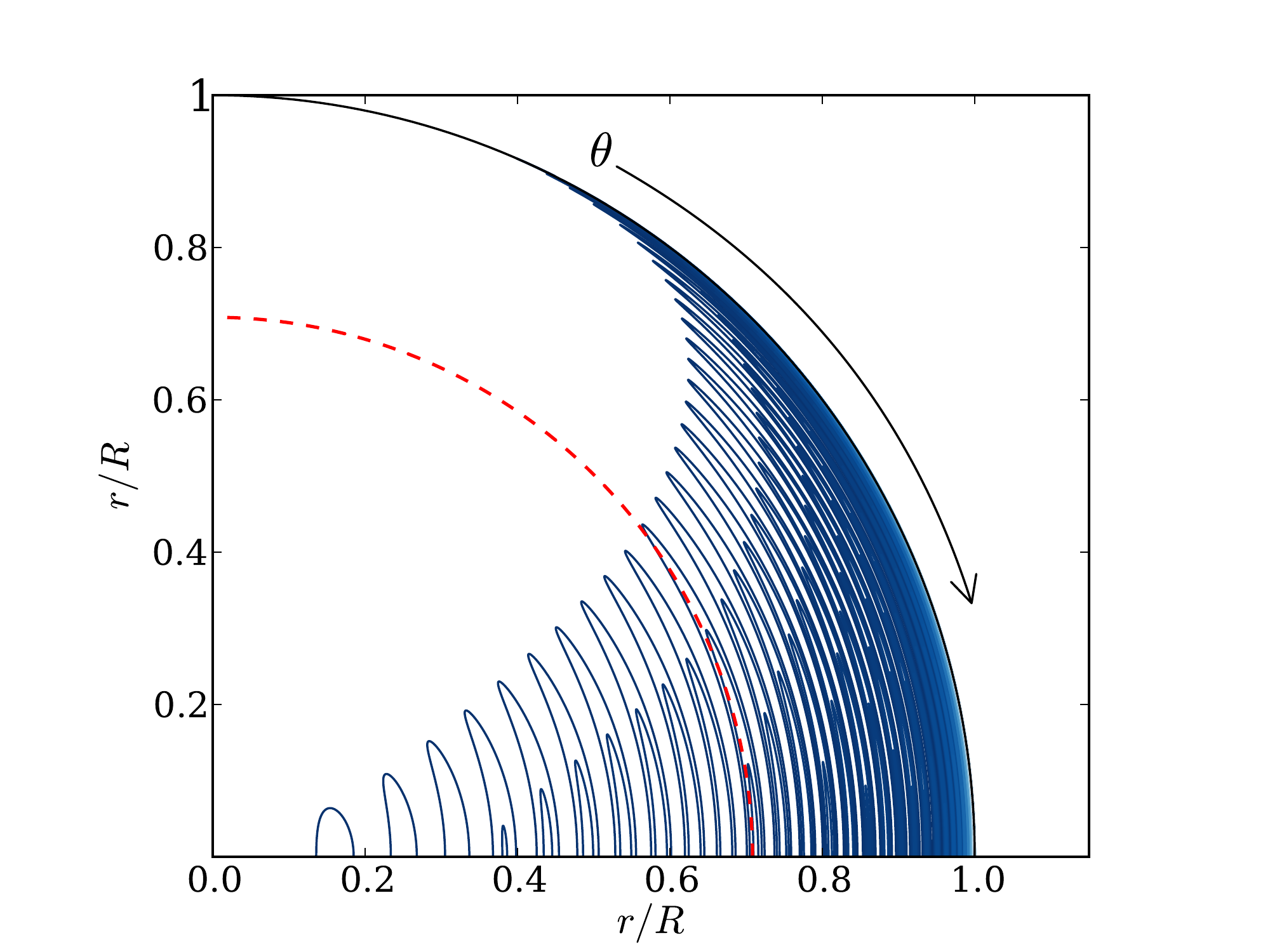}
   \label{fig:l2m2n20}
 }

\subfigure[$l,n,m$ = (3,20,1)]{
   \includegraphics[width=0.3\textwidth, trim = 16mm 0mm 16mm 0mm, clip] {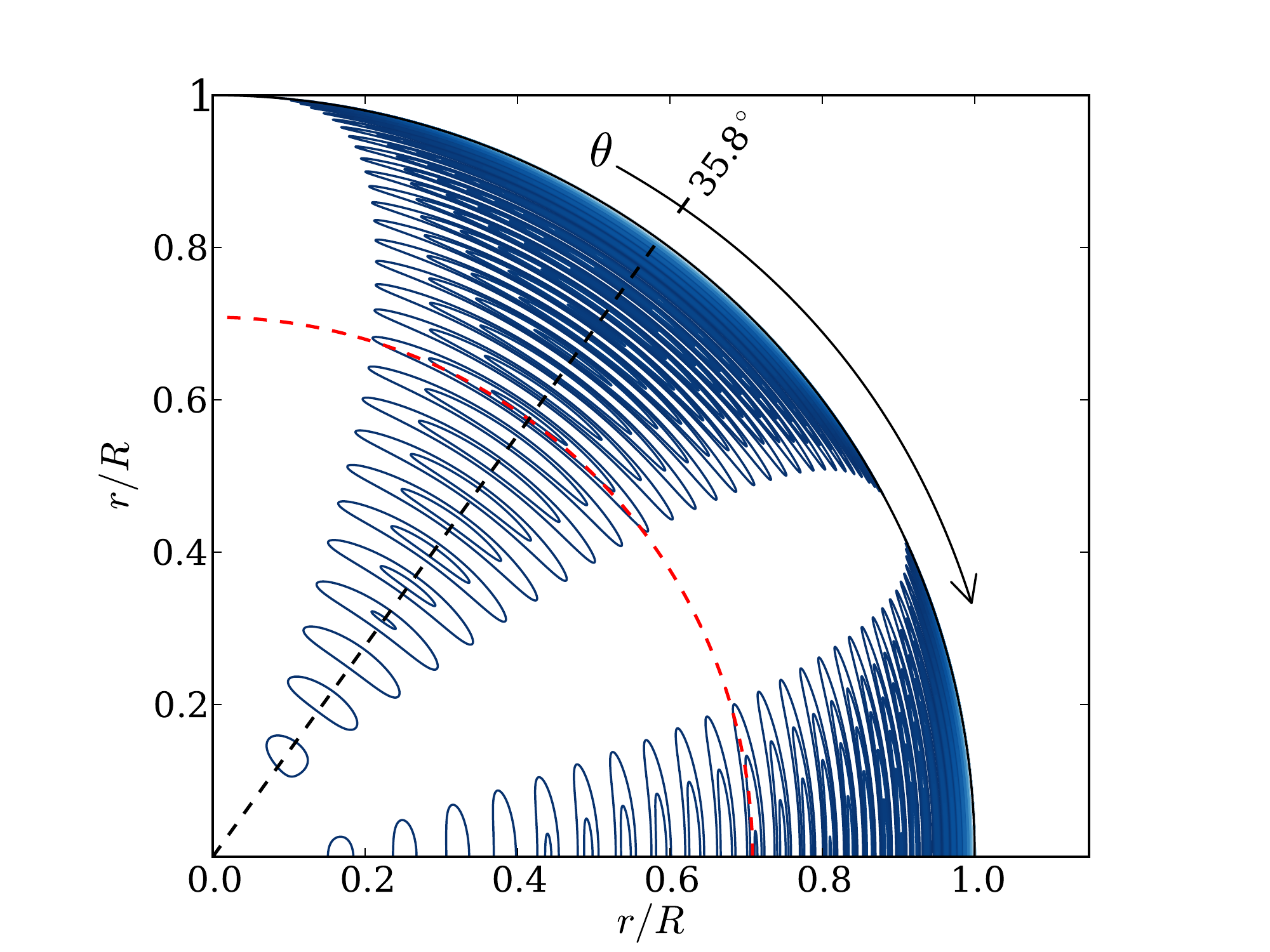}
   \label{fig:l3m1n20}
 }
 \subfigure[$l,n,m$ = (3,20,2)]{
   \includegraphics[width=0.3\textwidth, trim = 16mm 0mm 16mm 0mm, clip] {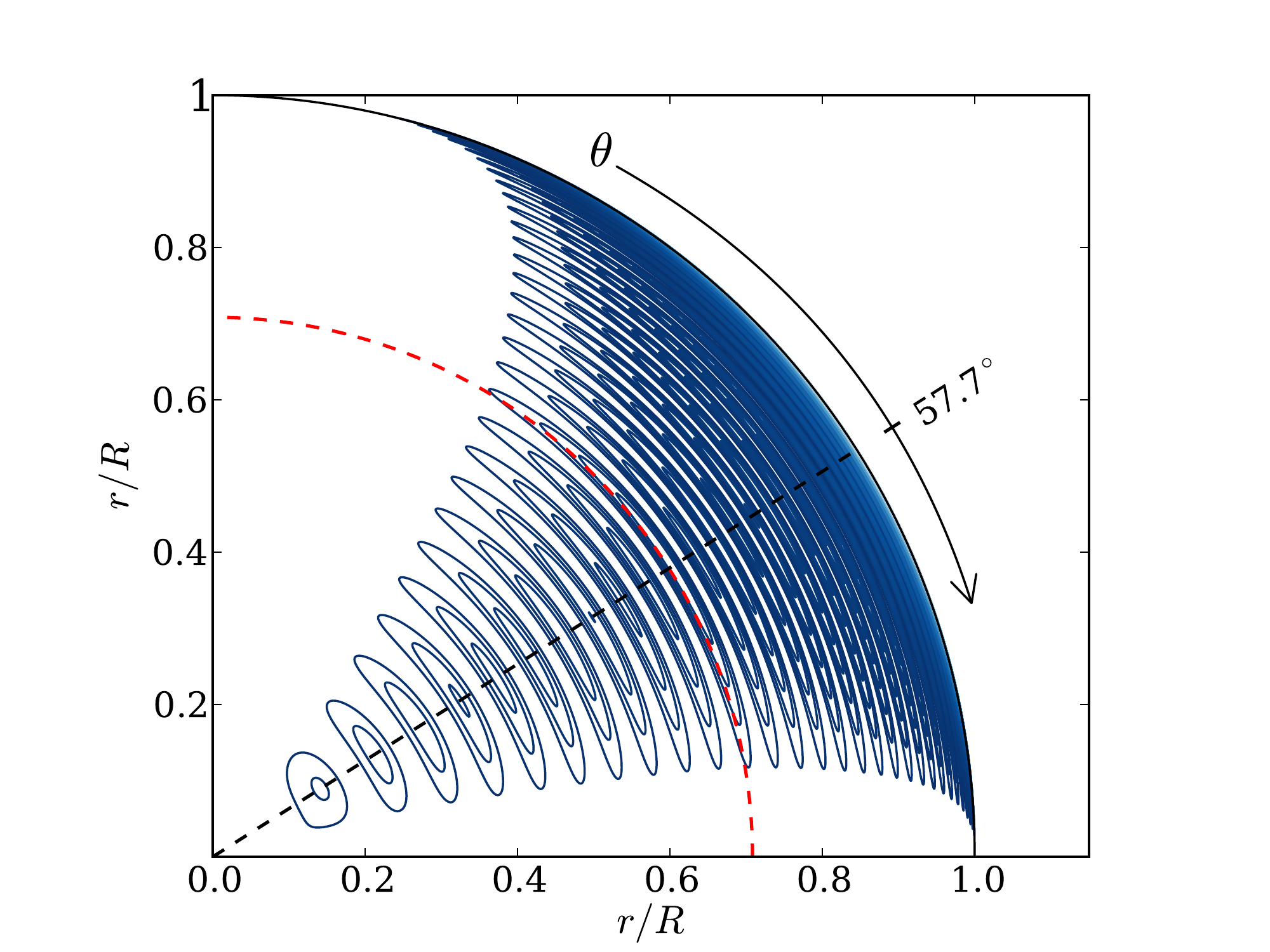}
   \label{fig:l3m2n20}
 }
 \subfigure[$l,n,m$ = (3,20,3)]{
   \includegraphics[width=0.3\textwidth, trim = 16mm 0mm 16mm 0mm, clip] {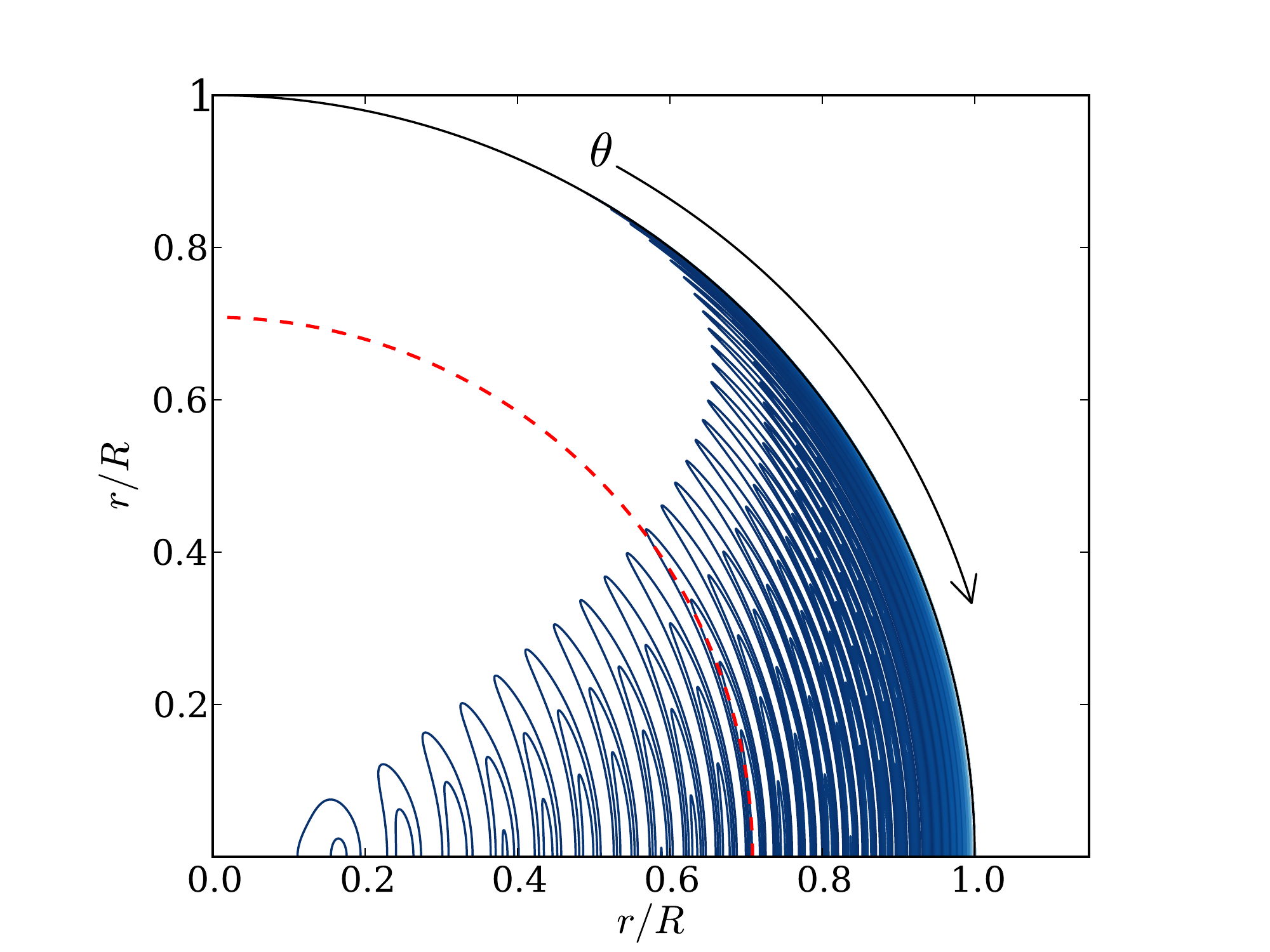}
   \label{fig:l3m3n20}
 }

\caption{\footnotesize Contour plots of the rotation kernels for modes of degree $l=$ 1, 2, 3, all with radial order $n=\,$20. Only one quadrant of the star is shown, and in units of the stellar radius. The displayed kernel may by mirrored in both axes. The dashed red circle indicates the base of the convection zone, $r_{\rm bcz}$, for the model considered. For kernels where the maximum in latitude is different from the equator ($\theta=\,$90$\, ^{\circ}$), a dashed line indicates the co-latitude of kernel maximum.}
\label{fig:kernels}
\end{figure*}

Much insight on the behavior of the splittings can be drawn directly from the rotation kernels. In \fref{fig:kernels} we show contour plots of the kernels from acoustic pulsation modes of a solar-like stellar model, all with radial order $n=20$ and $l=1,2,3$ (see \S~\ref{sec:model} and Table~\ref{tab:models} for details on the stellar model used). 
The base of the convection zone of the model, at $r_{\rm bcz} = 0.71\, R$, is given by the dashed red line. Note that only the first quadrant of the star has been shown, and so the pattern seen here is mirrored around the axes.

The kernels from different $(n,l,m)$ combinations are somewhat different with respect to where they have their highest sensitivity. In short, the \emph{sectoral} mode components (having $m= \pm l$) are sensitive to the equatorial region, while \emph{tesseral} components ($0<|m|<l$) are (at least) sensitive to higher latitudes. Also, the latitudinal extent of the kernels decreases with the degree of the mode. 
For all the kernels the minimum in the radial direction of the contour can to some extent be thought of as the $n$ nodal shells through the stellar radius. However, as the horizontal component of the fluid displacement also enters in the description of the kernel (see \eqref{eq:kernel_ref}), the minimum will generally not coincide with the nodal shells.  
\begin{figure*}[htb]
\centering
\includegraphics[scale=0.40]{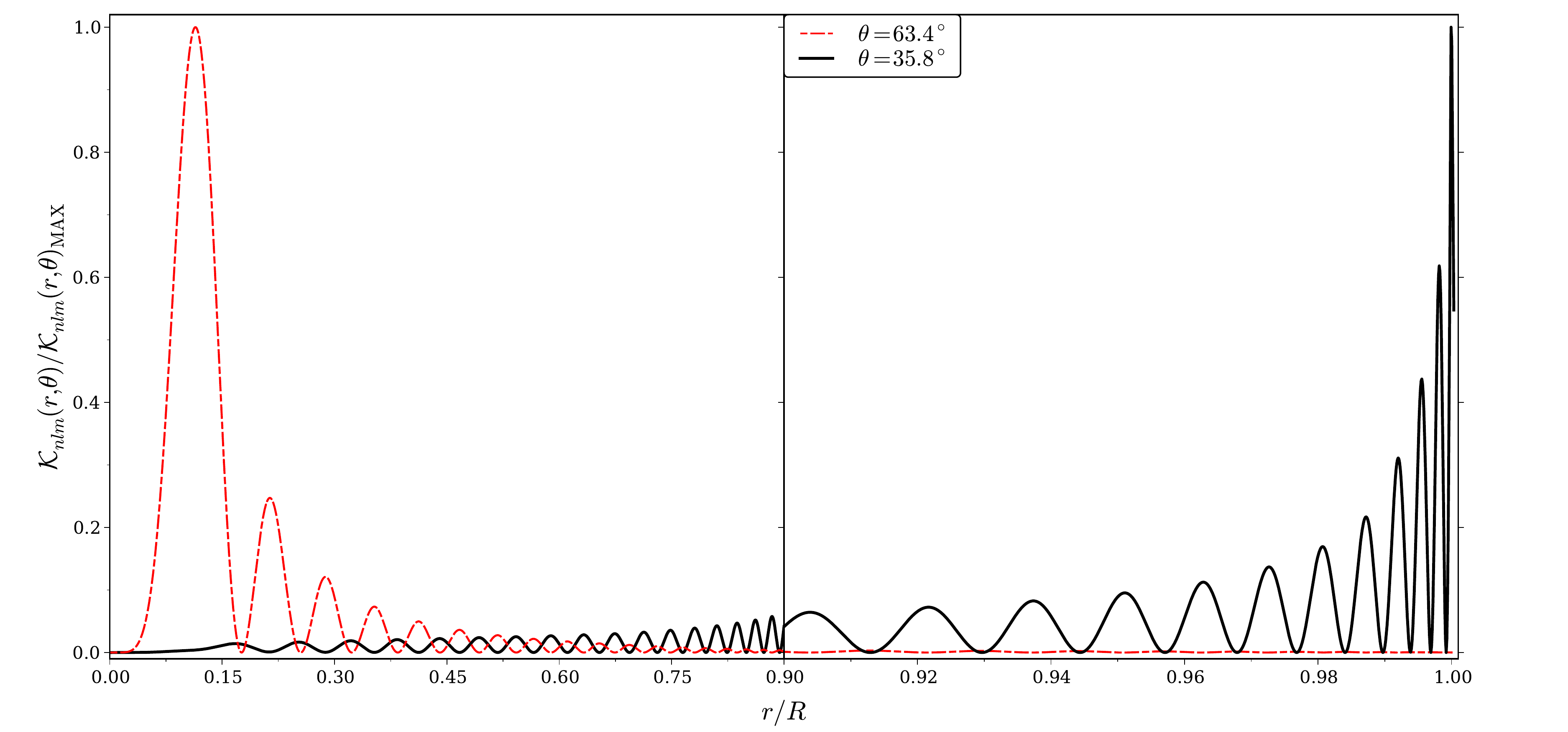}
\caption{\footnotesize Normalized (to maximum) radial profiles for the $l=\,$3, $m=\,$1, $n=\,$20 kernel at two specific co-latitudes. The profile for the kernel where it has its lowest latitudinal amplitude is given in dashed red, while the full black curve depicts the profile for the co-latitude of maximum kernel amplitude. The left part of the figure gives the profile up to $r=\,$0.9$\,R$, while the right part highlights the region from $r=\,$0.9$\,R$ to the surface. The absolute values for the maximum kernel amplitudes are 37.3 ($\theta=\,$35.8$^{\circ}$) and 0.33 ($\theta=\,$63.4$^{\circ}$).}
\label{fig:Specific_kernel}
\end{figure*}
It is clear for all these modes that the kernels are mainly sampling the outer parts of the star. 
In \fref{fig:Specific_kernel} we show the kernel of the $l=3$ $m=1$ mode at two specific co-latitudes, namely, the $\theta$ with highest ($\theta=35.8^{\circ}$) and lowest ($\theta=63.4^{\circ}$) amplitudes of the radius-integrated kernel and both normalized to a maximum value of 1. For the co-latitude with the highest amplitude we see that the bulk part of the kernel is indeed in the outer regions close to the stellar surface, while for the low-amplitude co-latitude a large portion of the kernel is in fact situated close to the stellar core. From \eqref{eq:kernel_fin} it can be found that at the co-latitude of lowest kernel amplitude (\ie, $\theta=63.4^{\circ}$) only the horizontal displacement, $\xi_h$, contributes, whereby the corresponding curve in \fref{fig:Specific_kernel} essentially shows a scaled version of the squared horizontal displacement, with $\xi_h$ being largest where the mode has its inner turning point\footnote{Located where the Lamb frequency equals the oscillation frequency $S_l(r_t)=\omega$, giving $r_t = c(r_t)\, \omega^{-1}\sqrt{l(l+1)}$, where $c$ is the local sound speed \citep[see Equation (3.189) and Figure 7.13 of][]{2010aste.book.....A}.}. For the latitude with highest kernel amplitude both the horizontal and radial displacement contributes, but with the radial displacement greatly dominating over the horizontal.

\begin{figure*}
\centering
\includegraphics[scale=0.33]{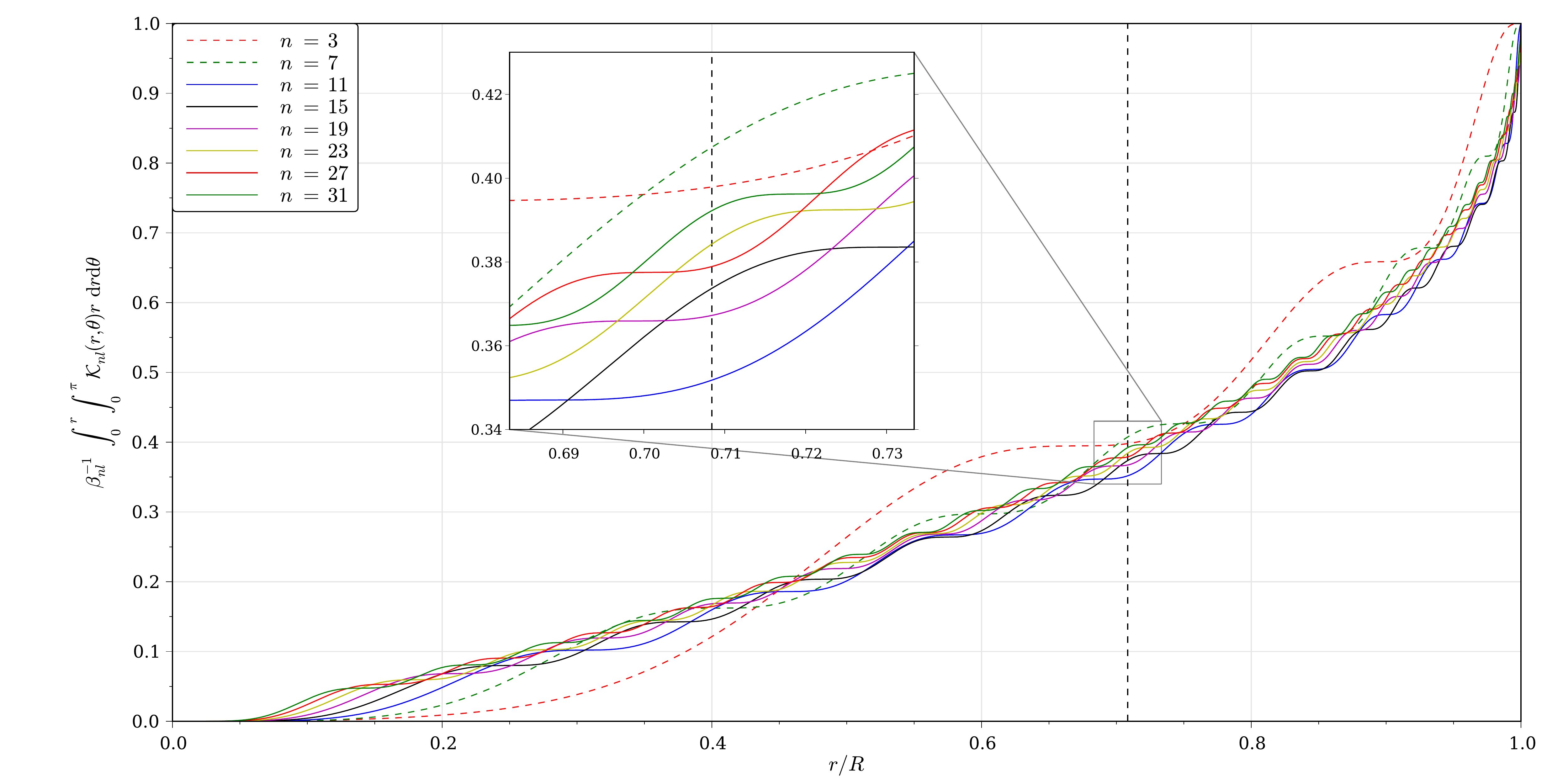}
\caption{\footnotesize Cumulative distribution function (maximum of 1) for the radius-weighted $l=\,$1, $m=\,$1 kernel as a function of radius for a number of radial orders (see legend). The dashed vertical line indicates for the stellar model used the division between radiative interior and the convective envelope, $r_{\rm bcz}$. The zoom box shows the behavior of the cumulative sum at $r_{\rm bcz}$.}
\label{fig:cumkern}
\end{figure*}

To see how the kernels are located in radius, we show in \fref{fig:cumkern} the integrated profile (IP) for the radius-weighted kernel where an integration has been performed over co-latitude. The base of the convection zone for the specific model considered is indicated by the vertical dashed line. The reason for highlighting the base of the convection zone is that for the Sun this base approximately divides the rotation profile into a solid-body rotation in the radiative zone beneath and a differentially rotating convection zone.
As seen, the kernels become increasingly concentrated toward the outer parts for the star with higher radial order. 
The subset of radial orders shown here have in general about $37\%$ of the radius-weighted kernel IP below the base of the convection zone.


The zoom box in this figure shows that the kernel IP within the base of the convection zone does not keep decreasing as one goes to higher and higher radial orders, but instead seems to have a minimum at around $n\sim15-23$ (for the shown subset of $n$-values) and then increases slightly for higher $n$-values.

\begin{figure*}[htb]
\centering
\includegraphics[scale=0.35]{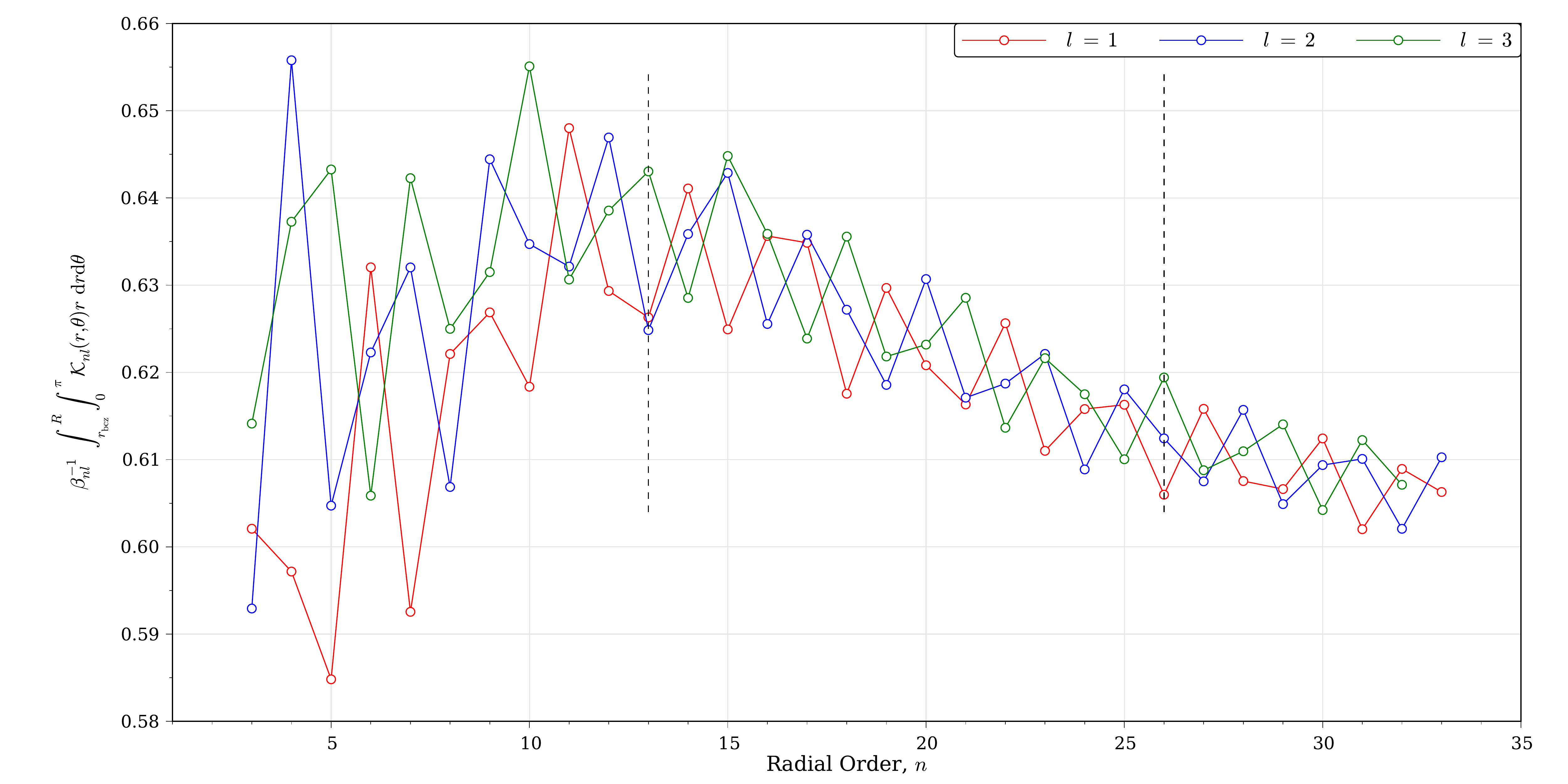}
\caption{\footnotesize Radius-weighted kernel IP confined in the convective envelope, \ie, $r \geq r_{\rm bcz}$, for rotation kernels of degree $l=$ 1, 2, 3 as a function of radial order. The frequency span for the modes can be seen in \fref{fig:models}. The vertical dashed lines indicate the range of radial orders ($n\sim$ 13$-$26) where modes are generally detectable with high enough S/N that they can be used in an asteroseismic analysis.}
\label{fig:cumkern_vol}
\end{figure*}

This minimum in $n$-value is seen more clearly in \fref{fig:cumkern_vol} (appearing here as a maximum), where the total radius-weighted kernel IP in the convective envelope has been computed as a function of radial order, and for the different mode degrees. The same pattern is seen as in \fref{fig:cumkern}, namely relatively little kernel IP in the envelope at low radial order, which rises steeply to a peak (equivalent to minimum below the base of convection zone) at around $n\sim13$, followed by a gradual decrease toward higher radial orders. The lowest radial orders are, however, not within reach in the present asteroseismology data, where a value of $n\sim 13$ in general sets a lower limit on the radial order and $n\sim 26$ sets the upper limit for solar-like stars \citep[see, \eg,][]{2012ApJ...748L..10M}. 

The lower limit for the available radial orders is to a large extent set by the noise background from stellar granulation \citep[see, \eg,][]{1985ESASP.235..199H,2009A&A...495..979M,2011ApJ...741..119M}, faculae \citep[][]{2012MNRAS.421.3170K}, and activity. These effects all increase toward low frequency, resulting in a decrease in the signal-to-noise ratio (S/N) for low radial order modes. 
In a star like the Sun the modulation of heights in the power spectrum as a function of frequency is generally well approximated by a Gaussian envelope centered on the frequency of maximum power, \numax\ \citep[see, \eg,][]{1988ApJ...334..510L,2008ApJ...687.1180A}. The modulation is set by the competition between mode excitation and mode damping. The mode damping can be attributed to many different effects, but the total combined damping rate is an overall increasing function of frequency \citep[see][and references therein]{1999A&A...351..582H}. The modes of solar-like oscillators are stochastically excited by turbulent convection in the outer envelope of these stars. Modes of different frequency propagate to different depths in the star, with the depth of the outer reflection point of the mode increasing with decreasing frequency. Also, properties such as the energy, size, and velocity of the turbulent convective eddies responsible for the excitation of the modes are dependent on their depth in the convection zone. Stated in a simplified manner \citep[][]{1990LNP...367...75O}, the outer reflection point for modes of decreasing frequency moves increasingly farther away from the depth at which efficient excitation by the turbulent eddies can be achieved -- so, even though the modes at low frequency and low $n$ are damped to a small extent, the lack of efficient excitation results in a decrease in the power of these modes. At high frequencies the modes propagate in the region where the eddies can excite, but the eddies are small and of little energy -- this, combined with the increased damping, results in a decrease in power at high frequencies \citep[][]{1992MNRAS.255..603B,1994ApJ...424..466G}.
At the frequency of \numax\ a depression is furthermore seen in the damping rate (and hence also in the mode line width $\Gamma_{nl}$) \citep[see, \eg,][]{1999A&A...351..582H}, and the excitation energy from turbulent eddies peaks. Both effects lead to an increase in power at \numax. The depression in the damping rate at \numax\ seems to stem from a resonance between the frequency of modes close to \numax\ and the thermal timescale of the gas \citep[see][and references therein]{2011A&A...530A.142B}.

The span in the radius-weighted kernel IP in the envelope in this interval ($n\sim 13-26$) of radial orders is about $2\%$, which directly tells us that inference on the radial structure of the differential rotation is difficult. It is furthermore clear that there is a small difference between the different degrees, with higher $l$ modes generally having more of their IP in the envelope. This is because higher degree modes have shallower propagation zones with the turning point for total internal refraction closer to the stellar surface.   

The wiggles in the curves in \fref{fig:cumkern_vol} come from the positioning of the base of the convection zone combined with the oscillations in the kernel IP. From the insert in \fref{fig:cumkern} it can be seen that if the base of the convection zone is shifted slightly up or down in radius, the pattern in the wiggles will change and possibly reverse; a change to $r_{\rm bcz}=0.69$ would, \eg, make the IP of the $n=23$ kernel larger than that of the $n=19$ kernel at $r_{\rm bcz}$; the opposite is the case for the value of the $r_{\rm bcz}$ indicated by the dashed vertical line.

The fact that different modes are sensitive to different regions of the star is what enables inference of the differential rotation. It is, however, clear that many modes probe very similar regions of the star, so the combined information from the splittings of these modes will give very little information on the differential rotation.  The difference in the value of $\theta$ where each mode peaks, illustrated in \fref{fig:kernels}, is what enables the extraction of the latitudinal differential rotation. 

Note that the minimum co-latitude $\theta$ for kernel maximum is found for the $l=3, m=1$ modes, and hence information from co-latitudes lower than $\theta\sim36^{\circ}$ is generally not possible from asteroseismic data. 
The very small differences in the positioning in radius of the mode kernels, illustrated by the kernel IP in the envelope versus the radiative zone in \fref{fig:cumkern_vol}, is what might enable information on the radial differential rotation to be extracted. Due to the small difference, it might only be possible to obtain a measure of the amplitude of the variation or the sign of the gradient between the radiative zone and the convection zone.


To better understand from which regions of the star we might obtain information on the differential rotation and which data should be used to accomplish this, we now take a look at the differences in the $S_{nl m}$ values. The value of a specific $S_{nl m}$ can be associated with the rotation at a given region in the star, and so the difference between these probed regions can give us the difference in rotation, \ie, the differential rotation.

\begin{figure*}[t]
\centering
\includegraphics[scale=0.45]{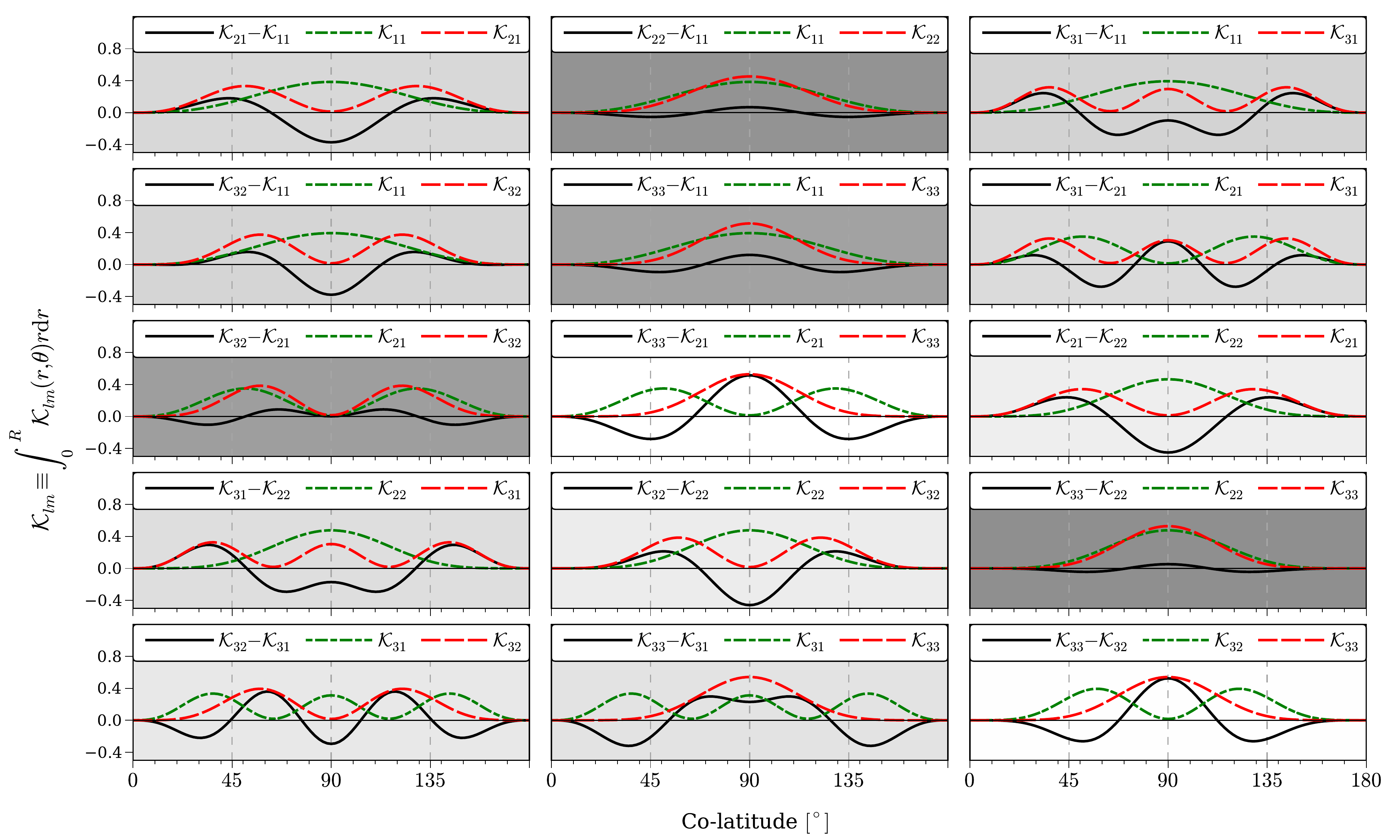}
\caption{\footnotesize Differences (black) between all pairs (red and green) of radially integrated kernels as a function of co-latitude. The legends indicate the $(l,m)$ combination for the kernel, $n=\,$20 for all kernels. The gray scale of the background indicates the peak-to-peak difference between the respective pairs of kernels, going from white (large difference) to dark gray (small difference). }
\label{fig:cumkern_comp}
\end{figure*}

We can see which latitudes a given kernel is sensitive to simply by displaying the kernel as a function of co-latitude and at each angle integrated over radius. The best sets of splittings to measure will be the ones where the difference between the radius-integrated kernels is the greatest.
In \fref{fig:cumkern_comp} all combinations of such kernel differences are given (for a degree up to $l=3$), with the gray scale of the background indicating the size of the peak-to-peak value in the difference between the two radially integrated kernels (black curve), going from dark (small difference) to light (large difference). Thus, the mode pairings indicated with light backgrounds are the optimal parings for inferring the latitudinal differential rotation.

Splitting differences of the kind $S_{n,l,l}-S_{n,l',l'}$ give very little information (\fref{fig:cumkern_comp}) as both kernels are concentrated to the equatorial region (\fref{fig:kernels}). The most information on latitudinal differential rotation can be obtained from the differences in splittings given by $S_{n,3,3}-S_{n,2,1}$ and $S_{n,3,3}-S_{n,3,2}$. The reason why the $l=3, m=3$ mode is so well represented is that it is the equatorially concentrated kernel that has the smallest spread in $\theta$. The $S_{n,2,1}$ kernel has a broader peak and thereby covers more $\theta$ values than the $S_{n,3,2}$ kernel, but on the other hand it is concentrated at $\theta$ values farther away from the equator than the $S_{n,3,2}$ kernel.

In the case in which no $l=3$ modes can be detected, which often would be the case, the optimum splitting pair is $S_{n,2,1}-S_{n,2,2}$. This is followed by $S_{n,2,1}-S_{n,1,1}$ and then finally $S_{n,2,2}-S_{n,1,1}$, where virtually no information is available.

From the above we see that in the case where $l=3$ modes can be observed one can potentially get information on the differential rotation between the equator and a latitude of ${\sim}54^{\circ}$ (co-latitude of $35.8^{\circ}$ for $S_{n,3,1}$), while the maximum latitude reachable is ${\sim}40^{\circ}$ when only $l=1,2$ modes are observable.

\subsection{Optimum Inclination Angles}
\label{sec:optinc}

In the previous section we determined which sets of modes contain the most information on the latitudinal differential rotation.
It is worthwhile to see how easily the splittings of these modes can actually be observed, here focusing on the impact of the stellar inclination angle. The heights in the power spectrum of individual azimuthal components ($m$) within a rotationally split multiplet can, when assuming equipartition of energy, be given as follows:
\begin{equation}
H_{nl m} = \mathcal{E}_{l m}(i)H_{nl} = \mathcal{E}_{l m}(i) V_{l}^2 \alpha_{nl} \, ,
\label{eq:vis}
\end{equation}
where the factor $V_{l}^2$ is the relative visibility (in power) of a mode with a given degree relative to the radial and non-split $l=0$ modes \citep[see][]{1977AcA....27..203D}. The parameter $\alpha_{nl}$ represents the (approximate Gaussian) frequency-dependent amplitude modulation (\S~\ref{sec:ker}).
The function $\mathcal{E}_{l m}(i)$ describes how the stellar orientation, represented as the inclination of the stellar rotation axis relative to the observers' line of sight, modulates the relative heights of the different azimuthal components in a rotationally split multiplet. Once again assuming equipartition of power between the modes in the multiplet, the following expression for the function applies \citep{1977AcA....27..203D, 2003ApJ...589.1009G}:

\begin{equation}
\mathcal{E}_{l m}(i) = \frac{(l - |m|)!}{(l + |m|)!}\left[P^{|m|}_{l} (\cos i) \right]^2 \, .
\label{eq:epsilon}
\end{equation}
Here $P^{|m|}_{l} (\cos i)$ are the associated Legendre functions and $i$ is the inclination, going from $i=0^{\circ}$ for an pole-on view to $i=90^{\circ}$ for a equator-on view \citep[see, \eg,][Appendix B]{2010aste.book.....A}.

\begin{figure*}[t]
\centering
\includegraphics[scale=0.45]{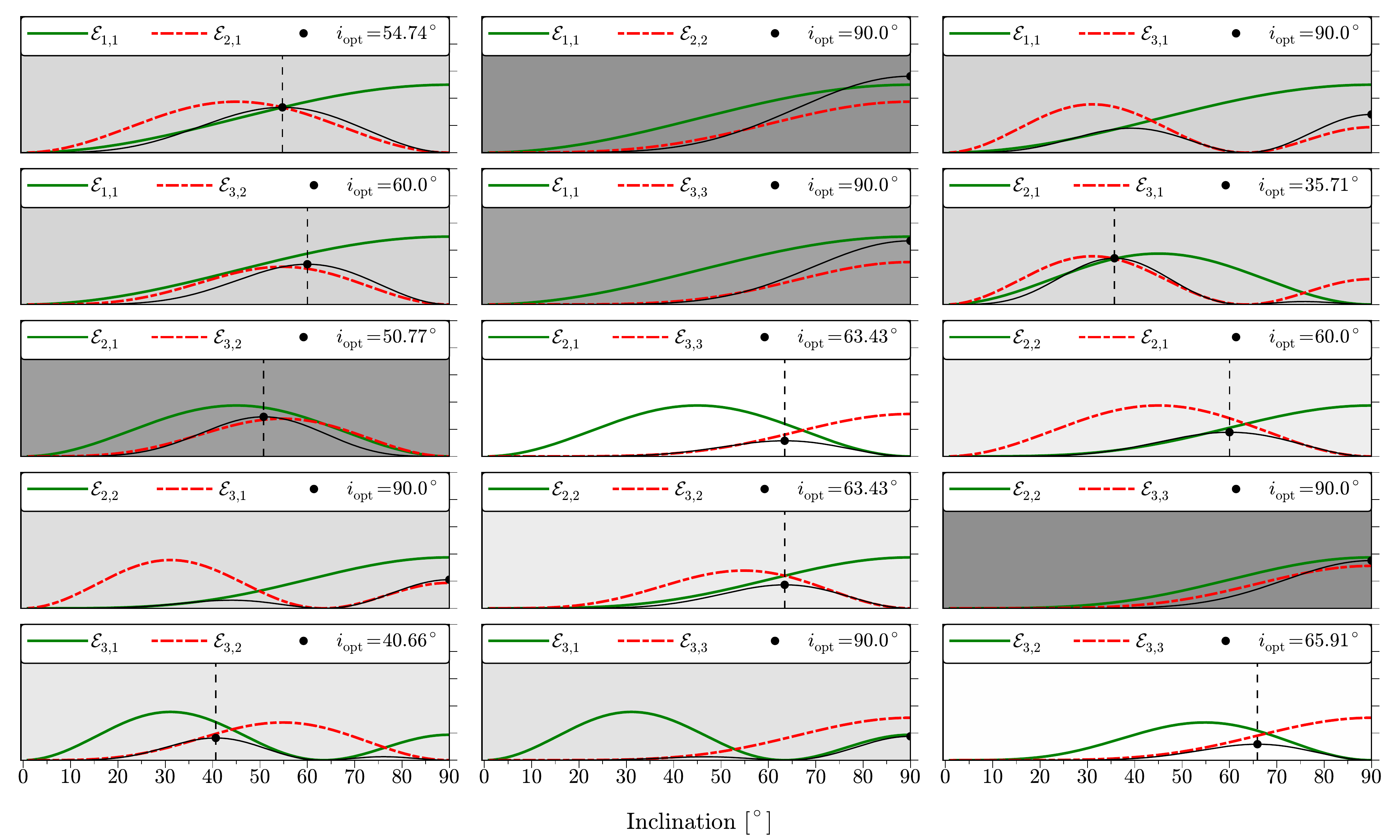}
\caption{\footnotesize Product of $\mathcal{E}_{l m}(i)$ for the different splitting combinations in arbitrary units as a function of the stellar inclination angle. As in \fref{fig:cumkern_comp}, the background gray scale indicates the available information content on latitudinal differential rotation for the respective splitting combinations, with more information the lighter the gray scale. The product curve is given in black (multiplied by a factor of three), while the curves for the individual $(l,m)$-modes are given in either red or green (see legends). The optimal inclinations, given by the peak of the respective product curves, are indicated by a black dot and a vertical dashed line. The scale on the ordinate goes from 0 to 1 in all tiles. \\}
\label{fig:geo_vis2}
\end{figure*}

To find the optimum inclination angle for the observation of a specific splitting difference, we find for all the combinations (\fref{fig:cumkern_comp}) the value of  $\mathcal{E}_{l m}(i)$ and then compute the product of this value from the two components entering the splitting difference. The optimum angle is then taken to be the inclination at maximum of the resulting product curve. The reason for taking the product of the curves from the two components is that both curves need to contribute, such that an inclination with a high signal for one component but with little signal for the other will be down-weighted in the product.
The products of $\mathcal{E}_{l m}(i)$ for the respective combinations are shown in \fref{fig:geo_vis2}, where the color of the background again, as in \fref{fig:cumkern_comp}, illustrates the information content on differential rotation from the specific mode combination, with light backgrounds having the most information. The optimum inclination angle for all tiles is given by a black circle and a dashed vertical line.

The optimum inclination angles for measurements of the splittings with the most information content are $i\sim 63.4^{\circ}$ ($S_{n,3,3}$-$S_{n,2,1}$) and $i\sim 65.9^{\circ}$ ($S_{n,3,3}$-$S_{n,3,2}$). The same region in inclination around $i=60^{\circ}-70^{\circ}$ holds for the two splitting combinations next in line, namely, $S_{n,2,2}$-$S_{n,3,2}$ ($i\sim 63.4^{\circ}$) and $S_{n,2,2}$-$S_{n,2,1}$ ($i\sim 60^{\circ}$). Here $S_{n,2,2}$-$S_{n,2,1}$ is again the best combination when no $l=3$ modes can be used in the measurement.

If we take as a premise that stars are randomly oriented in space, the distribution of inclination angles will be isotropic and uniform in $\cos i$, and we can assert the probability of an inclination angle that by chance lies in the interval $i=60^{\circ}-70^{\circ}$. The probability $P(60^{\circ}<i<70^{\circ})$ will be given by $P(i>60^{\circ}) P(i<70^{\circ})=\cos 60^{\circ} (1-\cos 70 ^{\circ})\approx 0.33$; thus, the stars will, by chance, be oriented in a close-to-optimal way ${\sim}33\%$ of the time.  

In the case where the mean reduced rotational splitting, $\langle S_{nlm} \rangle$, is very low, the above optimum inclinations could change slightly. The problem arises when the rotational splitting becomes comparable to the line width of the specific mode. For $l=3$ modes the $m=0$ component has a maximum in the same regime in inclination where the best information on differential rotation can be gained, that is to say, around $i\sim 65^{\circ}$ (see \fref{fig:geo_vis2}). However, as the mode that mainly will be affected by the $m=0$ peak, namely, the $m=1$ component ($S_{n,3,1}$), does not enter into any of the splitting combinations with much information on differential rotation, the problem is not great for $l=3$ modes. The same goes for $l=2$, where the $m=0$ component has a minimum at $i\sim 55^{\circ}$, close to the optimum angles found above. In all circumstances it will still be true that as the splitting becomes comparable to the mode line widths it will be increasingly difficult to measure the desired splittings \citep[see, \eg,][]{2008A&A...486..867B}.   

When dealing with the measurement of rotational splittings to get an overall estimate of the stellar rotation rate, the best situation would generally be an inclination of $i=90^{\circ}$ \citep[][]{2003ApJ...589.1009G}. Here the sectoral modes ($m=\pm l $) of $l=1$ show their maximum heights, the $m=0$ component is not visible, and the splitting is easier to measure. Furthermore, second-order effects on the splittings can largely be ignored in this scenario \citep[][]{2010AN....331..933B,2014ApJ...782...14V}. However, from the above we see that when the aim is to measure differential rotation from the splitting one should not hope for an inclination angle of the star close to equator-on as very little information will be available here; rather, one should opt for targets with inclinations in the range $i=60^{\circ}-70^{\circ}$.


\section{Stellar Model}
\label{sec:model}

For the model used in the calculation of kernels we have opted for a model that is close to the Sun, \ie, a star with roughly the same mass and radius.
The stellar evolution model used in this study was calculated using the Aarhus STellar Evolution Code \citep[ASTEC;][]{2008Ap&SS.316...13C}. In the evolution of the model no magnetic fields, rotation, or mass loss were included.
The OPAL equation of state \citep{1996ApJ...456..902R} was used in the computations along with the NACRE \citep{1999NuPhA.656....3A} reaction rates. At temperatures above $10^4$ K the OPAL opacity tables were used \citep{1996ApJ...464..943I}; at lower temperatures the \citet{2005ApJ...623..585F} opacities were adopted.
Convection is treated according to the mixing-length theory \citep[MLT;][]{1958ZA.....46..108B} using a mixing-length parameter of $\rm \alpha_{ML} = 2.0$ in units of the pressure scale height. 
Diffusion and settling of helium and heavy elements were not taken into account.
For the initial abundance (in mass units) of hydrogen (\textit{X}$_{\rm 0}$) and heavy elements (\textit{Z}$_{\rm 0}$) we use \textit{X}$\rm _0=0.72$ and \textit{Z}$\rm _0 = 0.02$. 
In Table~\ref{tab:models} the corresponding $\rm [Fe/H]$ value is given using the \cite{2009ARA&A..47..481A} reference value ($X_{\sun}/Z_{\sun}=0.0181$) for the present-day solar photosphere.

At a selected evolutionary step oscillation frequencies were calculated with the Aarhus adiabatic oscillation code \cite[ADIPLS;][]{2008Ap&SS.316..113C}.
From this adiabatic calculation we obtain the displacement eigenfunctions (density weighted) that are used in the computation of the rotation kernels for the respective modes (see Appendix~\ref{app:ker}). 
The information for the model considered can be found in Table~\ref{tab:models}.

\begin{figure*}
\centering

\subfigure{
   \includegraphics[width=0.44\textwidth] {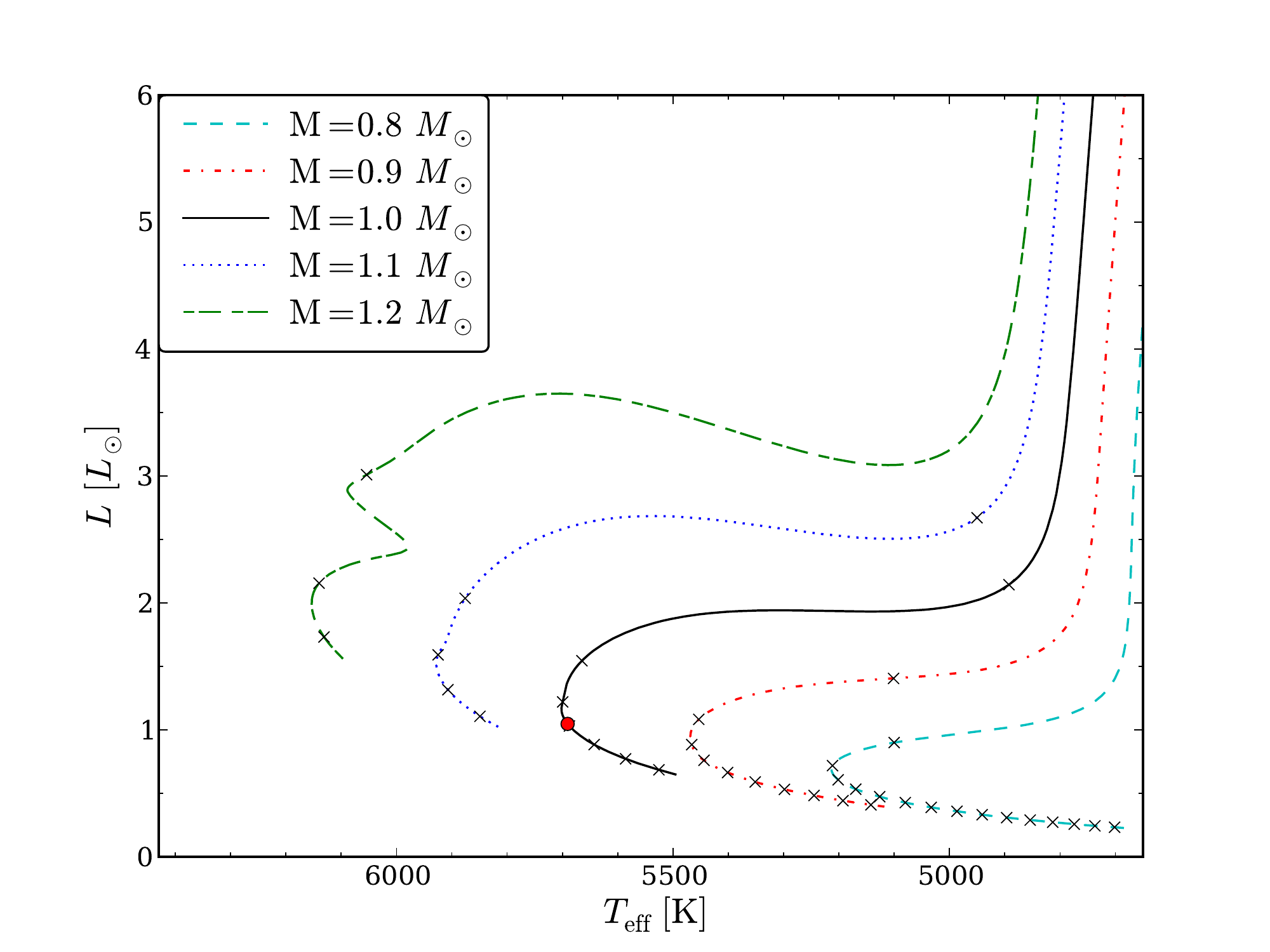}
   \label{fig:evol_track}
 }
 \subfigure{
   \includegraphics[width=0.44\textwidth] {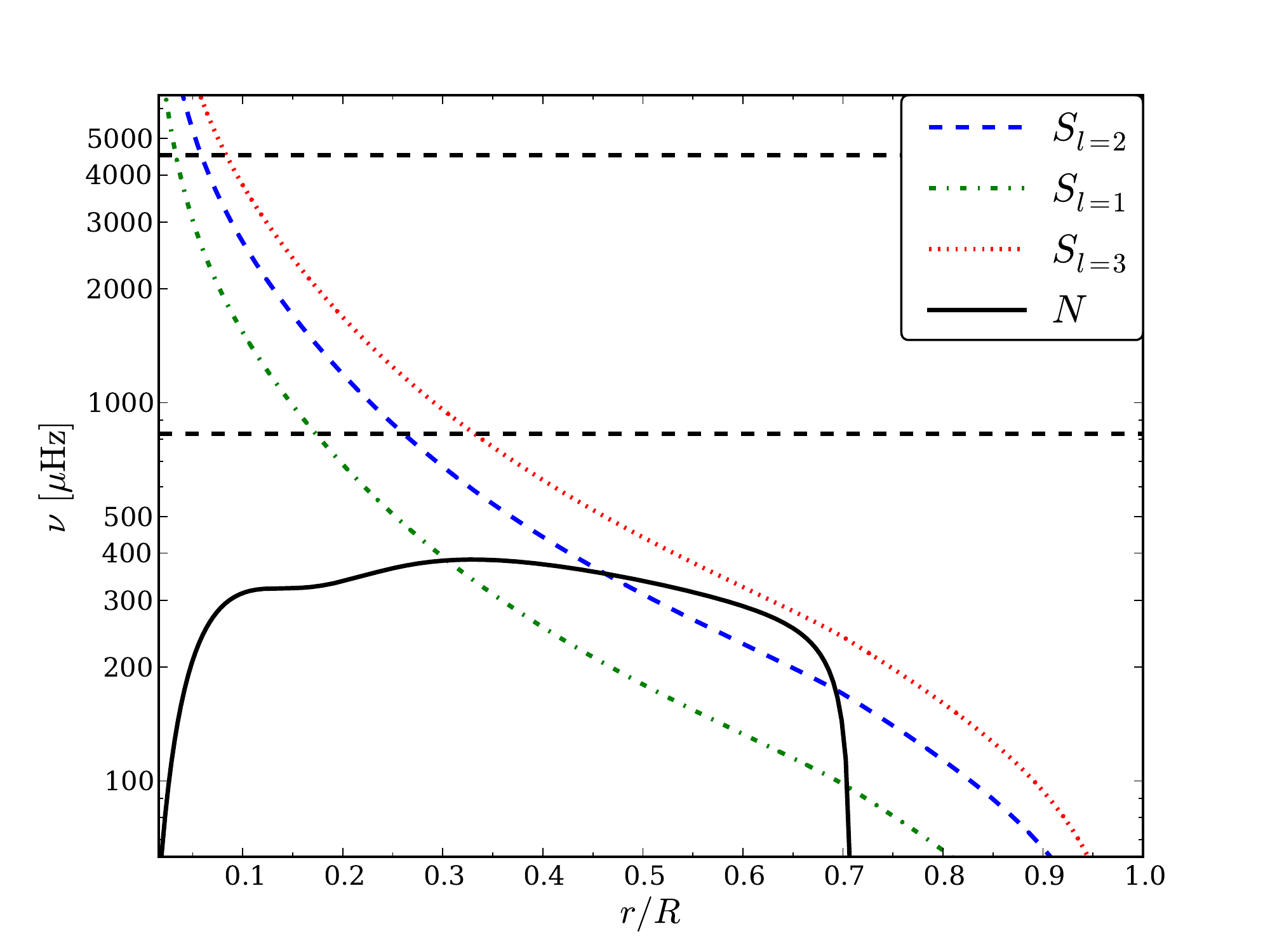}
   \label{fig:mode_proba}
 }

\label{fig:models}
\caption{\footnotesize {\rm \textbf{Left:}} Evolutionary tracks for a set of stellar masses. The track for the considered model ($M=\,$1.0$\, M_{\odot}$) is given by the solid black line, while the specific timestep for the model is given by the red dot. The crosses overlain on the tracks are separated in time by 2 Gyr, starting at 1 Gyr. {\rm \textbf{Right:}} Propagation diagram for the selected model. The three colored lines give the Lamb frequencies, $S$, for $l=\,$1 (dashed), $l=\,$2 (dash-dotted), and $l=\,$3 (dotted). The Brunt--V\"{a}is\"{a}l\"{a} frequency, $N$, is given by the solid black line. The two horizontal dashed lines indicate the region in frequency where oscillation modes from the model are calculated.}
\end{figure*}

The position of the selected model in the Hertzsprung--Russell diagram can be seen in the left panel of \fref{fig:models}. In the right panel of this figure the propagation diagram for the model is shown. From the horizontal dashed lines, which give the frequency range where we have modes from the model, one can see that the modes are all pure \emph{p}-modes (pressure-dominated).

\begin{table*}
\begin{center}
\begin{threeparttable}
\caption{\footnotesize Stellar Model Parameters}
\begin{tabular}{cccccccccccc}
\toprule \\ [-0.3cm]  
$M$ $(M_{\sun})$ & $R$ $(R_{\odot})$ & $T_{\rm eff}$ $($K$)$ & Age $($Gyr$)$ & \textit{X}$_0$ & \textit{Z}$_0$ & $X_c$ & [Fe/H] & $\log g$ $(\rm cm\,s^{-2})$ & $r_{\rm bcz}$ $(r/R)$ & $L\, (L_{\odot})$ & $\nu_{\rm max}$ $(\rm\mu Hz)$ \\ 
\hline\\[-0.3cm]
1.000 & 1.055 & 5690.59  & 7.189  & 0.720  & 0.020 & 0.205  & 0.186  & 4.390 & 0.708 & 1.047 & 2851.5  \\[0.1cm]
\hline 
\end{tabular}
\label{tab:models}
\begin{tablenotes}
	\footnotesize
	\item \textbf{Note.} The value of $\nu_{\rm max}$ is found using the expression of \citet{1995A&A...293...87K}.
\end{tablenotes}
\end{threeparttable}
\end{center}
\end{table*}

\section{Rotation Profiles and Rates}
\label{sec:rates}

\subsection{Stellar Rotation and Differential Rotation}
\label{sec:overview}

Stars are normally born with rapid rotation \citep{2003ApJ...586..464B}. In the initial stages of star formation and evolution the interaction of the stellar magnetic field with the circumstellar accretion disk is likely an important factor in establishing the initial angular velocity distribution \citep[at least for late type stars, see, \eg,][]{1994ApJ...429..781S,1997A&A...326.1023B,2013ApJ...776...67V}.
Late-type stars with a surface convection zone will subsequently be slowed down in the outer layers by magnetic braking from the interaction between the stellar magnetic field and the magnetized wind from the surface \citep[][]{1988ApJ...333..236K, 2012ApJ...746...43R}.
The interaction and transfer of angular momentum between the interior and the envelope, and hence the response of the interior to the spin-down of the envelope, are commonly parametrized by the core-envelope coupling timescale $\tau_{ce}$, given as the time needed for angular momentum equalization of the two regions \citep[][]{2010ApJ...716.1269D}. During the main-sequence phase of a star like the Sun, the evolutionary change of the relative sizes of the core and envelope regions is sufficiently slow, compared to $\tau_{ce}$, to produce only a modest difference between the rotation of the interior and envelope (the regions are relatively "well coupled"). As the star evolves off the main-sequence, the interior and envelope will to some extent decouple as the core contracts and spins up and the envelope expands and spins down. One thus expects a large difference between the mean rotation of the core compared to the envelope in subgiants and an even greater difference in red giants.

\begin{figure}[h]
	\centering
	\includegraphics[scale=0.45]{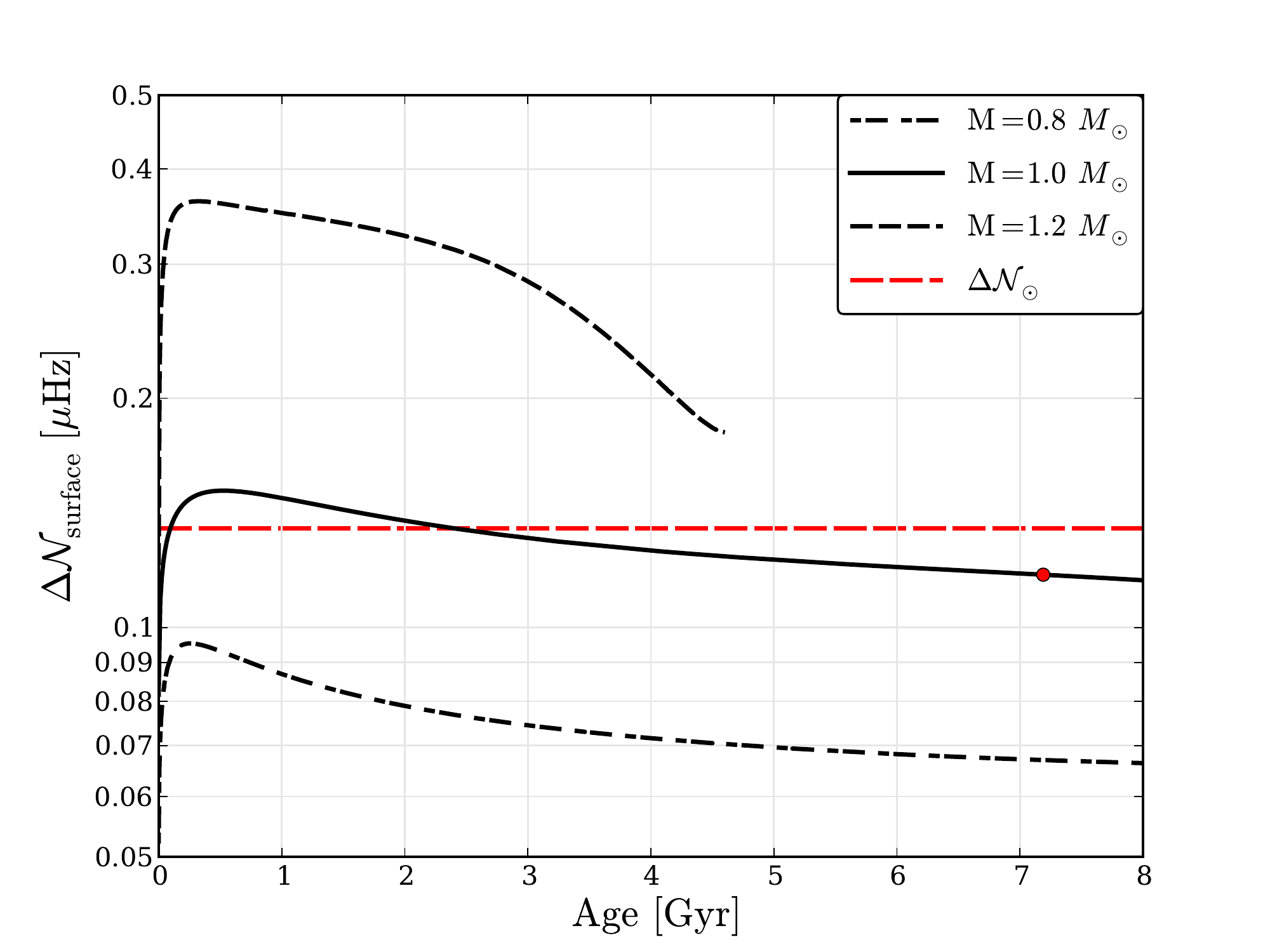}
	\caption{\footnotesize Evolution of the surface differential rotation with age for three stellar masses when using the rotation period predicted by gyrochronology \citep{2007ApJ...669.1167B} (\eqref{eq:prot}) with the description of the surface differential rotation as found from mean-field theory in \citet{2012MNRAS.423.3344K} (\eqref{eq:shear}). The compositions for all models are as in Table~\ref{tab:models}. The dashed horizontal line indicates the present-day value for the solar surface differential rotation.\\}
\label{fig:gyro}
\end{figure}

This process of stellar spin-down yields a correlation between rotation period $P_{\rm rot}$ (as measured at the stellar surface) and age $t$ known as gyrochronology \citep[see, \eg,][]{1972ApJ...171..565S,2009ApJ...695..679M,2014ApJ...780..159E}. In what follows we will use the gyrochronic relation by \citet{2007ApJ...669.1167B}:
\begin{equation}
P_{\rm rot} = t^a \, b \, [(B-V)_0 - c]^d\, \text{days}.
\label{eq:prot}
\end{equation}
Here $t$ is the stellar age in Myr (see Table~\ref{tab:models}), and the remainder of the parameters are $a = 0.519$, $b=0.773$, $c=0.4$, and $d=0.601$. For the computation of the $(B-V)_0$ color as a function of $\rm [Fe/H]$, $T_{\rm eff}$, and $g$, we use the relation of \citet{2000AJ....120.1072S}.
In \fref{fig:gyro} we show the evolution of the surface shear as a function of age for three stellar models when combining Equations~\ref{eq:prot}~and~\ref{eq:shear} below. The position of the specific model (\S~\ref{sec:model}) used for the calculation of splittings is indicated by a red dot.

Late-type stars also exhibit a differential rotation that can be measured from analysis of the time series modulation due to stellar spots \citep[see, \eg,][]{1995ApJS...97..513H}, from Doppler imaging \citep[see, \eg,][]{2005MNRAS.357L...1B,2013A&A...553A..27K}, from chromospheric (magnetic) activity \citep[see, \eg,][]{1996ApJ...466..384D,2000A&A...360.1067K}, and from spectroscopy such as the Fourier transform method or simply from the broadening of spectral lines \citep[][]{1976oasp.book.....G,2005oasp.book.....G}. In the power spectrum, also used to extract the frequency splittings, the presence of co-rotating surface spots will impart a signal at very low frequencies, which also can be used to get information on the mean rotation rate \citep[see, \eg,][]{2013A&A...557L..10N} \citep[see also][and references therein]{2013arXiv1307.4163G} and can possibly be used to get some information on differential rotation \citep[see, \eg,][]{2013A&A...557A..11R,2013A&A...560A...4R}. Another way of obtaining the mean rotation rate of the star is from the autocorrelation of the time series \citep[see, \eg,][]{2013MNRAS.432.1203M,2014ApJS..211...24M}. Naturally, these methods are limited to the study of stellar surface differential rotation.
These observational studies reveal that the surface differential rotation $\rm \Delta\Omega \equiv \Omega_{\rm eq.} - \Omega_{\rm pole}$\footnote{The strength of surface differential rotation is also often expressed via the \emph{lapping time} between the equator and the pole, given by $P_{\rm lap} = 2\pi/ \Delta\Omega$.} depends on the star's rotation rate $P_{\rm rot}$, as well as its mass and metallicity.  The latter two dependencies can be jointly quantified by the effective surface temperature $T_{\rm eff}$.  From observations, solar-like differential rotation has in general been observed, in the sense that the equator rotates faster than the poles \citep[see, \eg,][]{1997MNRAS.291....1D, 2003A&A...398..647R, 2004A&A...417.1047K,2012A&A...542A..99O}. Ensemble studies encompassing many stars indicate a rather weak increase of the differential rotation $\Delta\Omega$ with the mean rotation rate $\Omega$ (or equivalently a decrease with $P_{\rm rot}$) that is often expressed as a power-law; $\Delta\Omega \propto \Omega^\kappa$. There are, however, differences in the reported value of the power law exponent from different observers, ranging from $\kappa\sim 0.15\pm 0.1 $ by \cite{2005MNRAS.357L...1B} to $\kappa \sim 0.66 \pm 0.26 $ by \cite{2003A&A...398..647R}, $\kappa \sim 0.7\pm0.1$ by \cite{1996ApJ...466..384D}, and $\kappa \sim 1$ by \cite{2006A&A...446..267R} (for $P_{\rm rot} > 3\, \mathrm{days}$).  Convection simulations and mean-field models give similar results, suggesting values of $\kappa$ between $0.4$ and $0.6$ \citep[see, \eg,][]{2008ApJ...689.1354B,2012ApJ...756..169A,2012MNRAS.423.3344K}.

This power-law relationship is valid for moderate rotation rates.  At rapid rotation rates ($P_{\rm rot} \lesssim 2$ days, depending on stellar type), observations and models indicate that $\rm \Delta\Omega$ decreases with $\Omega$, implying a negative value of $\kappa$, albeit with much scatter in the data \citep[see, \eg,][]{1999A&A...344..911K,2005AN....326..265K,2006A&A...446..267R,2013IAUS..294..399K}.  In this paper we use the following parameterization suggested by 
\citet{2012MNRAS.423.3344K} based on mean-field models:
\begin{equation}
\Delta\Omega = W_0 + \sum\limits_{j=1}^{6}  W_j \times \tanh(w_j + w_j^T T_{\rm eff} + w_j^P P_{\rm rot}).
\label{eq:shear}
\end{equation}

In the limit of slow rotation, convection simulations predict a transition to anti-solar differential rotation ($\rm \Delta \Omega < 0$), with fast poles and a slower equator. However, observations are ambiguous on this issue, and mean-field models tell a different story.  We will return to this issue in \S~\ref{sec:drcz}.

It is important to remember that a measure of the rotation rate obtained from frequency splittings in general cannot readily be compared to values from these other techniques \citep[for a review over the comparison of different measures of differential rotation see][]{2000SoPh..191...47B}. This is because the value of the frequency splitting will, to some extent, be affected by the whole star, albeit strongly weighted toward the surface, while the value from, for instance, the broadening of spectral lines is only sensitive to the near-surface photospheric layers where the lines are formed and will furthermore be sensitive to velocity fields other than that of rotation. The difference will be even more pronounced if the star has a near-surface shear layer, as found in the Sun \citep[see, \eg,][]{1996Sci...272.1300T,1998ApJ...505..390S}. For the measurement of surface tracers such as starspots and magnetic features the value obtained for the rotation depends on many quantities, such as the depth at which the tracer is anchored \citep[][]{1972ApJ...173..439F} and the size-distribution of the starspots \citep[see, \eg,][]{1984ApJ...283..373H}, and, in assuming a migration of activity belts over a stellar cycle (if any), the phase of the cycle at the time of the observation would also influence the measured rotation rate. An anchoring below a potential near-surface shear layer may result in higher rotation rates than found from measurements of spectral lines or frequency splittings \citep[see, \eg,][]{1970SoPh...13..251W,1989ApJ...343..526B}. The effect would be even more enhanced for a very deep anchoring close to the tachocline. It should be noted, however, that for very Sun-like stars, it does not currently seem possible to discriminate among the estimates of the rotation rate from different sources \citep[see, \eg,][]{2013PNAS..11013267G, 2013ApJ...766..101C}. Also, the latitudinal distribution of the specific tracer becomes important if the observations of the star are unresolved, as is currently the case for asteroseismic data.

Thus, few observational constraints have been placed on the internal differential rotation profile for stars other than the Sun. Still, insight on the nature of the internal differential rotation profiles we might expect to detect from asteroseismology can come from theory and modeling, which we now consider.


\subsection{Internal Rotation of Stellar Convection Zones }\label{sec:drcz}

Differential rotation in the envelopes of late-type main-sequence stars is thought to arise from the interaction of turbulent convection and rotation. Coriolis-induced velocity correlations give rise to a convective Reynolds stress that redistributes angular momentum, establishing rotational shear \citep[see, \eg,][]{rudig89,2005LRSP....2....1M}. This shear feeds back on the convective structures and induces a meridional circulation that transports angular momentum and entropy, altering the rotation profile and potentially influencing the convective driving. The differential rotation profile that ultimately results from this nonlinear coupling depends on the nature of the turbulent transport and potentially the dynamics in the upper and lower boundary layers that straddle the convection zone \citep[see, \eg,][]{2005ApJ...622.1320R,2009AnRFM..41..317M,2011ApJ...743...79M,2013IAUS..294..399K}.

Turbulent transport by convective motions has been studied extensively both with mean-field models \citep[][]{rudig89} and with global 3D convection simulations. Both approaches have converged on a general picture of how differential rotation is maintained in the Sun, though some puzzles remain. The monotonic decrease of angular velocity $\Omega$ with latitude (fast equator, slow pole) is attributed to an equatorward angular momentum transport by convective structures that tend to align with the rotation axis. In the case of recent high-resolution convection simulations, this alignment is most readily apparent in coherent downflow lanes that exhibit a preferential north-south orientation at low latitudes \citep[see, \eg,][]{2008ApJ...673..557M,2011A&A...531A.162K,2013IAUS..294..417G}. These are the turbulent, stratified echoes of laminar {\em banana cells}, which are the linearly preferred modes of convection in global spherical shells \citep[see, \eg,][]{1970JFM....44..441B,1983ApJS...53..243G,1984JCoPh..55..461G,2009JFM...634..291J}.

Both mean-field models and convection simulations attribute the conical nature of the solar $\Omega$ profile (angular velocity nearly constant with radius in the bulk of the convection zone) to thermal gradients \citep[see, \eg,][]{1995A&A...299..446K,2009AnRFM..41..317M,2013IAUS..294..399K}. In particular, a poleward entropy gradient induces a baroclinic torque that sustains the conical $\Omega$ profile against a Coriolis-induced meridional flow that acts to establish cylindrical $\Omega$ isosurfaces in accordance with the Taylor--Proudman theorem ($\partial \Omega / \partial z = 0$, where $z$ is parallel to the rotation axis). In this so-called thermal wind balance, the mean meridional momentum equation can be written as
\begin{equation}\label{eq:twb}
\frac{\partial \Omega^2}{\partial z} = \frac{g}{r^2 \sin\theta \, C_P} \frac{\partial \langle S \rangle}{\partial \theta} \, ,
\end{equation}
where $g$ is the gravitational acceleration, $S$ is the specific entropy, and $C_P$ is the specific heat at constant pressure. The requisite thermal perturbations are small, about one part in $10^5$ for the Sun, corresponding to a pole-equator temperature difference of about 10 K \citep{2005LRSP....2....1M}. However, it is still somewhat uncertain how these thermal perturbations are established.  Possibilities include the influence of rotation on convective heat transport \citep[see, \eg,][]{1995A&A...299..446K,2001MNRAS.321..723R,2002ApJ...570..865B,2013IAUS..294..399K} and/or thermal coupling to the tachocline\footnote{Shear layer formed at the boundary between the envelope convection zone and the interior \citep[][]{1992A&A...265..106S}.} \citep[see, \eg,][]{2005ApJ...622.1320R,2006ApJ...641..618M}.
What does this imply about the rotation profile for more rapidly rotating stars? The influence of rotation is typically quantified by the Rossby number (inverse of the Coriolis number) \citep[see, \eg,][]{2013IAUS..294..399K,2012MNRAS.423.3344K}:
\begin{equation}
R_{o} \equiv  \frac{1}{2 \Omega_0 \tau_c} \, ,
\end{equation}
where $\Omega_0$ is the mean rotation rate and $\tau_c$ is a convection timescale. Small $R_o$ implies strong rotational influence, while vigorous convection with a small turnover time will produce a large Rossby number and in turn small differential rotation as the convection cells have little time to interact with and be affected by the rotation. Thus, the evolution of the differential rotation is also affected by stellar mass, due to differences in the turnover time of the convection and differences in the depth of the convection zone for different masses. If the baroclinic term on the right-hand side of \eqref{eq:twb} arises from a quasi-linear modification of the convective heat flux by the Coriolis force, one might expect it to scale as $R_o^{-1}$. This would then imply that $\partial \Omega / \partial z$ would be proportional to $\tau_c$ but independent of $\Omega_0$. However, convection simulations, mean-field models, and stellar observations all suggest that the surface differential rotation $\rm \Delta\Omega$ increases with $\Omega_0$, out to a rate of roughly $10\,\Omega_\odot$, after which it saturates (\S~\ref{sec:overview}). This is attributed to an increase and eventual suppression of the Reynolds stress, as well as the suppression of differential rotation by the Lorentz force in the most rapidly rotating, magnetically active stars. Thus, if $\boldsymbol\nabla \Omega$ increases with $\Omega_0$ while $\partial \Omega / \partial z$ remains the same, then this implies that rotation profiles should become more cylindrical as the rotation rate of the star increases. This intuitively agrees with what one would expect from the Taylor--Proudman theorem.

This simple argument that $\Omega$ profiles should become more cylindrical as $\Omega_0$ is increased is largely borne out by convection simulations and mean-field models \citep[see, \eg,][]{2008ApJ...689.1354B,2011ApJ...740...12H,2011A&A...531A.162K,2011A&A...530A..48K,2012MNRAS.423.3344K}, though the F-star convection simulations by \citet{2012ApJ...756..169A} remain conical down to Rossby numbers as small as 0.15 $(\Omega = 20 \Omega_\odot)$. This may be an indication that thermal coupling to the stable zone as described by \citet{2005ApJ...622.1320R} is playing an important role in the F star simulations, such that the baroclinic term on the right-hand side of \eqref{eq:twb} keeps pace with the inertial term on the left-hand side as $\Omega_0$ is increased.

In the slowly rotating regime $(R_o > 1)$, there are significant differences between mean-field models and convection simulations. Convection simulations suggest that there are two distinct mean flow regimes, with a transition to anti-solar differential rotation profiles (slow equator, fast pole) for $R_o \gtrsim 1$ (see, \eg, \citet[][]{1977GApFD...8...93G,2011A&A...531A.162K,2011ApJ...728..115B,2013Icar..225..156G,2014MNRAS.438L..76G}; {Featherstone \& Miesch} (2014, submitted)). Though they are far from the rapidly rotating regime where the Taylor--Proudman theorem applies, these anti-solar simulations typically exhibit cylindrical profiles, with $\boldsymbol\nabla \Omega$ directed toward the rotation axis. The transition is attributed to a change in the sense of the convective Reynolds stress, as convective motions with a weak rotational influence tend to conserve their angular momentum locally, speeding up as they approach the rotation axis. This shift in the sense of the angular momentum transport by the Reynolds stress, from equatorward to radially inward, is not exhibited by recent mean-field models \citep[see, \eg,][]{2004AN....325..496K}. Such models only yield anti-solar differential rotation when a strong, baroclinically driven meridional flow is present, potentially induced by non-convective phenomena such as a perturbation to the gravitational potential due to a binary companion or a suppression of convective heat transport by a large polar starspot. The difference can likely be attributed to the underlying assumptions of mean-field theory, which generally treats convective transport as due to a local, quasi-linear distortion of convective motions \citep{2013IAUS..294..399K}. Current global convection simulations, meanwhile, are dominated by large-scale coherent structures that emerge from nonlinear interactions that sense the spherical geometry and stratification.  Though these simulations robustly exhibit anti-solar differential rotation at high $R_o$, we must keep in mind the possibility that the coherent structures they exhibit may be quite different than those in the extreme parameter regimes of real stellar convection zones.

The presence of anti-solar differential rotation in slowly rotating convection simulations is linked to the idea of a near-surface shear layer (NSSL). The prototype for the NSSL is our own Sun, which exhibits an outward decrease in the angular velocity ($\partial \Omega / \partial r < 0$) of several percent from $0.95\, R$ to $R$ \citep{2003ARA&A..41..599T} at low and mid-latitudes. This behavior may also extend to higher latitudes, but current helioseismic inversions are inconclusive. The existence of the NSSL is thought to be linked to a depth variation of the Rossby number, from $R_o < 1$ in the deep interior, where the equatorward $\Omega$ gradient is established, to $R_o > 1$ in the solar surface layers, where the small density scale height and steep superadiabatic stratification drive small-scale convective motions ranging from granulation to supergranulation. Solar surface convection transports angular momentum radially inward as in global convection simulations with weak rotational influence \citep{1975ApJ...199L..71F}, but in order to sustain the NSSL, a change in the meridional force balance is also required \citep{2011ApJ...743...79M}. An upper limit on the $\Omega$ drop across the NSSL can be obtained by assuming that the specific angular momentum is perfectly mixed throughout the layer \citep{2013Icar..225..156G}. This gives a fractional drop in $\Omega$ of $\delta (2-\delta) (1-\delta)^{-1}$, where $\delta = d/R$ is the normalized thickness of the NSSL. For the Sun this upper limit gives 11\%, about a factor of $2.5-3$ larger than the actual value.

Since the base of the NSSL is determined by the condition that $R_o \sim 1$, we expect that its extent will vary with stellar type and rotation rate. In G-type stars rotating more rapidly than the Sun it should be narrower, commensurate with the lower Rossby number throughout. Similarly, the NSSL should be shallower and deeper for more and less massive stars, respectively, due to the shorter and longer convection timescales.

In summary, we expect rapidly rotating stars ($R_o \ll 1$) to exhibit cylindrical $\Omega$ profiles with a solar-like (positive) $\Delta \Omega$ and a thin NSSL (thinner than the Sun). Meanwhile, slow rotators ($R_o > 1$) may exhibit anti-solar differential rotation with a cylindrically inward $\Omega$ gradient (toward the rotation axis). However, this slowly rotating regime is not as well understood, with mixed results from mean-field models and convection simulations. The Sun is thought to be near the transition for G-type stars ($R_o \sim 1$), with conical $\Omega$ surfaces and an equatorward $\Omega$ gradient.


\subsection{Rotation of the Radiative Interior}

Nothing can be inferred from mean-field or convective-shell models concerning the rotation of the interior radiative zone of a star like the Sun.
Observationally, the internal rotation of the Sun has been studied down to a depth of about $R\sim 0.2R_{\sun}$, and it is found that the radiative zone at least down to this limit rotates as a solid body \citep[see, \eg,][]{1995ApJ...448L..57T,1996Sci...272.1300T,1998ApJ...505..390S,1999MNRAS.308..405C}. Probing the very core of the Sun is still a great challenge as only low-degree \emph{p}-modes%
\footnote{or \emph{g}-modes, although a firm and unequivocal detection of these in the Sun is still lacking \citep[see, \eg,][]{2007Sci...316.1591G,2010arXiv1007.4445G,2010A&ARv..18..197A}.} penetrating this region can be used, and here one needs as many radial orders as possible in order to obtain better spatial resolution in the inversion of the rotation profile 
\citep[see, \eg,][]{2001MNRAS.327.1127C,2004MNRAS.355..535C}. Meanwhile, the information needed from these low-degree modes is difficult to obtain at high and low radial orders due to effects like the frequency dependence of the mode line width, the stellar noise background, etc. Some early inversions seemed to indicate a small decrease in the rotation rate of the deep core \citep[see, \eg,][]{1995Natur.376..669E, 1998ESASP.418..685E,2009LRSP....6....1H}, but generally the inversions are all consistent (\ie, within errors) with solid-body rotation for the whole radiative interior \citep[see, \eg,][]{1995ApJ...448L..57T,1998ApJ...496.1015C,1999MNRAS.308..405C,2013SoPh..287...43E}.

The net angular momentum in the radiative zone of a Sun-like star is determined by stellar spin-down and the coupling timescale $\tau_{ce}$ discussed in \S ~\ref{sec:overview}. However, the physical processes responsible for the angular momentum transport and hence defining $\tau_{ce}$ are still uncertain.
The angular momentum loss through the torques from a magnetized wind at the surface establishes 
steep gradients in the rotation rate, giving rise to hydrodynamical instabilities that
can cause redistribution of angular momentum, \eg, via diffusion by turbulent viscosity \citep[see, \eg,][]{1989ApJ...338..424P}.
The angular momentum loss may also establish meridional circulation \citep[see, \eg,][]{1998A&A...334.1000M},
leading to advective and diffusive transport of chemical species and angular momentum.
However, these processes are apparently not sufficiently efficient to account for the slow rotation of the solar interior.
One potentially important process is that of so-called internal gravity waves (IGWs), which are excited mainly by overshoot at the borders of convective zones. These waves can extract and transport angular momentum, which is then deposited in the damping region for the waves \citep[see, \eg,][]{1993A&A...279..431S,1997A&A...322..320Z,1997ApJ...475L.143K,2003A&A...405.1025T,2005A&A...440..981T,2008A&A...482..597T,2009A&A...506..811M,2013ApJ...772...21R}. The early analysis of IGW transport by \citet{1997A&A...322..320Z} was criticized by \citet{1998Natur.394..755G} for neglecting the complexity of the interaction between gravity waves and rotation as also observed in the laboratory \citep[][]{Plumb:1978}. This objection can perhaps be overcome through ``asymmetric filtering" established in a layer just beneath the convection zone \citep[][]{2002ApJ...574L.175T}, and hence for a star like the Sun IGWs may be an efficient means of redistributing angular momentum during the main-sequence \citep[][]{2003A&A...405.1025T}. 
Alternatively, a magnetic field can cause strong coupling throughout the radiative interior and hence the
required angular momentum transport; the field could either be fossil \citep[see, \eg,][]{1993ApJ...417..762C,1998Natur.394..755G}
or generated through a putative dynamo action in the radiative interior \citep{1973MNRAS.161..365T,1999A&A...349..189S,2002A&A...381..923S,2004A&A...422..225M,2005A&A...440L...9E}. 

From observation of \emph{Kepler} red giants by, \eg, \citet[][]{2012Natur.481...55B,2012A&A...540A.143M,2012A&A...548A..10M} and subgiants by \citet[][]{2012ApJ...756...19D,2014A&A...564A..27D}, it is generally found that the ratio of the core-to-envelope rotation rate is in the range of 5--20. Comparisons of stellar evolutionary models including various mechanisms for angular momentum redistribution and the observations from \emph{Kepler} made by \citet[][]{2012A&A...544L...4E}, \citet[][]{2013A&A...549A..74M}, and \citet[][]{2014ApJ...788...93C} find that the core-to-envelope ratios from such models are too high to conform to the observations \citep[see also][]{2000ApJ...540..489S,2010ApJ...716.1269D}. In these studies the authors find core-to-envelope ratios of the order $100-1000$, \ie, around $1-2$ orders of magnitude larger than inferred ratios from \emph{Kepler}. Thus, additional agents for core-envelope angular momentum coupling are still needed for a successful description of the observations from \emph{Kepler} red giants. This discrepancy also raises the question of whether our current ideas on the radial rotation profile of a solar-like star on the main-sequence (such as the one studied in this work) should be revised.

We have for all rotation models in the following assumed differential rotation to reside in the stellar convection zone, while the radiation zone, \ie, for radii below the base of the convection zone, $r_{\rm bcz}$, is assumed to rotate as a solid body.


\section{Frequency splittings for differential rotation profiles}
\label{sec:profiles}

As mentioned in \S~\ref{sec:intro}, we will in the following see how the splittings of oscillation modes behave for a number of different rotation profiles. 
This approach defines the so-called \emph{forward problem}, where the kernels and rotation profile are known upon which the splittings are calculated. In the reverse approach, namely, the \emph{inverse problem}, one instead starts with computed kernels and observed splittings and then attempts to infer the underlying rotation profile.
We will start out with the most simple profile for rotation, to wit a solid-body rotation, and gradually add the complexity of differential rotation.


\subsection{Solid-body Rotation}

\begin{figure*}[ht]
\centering

\subfigure{
   \includegraphics[width=0.47\textwidth] {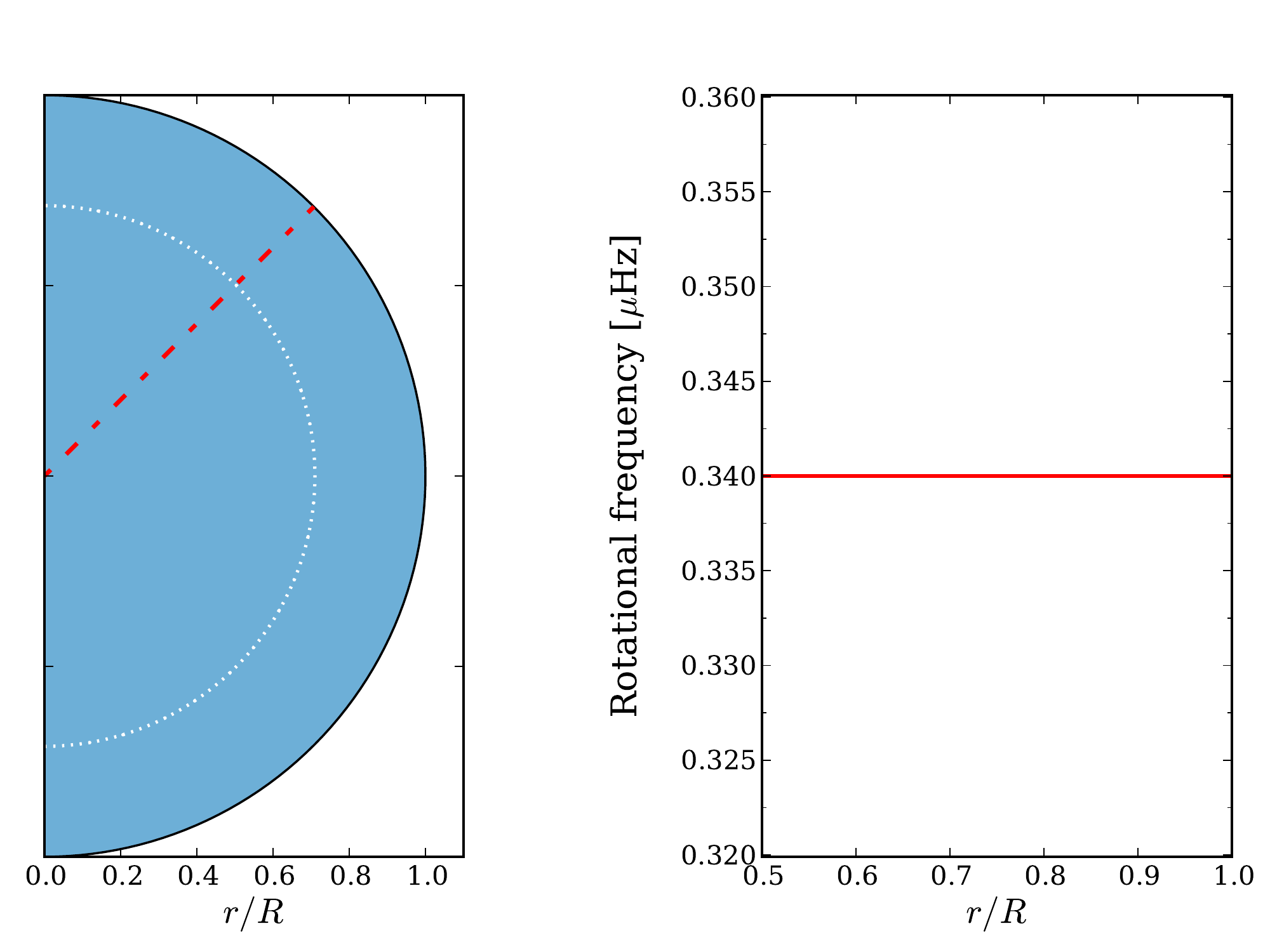}
 }\qquad
 \subfigure{
   \includegraphics[width=0.47\textwidth] {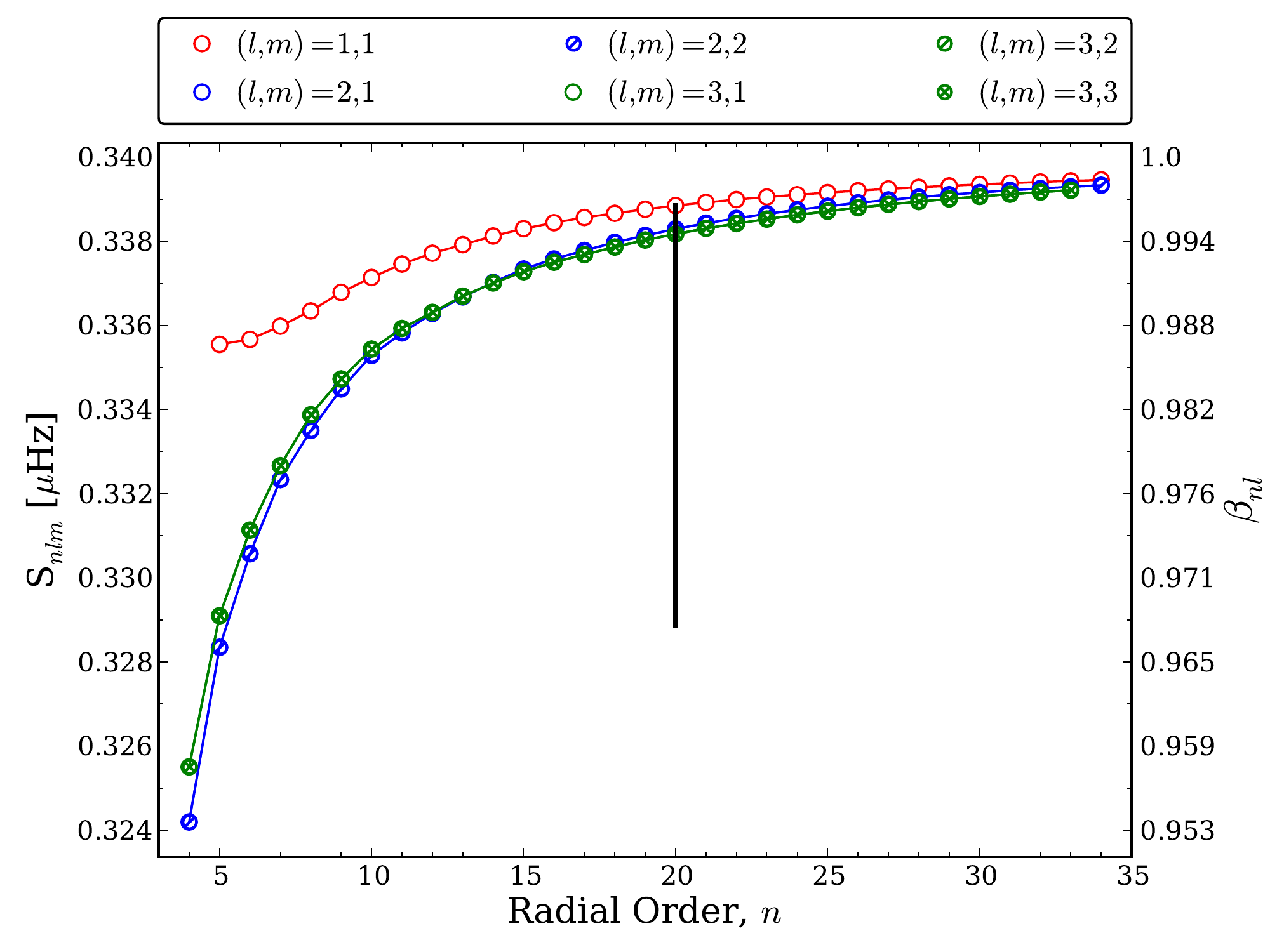}
 }
\caption{\footnotesize {\rm \textbf{Left:}} Rotation profile for a star rotating as a solid body. The small left panel gives the 2D rotation profile of the star, while the 1D rotation profile is given by the small right panel for the co-latitude indicated on the 2D rotation profile. {\rm \textbf{Right:}} Calculated frequency splittings as a function of radial order for the solid-body rotation profile in the left panel and the rotation kernels computed from the adopted model star. The solid vertical reference bar at $n=\,$20 shows the extent of a 0.01$\rm \, \mu Hz$ frequency difference.}
\label{fig:rot_const2}
\end{figure*}

The simplest form of rotation is the solid-body rotation (\fref{fig:rot_const2}). While not being a physically realistic scenario for a real star, it is a good starting point for our analysis, and often this is also the type of rotation assumed when using frequency splittings in asteroseismology to estimate rotation rates. It is clear that with a constant rotation rate/frequency, $\mathcal{N}(r,\theta)=const.$ can be taken outside the integrals in \eqref{eq:split} and the splitting amounts to

\begin{align}\label{eq:solid_body}
\delta\nu_{nl m} = \delta\nu_{nl} &= \mathcal{N}\int_0^{\pi}\int_0^R \mathcal{K}_{nl m}(r, \theta) r dr d\theta\\ \nonumber
				&= \mathcal{N} \beta_{nl}\\ \nonumber
				&= \mathcal{N} (1-C_{nl})\, .\\ \nonumber
\end{align}  
The $\beta_{nl}$ here simply gives a measure of the total kernel integral weighted by $r$. The other popular way of writing $\beta_{nl}$ is as $(1-C_{nl})$, with the parameter $C_{nl}$ being a so-called \emph{Ledoux} constant \citep[][]{1951ApJ...114..373L}. In this description $C_{nl}$ measures the effect of the Coriolis force, acting against the rotation for a mode advected in the prograde direction. Another frequent approximation in asteroseismology when adopting this approach is to assume $\beta_{nl}\sim 1$, whereby the constant rotation rate can be computed directly from the frequency splittings. In \fref{fig:rot_const2} the frequency splittings are shown assuming a solid-body rotation, with the $\beta_{nl}$ value indicated on the right ordinate. The value adopted for the rotation rate is the latitudinal mean of the rotation rate at the surface estimated for our stellar model (see Table~\ref{tab:models3}). First, it is clear that the splittings are independent of the azimuthal order $m$, which is expected as there is no latitudinal change in the rotation rate and the calculation is done to first order. Secondly, we see that $\beta_{nl}\sim 1$ seems to be a fair approximation with about $1\%$ offset at $n=10$, in other words, the modes are dominated by advection. One should, however, be aware that when adopting this approach the rotation rate extracted if using $\beta_{nl}= 1$ is unavoidably underestimated, and more so if low-radial-order modes are included in the estimation \citep[see, \eg,][]{2009ASPC..416..385K}.

\subsection{Piecewise Constant Differential Rotation}
\label{sec:pwc}

The second-most simple configuration conceivable could then be a piecewise constant rotation, still without any latitudinal dependence. Such a rotation profile can be written as
\begin{equation}\label{eq:pwc}
\mathcal{N}(r, \theta) = \mathcal{N}(r) = \left\{ 
  \begin{array}{l l}
    \mathcal{N}_1 & \quad r>r_{\rm break}\\
    \mathcal{N}_0 & \quad r<r_{\rm break}\\
  \end{array} \right. 
\end{equation}
In \fref{fig:pwc_rot_prof2} such rotation profiles are illustrated, the first of these having the outer shell rotating at twice the rate of the interior, while the opposite situation applies to the second profile. For both profiles the surface rotation is equal to the mean surface rotation rate for the model (see Table~\ref{tab:models3}). It is easy to see that the only modification needed in describing the splittings when going from the solid-body profile to a piecewise constant profile is to split the integral in radius in two parts (or the number of adopted shells):

\begin{figure*}[ht]
\centering

\subfigure{
   \includegraphics[width=0.47\textwidth] {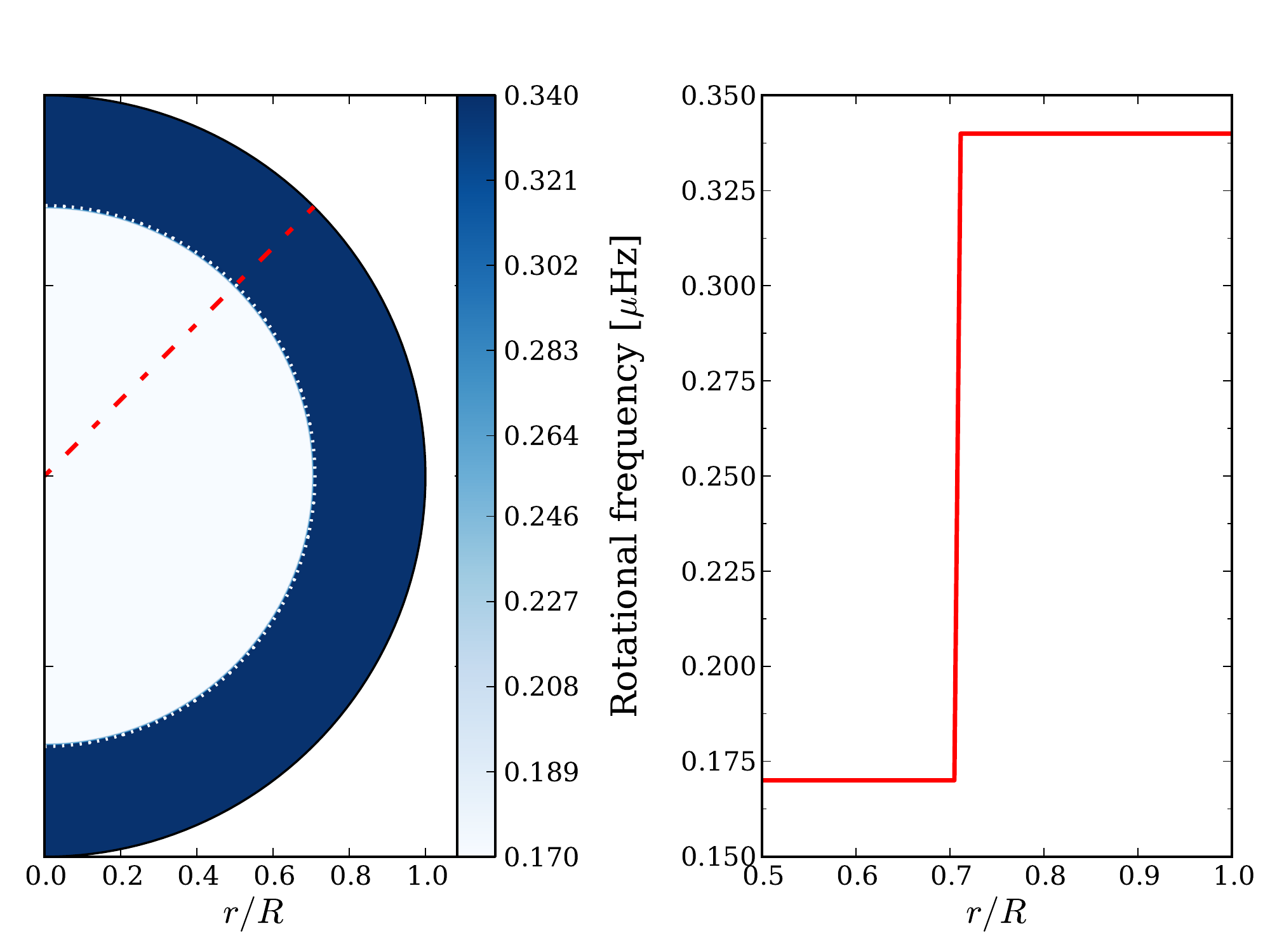}
 }\qquad
 \subfigure{
   \includegraphics[width=0.47\textwidth] {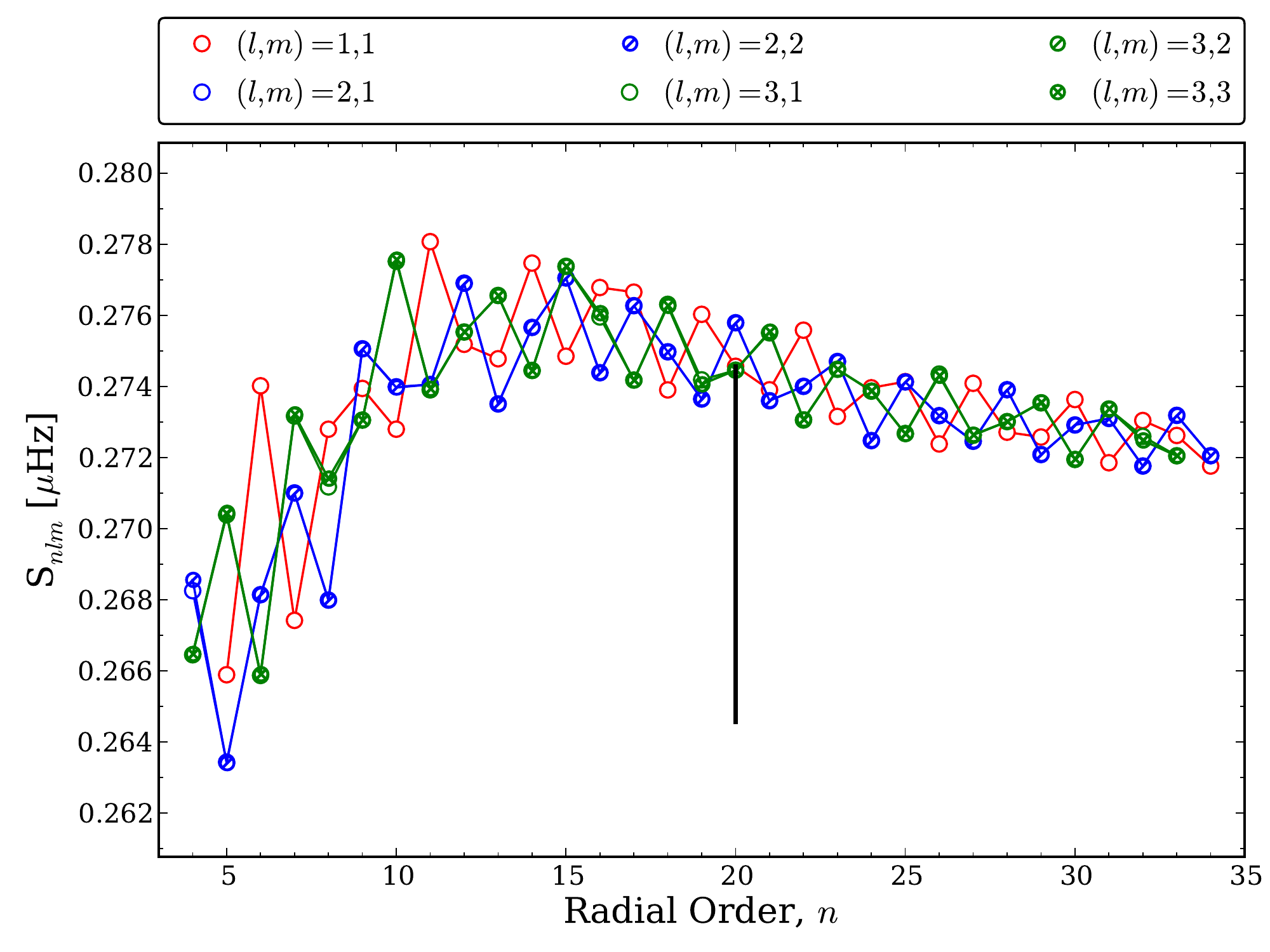}
 }
\subfigure{
   \includegraphics[width=0.47\textwidth] {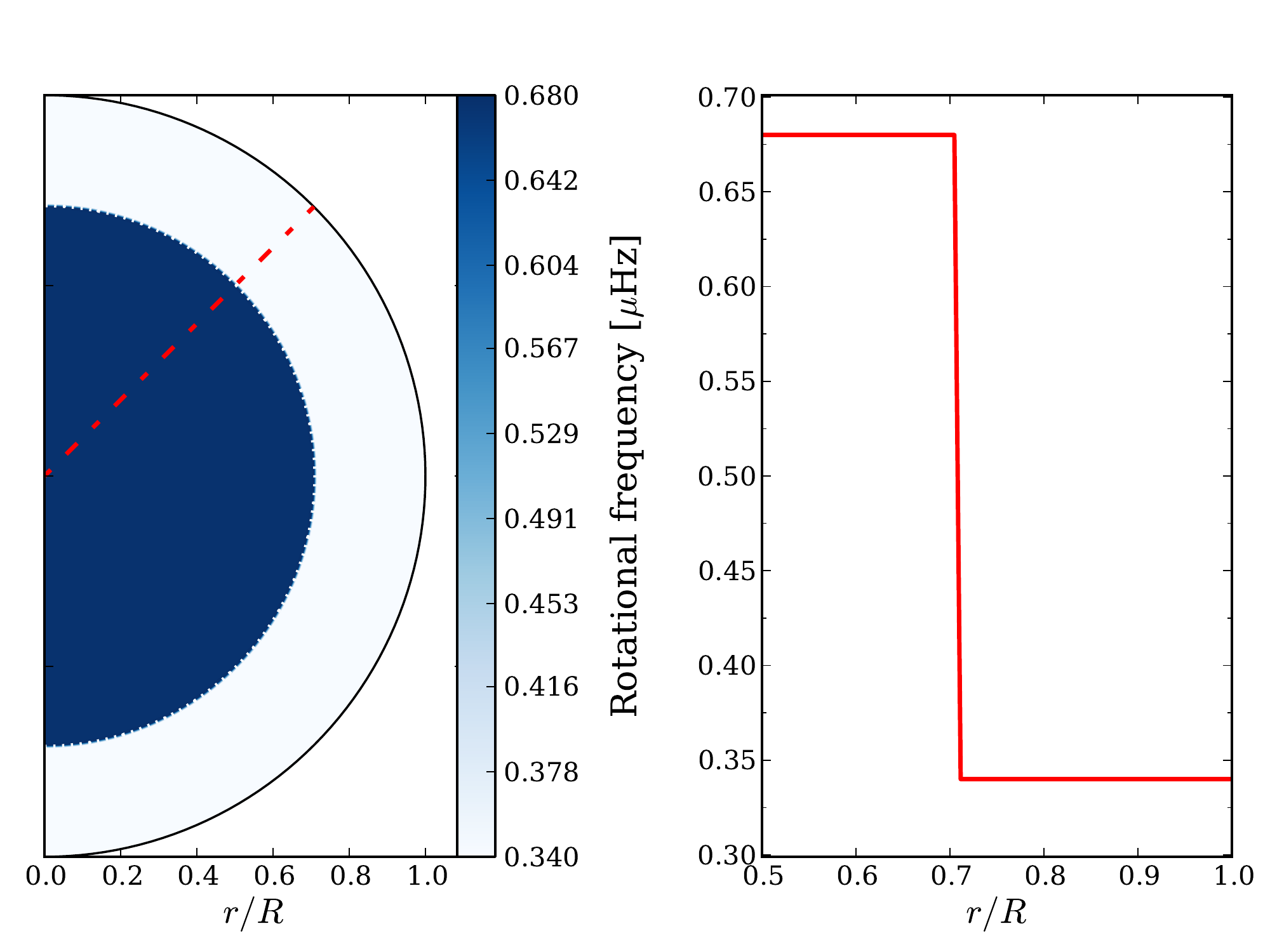}
 }\qquad
 \subfigure{
   \includegraphics[width=0.47\textwidth] {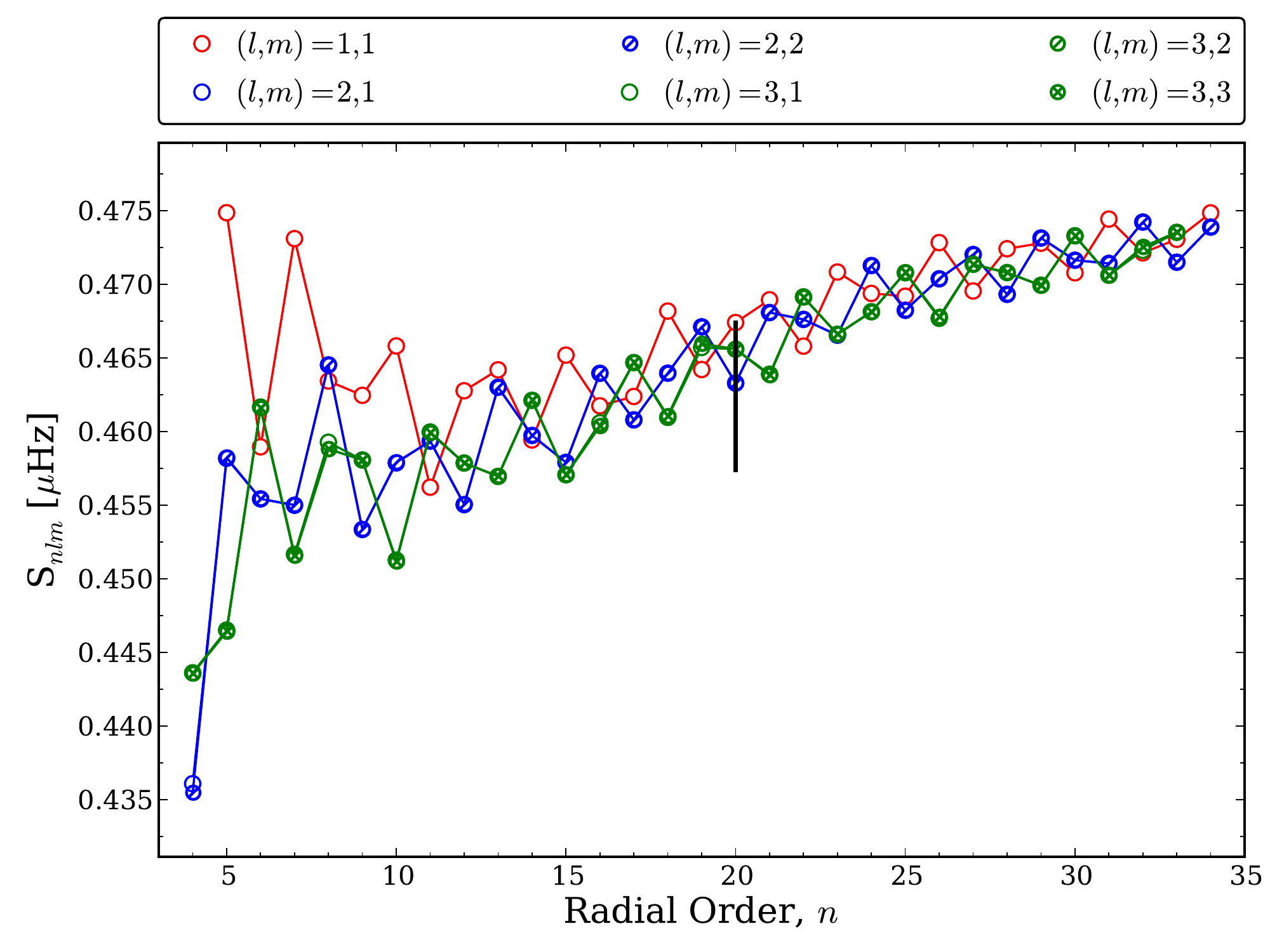}
 }
\caption{\footnotesize {\rm \textbf{Left:}} Rotation profiles for a star rotating with a piecewise constant configuration. The small left panels give the 2D rotation profiles of the star, while the 1D rotation profile is given in the small right panel for the co-latitude indicated on the 2D rotation profiles. {\rm \textbf{Right:}} Calculated frequency splittings as a function of radial order for the piecewise constant rotation profiles in the left panels and the rotation kernels computed from the adopted model star. The solid vertical reference bar at $n=\,$20 shows the extent of a 0.01$\rm \, \mu Hz$ frequency difference.}
\label{fig:pwc_rot_prof2}
\end{figure*}

\begin{align}\label{eq:pwc2}
\delta\nu_{nlm} = &\mathcal{N}_0\int_0^{r_{\rm break}}\int_0^{\pi} \mathcal{K}_{nlm}(r, \theta) rdrd\theta\\ 
+ &\mathcal{N}_1\int_{r_{\rm break}}^R\int_0^{\pi} \mathcal{K}_{nlm}(r, \theta)r dr d\theta \, .\notag
\end{align}
The simplicity in the calculations of the splittings for such a profile also makes for a quite simple inverse problem as the kernel integrals only need to be calculated once. In stars where very little latitudinal differential rotation is expected, the piecewise constant profile can be used to get an idea of the mean rotation rates of the two (or more) radial intervals. This approach was, \eg, adopted in the estimations of the core-to-envelope rotation rates in \emph{Kepler} red giants \citep[see, \eg,][]{2012Natur.481...55B,2012A&A...548A..10M} and subgiants \citep[see][]{2012ApJ...756...19D,2014A&A...564A..27D} \citep[see also][]{2013A&A...549A..74M,2013A&A...549A..75G}. In the right panels of \fref{fig:pwc_rot_prof2} the splittings for the two piecewise constant profiles are shown. For both we still have no dependence on the azimuthal order as the rotation has no dependence on latitude. For the one where the envelope rotates faster than the interior (top) we see a behavior in the splittings that resembles the behavior of the radius-weighted kernel IP in the envelope (\fref{fig:cumkern_vol}). This is to be expected as the kernel IP in the envelope and thereby the last integral in \eqref{eq:pwc2} dominate the contribution to the splittings. So, what is seen in the splittings is a constant factor and $\beta_{nl}$ times the kernel IP in the envelope plus a small contribution from the interior. The contribution from the inner part of the profile to the splittings is largest at radial orders away from the peak in the radius-weighted kernel IP in the envelope (\fref{fig:cumkern_vol}), with the effect that the variation in splittings as a function of radial order will progressively approach a straight line the higher the contribution from the inner part gets. This effect can be seen in the bottom right panel of \fref{fig:pwc_rot_prof2}, where the splittings now follow more of a straight line. For a rotation rate in the inner parts much higher than in the envelope the high and low regimes of radial order would start to show the highest splittings simply because these regimes are the ones most sensitive to the inner part. The zigzag pattern in the splittings originates from the sharp transition between the core and envelope and becomes more prominent as the angular velocity difference increases.


\subsection{Shellular Differential Rotation}

The natural next step to take in describing an increasingly realistic stellar rotation is to allow for a gradual variation in the rotation rate with radius in the outer convection zone. This can be given as  
\begin{equation}\label{eq:shellular}
\mathcal{N}(r, \theta) = \mathcal{N}(r) =\left\{  
  \begin{array}{l l}
    \mathcal{N}_1+\gamma(r-r_{\rm bcz})/R   & \quad r>r_{\rm bcz}\\
    \mathcal{N}_0 & \quad r < r_{\rm bcz}\\
  \end{array} \right.
\end{equation}
This describes a profile with solid-body rotation in the interior and a linear change with radius in the convection zone. The change in radius could, however, also be given by a smoothly changing function. The use of $\mathcal{N}_1$ in the convection zone instead of simply $\mathcal{N}_0$ allows for a sharp transition between the two regions. This type of profile has been used often in the modeling of rotation in, \eg, $\delta$ Scuti stars \citep[see, \eg,][]{1996A&A...305..487G,2006A&A...449..673S}, $\beta$ Cep stars \citep[see, \eg,][]{2003Sci...300.1926A,2004A&A...415..251D,2004MNRAS.350.1022P}, sdB stars \citep[see, \eg,][]{2005ApJ...621..432K}, and white dwarfs \citep[see, \eg,][]{1999ApJ...516..349K,2009ASPC..416..385K}.

The hypothesis of shellular rotation \citep[][]{1992A&A...265..115Z}, which is often the employed description in stellar models including evolution of rotation \citep[see][and references therein]{2012A&A...537A.146E}, states that the horizontal turbulence is very strong compared to the component of turbulence in the vertical direction. This anisotropy in turbulence with the horizontal direction dominating favors rotation constant on isobars. Note that we here use the term \emph{shellular} for a gradual radial variation in the rotation rate, while in principle a solid-body or piecewise constant profile also fulfills the formal definition of rotation constant on isobars. It is most applicable to stably stratified radiative zones where the buoyancy force resists vertical motions, but it may also occur in convection zones with a steep density stratification.
The assumption of shellular rotation is generally found valid at low rotation \citep[][]{2000ARA&A..38..143M}. The presence of shellular rotation in the extended convective envelope of an RGB star was also found by \citet[][]{2011EAS....44...81B} in a 3D simulation using the ASH (Anelastic Spherical Harmonic) code.

\begin{figure*}[ht]
\centering

\subfigure{
   \includegraphics[width=0.47\textwidth] {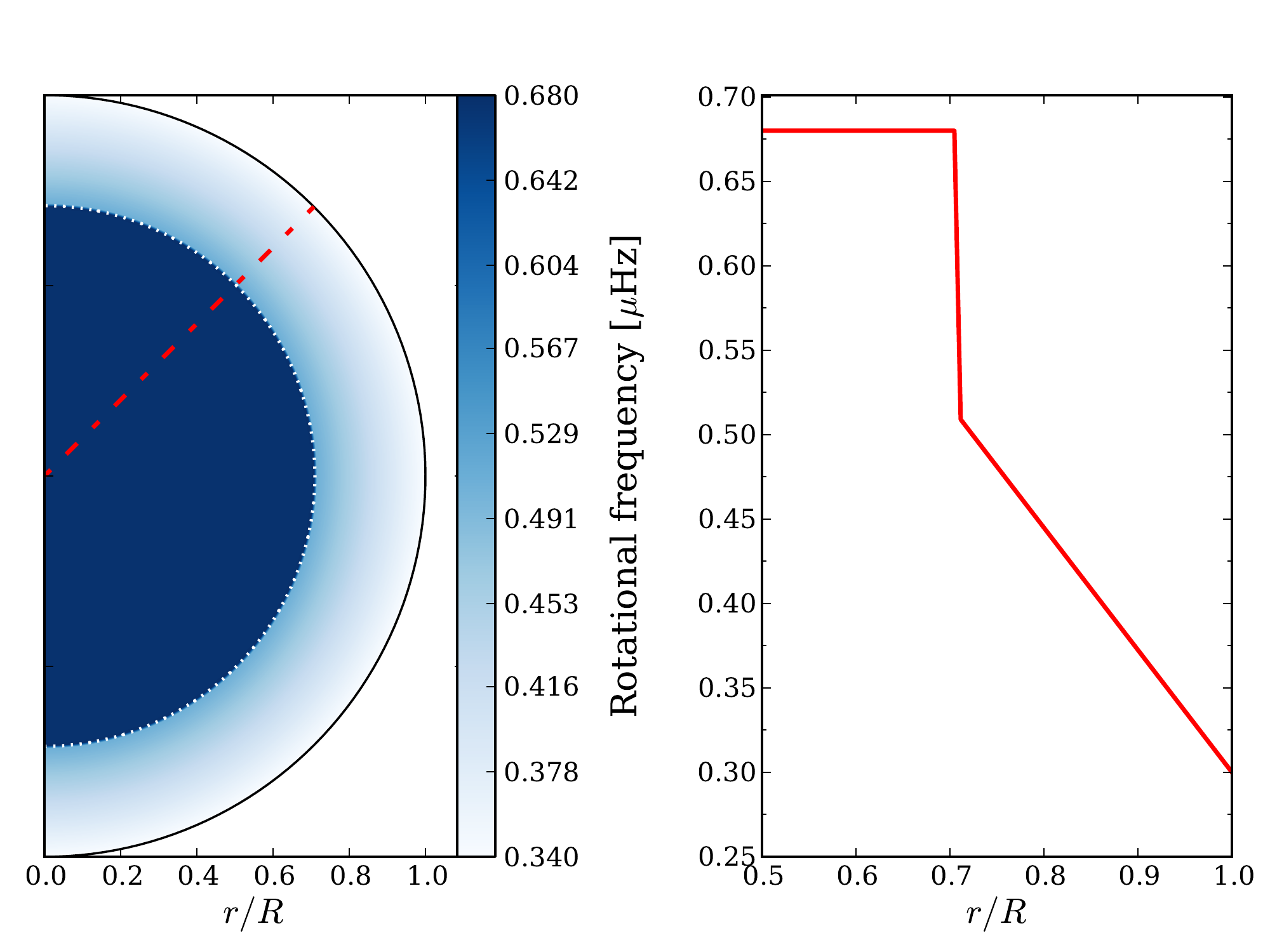}
 }\qquad
 \subfigure{
   \includegraphics[width=0.47\textwidth] {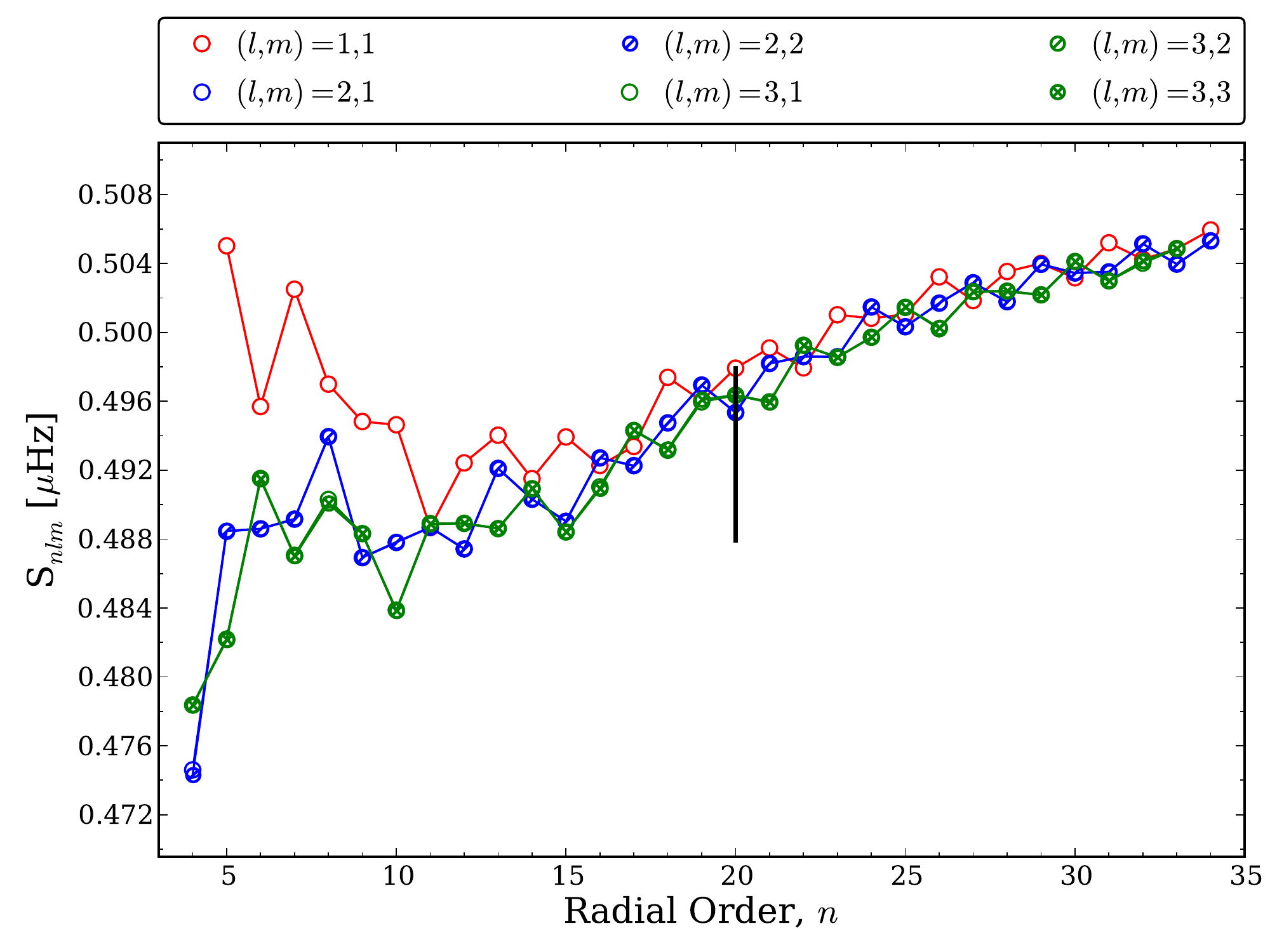}%
 }
\caption{\footnotesize {\rm \textbf{Left:}} Rotation profile for a star rotating with a solid-body interior ($r<r_{bcz}$) and a linearly decreasing rotation rate in the convection zone, \ie, a shellular rotation profile. The small left panel gives the 2D rotation profile of the star, while the 1D rotation profile is given in the small right panel for the co-latitudes indicated on the 2D rotation profile. {\rm \textbf{Right:}} Calculated frequency splittings as a function of radial order for the shellular rotation profile in the left panel and the rotation kernels computed from the adopted model star. The solid vertical reference bar at $n=\,$20 shows the extent of a 0.01$\rm \, \mu Hz$ frequency difference.}
\label{fig:rot_shel}
\end{figure*}

An example of a shellular rotation profile is shown in the left panel of \fref{fig:rot_shel}. The right panel of this figure shows the corresponding values obtained for the frequency splittings from this profile. For this profile the rotation rate at the surface equals the mean surface rotation rate for our model (see Table~\ref{tab:models3}), while the interior is set to rotate at twice the rate of the surface. There will for such a rotation profile again not be a dependence on the azimuthal order, as was also the case for the two above rotation profiles.
We can better understand the behavior seen in \fref{fig:rot_shel} by examining the individual terms making up the equation for the frequency splitting:

\begin{align}
\delta\nu_{nlm} = &\mathcal{N}_0\int\int_{0,0}^{r_{\rm bcz},\pi} \mathcal{K}_{nlm}(r, \theta) drd\theta\\ \notag
 + &\mathcal{N}_1\int\int_{r_{\rm bcz},0}^{R,\pi} \mathcal{K}_{nlm}(r, \theta)r dr d\theta \\ \notag
 + &\gamma/R\int\int_{r_{\rm bcz},0}^{R,\pi} (r-r_{\rm bcz}) \mathcal{K}_{nlm}(r, \theta)r dr d\theta\, . 
\label{eq:shellsplit}
\end{align}
The first two terms are similar to a piecewise constant profile, here with the interior rotating faster than the envelope ($\mathcal{N}_0=0.68 \, \rm \mu Hz > \mathcal{N}_1=0.51 \,\rm \mu Hz$), and indeed the splittings behave in much the same way as for this profile. The last term (with $\gamma=-0.58 \,\rm \mu Hz$) is essentially the same as shown in \fref{fig:cumkern_vol} (however not normalized), and as $\gamma$ is negative the modes with the highest kernel IP in the envelope will be reduced the most, making the variation with $n$ slightly more convex.

\subsection{Solar-like Differential Rotation}

\begin{table*}[hbt]
\begin{center}
\begin{threeparttable}
\caption{\footnotesize Rotation Profile Components Estimated from Gyrochronologically Expected Mean Rotation Period (\eqref{eq:prot}) Combined with the Description for the Surface Differential Rotation (\eqref{eq:shear})}
\begin{tabular}{ccccccccc}
\hline \hline
& & &\multicolumn{4}{c}{Solar Rotation} &\multicolumn{2}{c}{Axial Rotation} \\[0.07cm]
\cline{4-7}\cline{8-9}\\[-0.3cm] 
$\langle P_{\rm rot} \rangle$ (days) & $\langle \mathcal{N}(R) \rangle_{\theta}$ $(\rm \mu Hz)$\tnote{a} & $\Delta\mathcal{N}_{\rm surf}$ $(\rm \mu Hz)$ & $\mathcal{A}$ $(\rm \mu Hz)$ & $\mathcal{B}$ $(\rm \mu Hz)$ & $\mathcal{C}$ $(\rm \mu Hz)$ & $\mathcal{D}$ $(\rm \mu Hz)$ & $\alpha$ $(\rm \mu Hz)$ & $\beta$ $(\rm \mu Hz)$ \\[0.07cm]
\hline\\[-0.3cm]
34.03  & 0.3401 &  0.1173 & 0.390 &  -0.049  & -0.068 & 0.373 & 0.265 & 0.117\\[0.07cm]
\hline
\end{tabular}
\label{tab:models3}
\begin{tablenotes}
	\footnotesize
	\item \textbf{Note.} The values of the solar-like rotation parameters (\ie, $\mathcal{A},\mathcal{B},\mathcal{C},\mathcal{D}$), relate to each other as the equivalent values for the Sun in using the solar parameters given in \citet{2004SoPh..220..169G} (see Appendix~\ref{app:param} for the derivation of these values).
	\item [a] Mean rotation frequency in latitude at the stellar surface, see \eqref{eq:meanrotangle} in Appendix~\ref{app:param}.
\end{tablenotes}
\end{threeparttable}
\end{center}
\end{table*}

A solar-like differential rotation profile is naturally of great interest when studying solar-like oscillators, in particular very Sun-like stars.
For the Sun it is found that the inner radiative zone, \ie, the region below the base of the convection zone at $r_{\rm bcz}\approx 0.71\, R_{\sun}$ \citep[][]{1991ApJ...378..413C}, rotates much like a solid body with a rotation period of about $27$ days (${\sim} 430\, \rm nHz$) \citep[][]{1998ApJ...505..390S}. In the convection zone  the rotation rate changes as a function of latitude, going from a rotation period of ${\sim} 25.4$ days (${\sim} 456\, \rm nHz$) at the equator to ${\sim} 36.1$ days (${\sim} 321\, \rm nHz$) at the pole (values found at $r=0.995\, R_{\sun}$ using a 2dRLS\footnote{2D Regularized Least Squares.} inversion), amounting to a difference of ${\sim} 30\%$ and a differential rotation of $\Delta\mathcal{N}\approx 135\,\rm nHz$  \citep[][]{1998ApJ...505..390S}. For each latitude in the convection zone the rotation rate is approximately constant in the radial direction, which is the simple description we will use. A more detailed look at the Sun suggests that a description in which the contours in the convection zone have an angle of about ${\sim}25^{\circ}$ to the rotation axis is actually more fitting \citep[see, \eg,][]{2003ESASP.517..283G,2005ApJ...634.1405H}, whereby the rotation profile in the convection zone can be seen as a mix between a radially invariant and a cylindrical profile.

The finer details of the solar rotation profile reveal a near-surface shear layer where the rotation rate in the outer ${\sim}5\%$ of the star is reduced by about $5\%$ \citep[see, \eg,][]{1991ApJ...367..649G,1996Sci...272.1300T,1997SoPh..170...43K,1998ApJ...505..390S,2002SoPh..205..211C}. To within ${\sim}20^{\circ}$ of the pole the rotation rate seems to deviate from the governing profile in the remainder of the convection zone, with the polar regions rotating even slower than expected from a simple parameterization \citep[see, \eg,][]{1998ApJ...503L.187B,1998ApJ...505..390S}. However, as noted in \S~\ref{sec:ker} the splittings of low-degree modes are very insensitive to such high latitudes. Additional fine-scale details include, for instance, torsional oscillations \citep[see, \eg,][]{2002Sci...296..101V,2003ApJ...585..553B,2005ApJ...634.1405H,2006ApJ...649.1155H}.

Adopting the simplest parts from the above description, the solar rotation profile can to a good approximation be given as \citep{1983ApJ...270..288S,2004SoPh..220..169G}
\begin{equation}\label{eq:solar_rot_profile}
\mathcal{N}(r, \theta) = \left\{ 
\begin{array}{l l}
\mathcal{A}+\mathcal{B} \cos^2\theta   +\mathcal{C} \cos^4\theta  & \quad r>r_{\rm bcz}\, ,\\
\mathcal{D} & \quad r < r_{\rm bcz}\\ 
\end{array} \right.
\end{equation}
with $\theta$ denoting the co-latitude. This description gives a profile with latitudinal differential rotation in the outer convection zone, while the inner radiative zone rotates as a solid body. For each co-latitude the radial change in rotation rate is given by a step function, \ie, there is no radial component of differential rotation in the envelope. In Appendix~\ref{app:param} it is shown how the values of the different parameters (see Table~\ref{tab:models3}) are found using the relationship between these parameters for the Sun (as found from Doppler measurements of the Sun by \citet[][]{1984SoPh...90..199S}; see also \citet[][]{2004SoPh..220..169G}). 

\begin{figure*}[ht]
\centering

\subfigure{
   \includegraphics[width=0.47\textwidth] {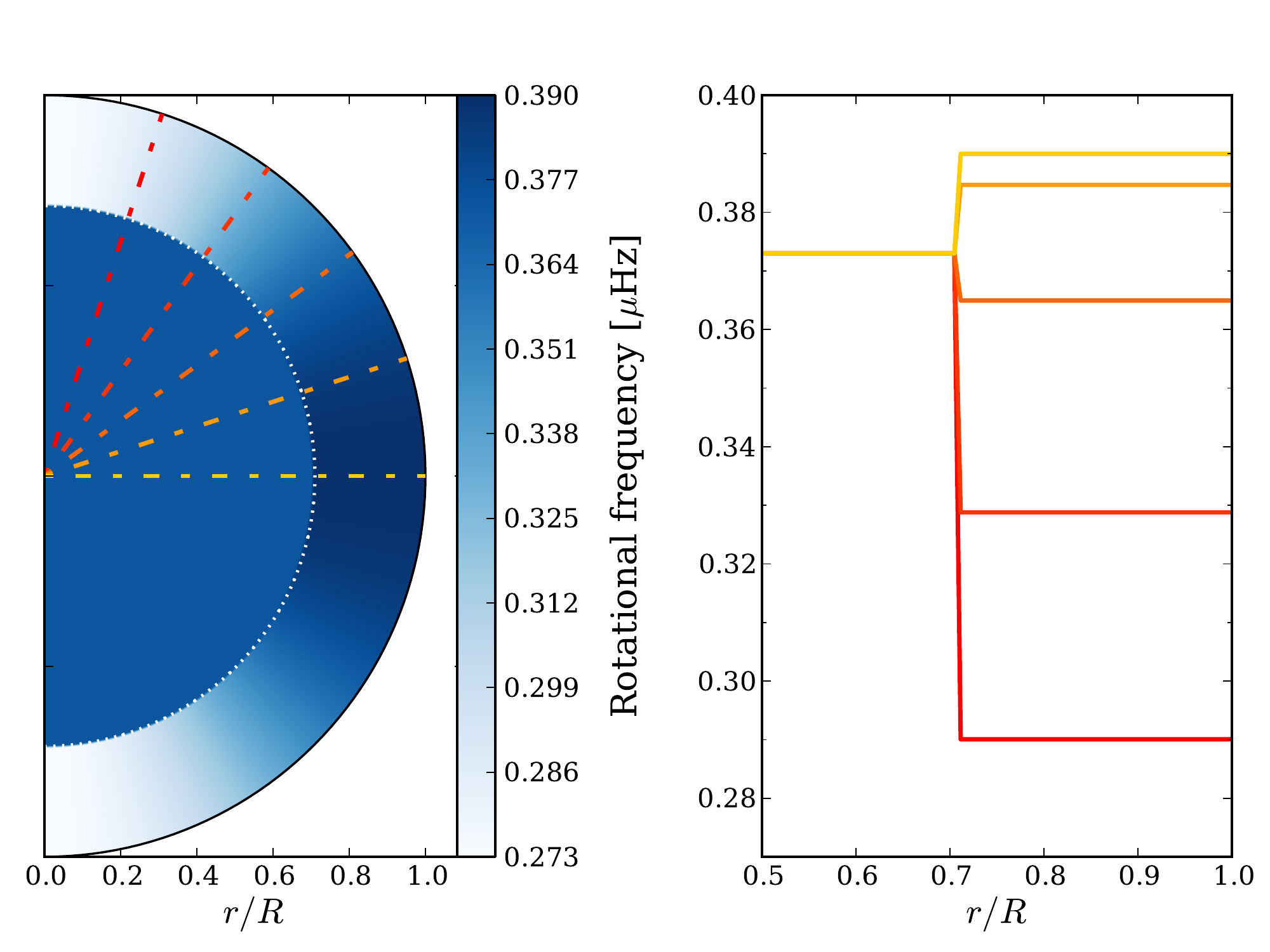}
 }\qquad
 \subfigure{
   \includegraphics[width=0.47\textwidth] {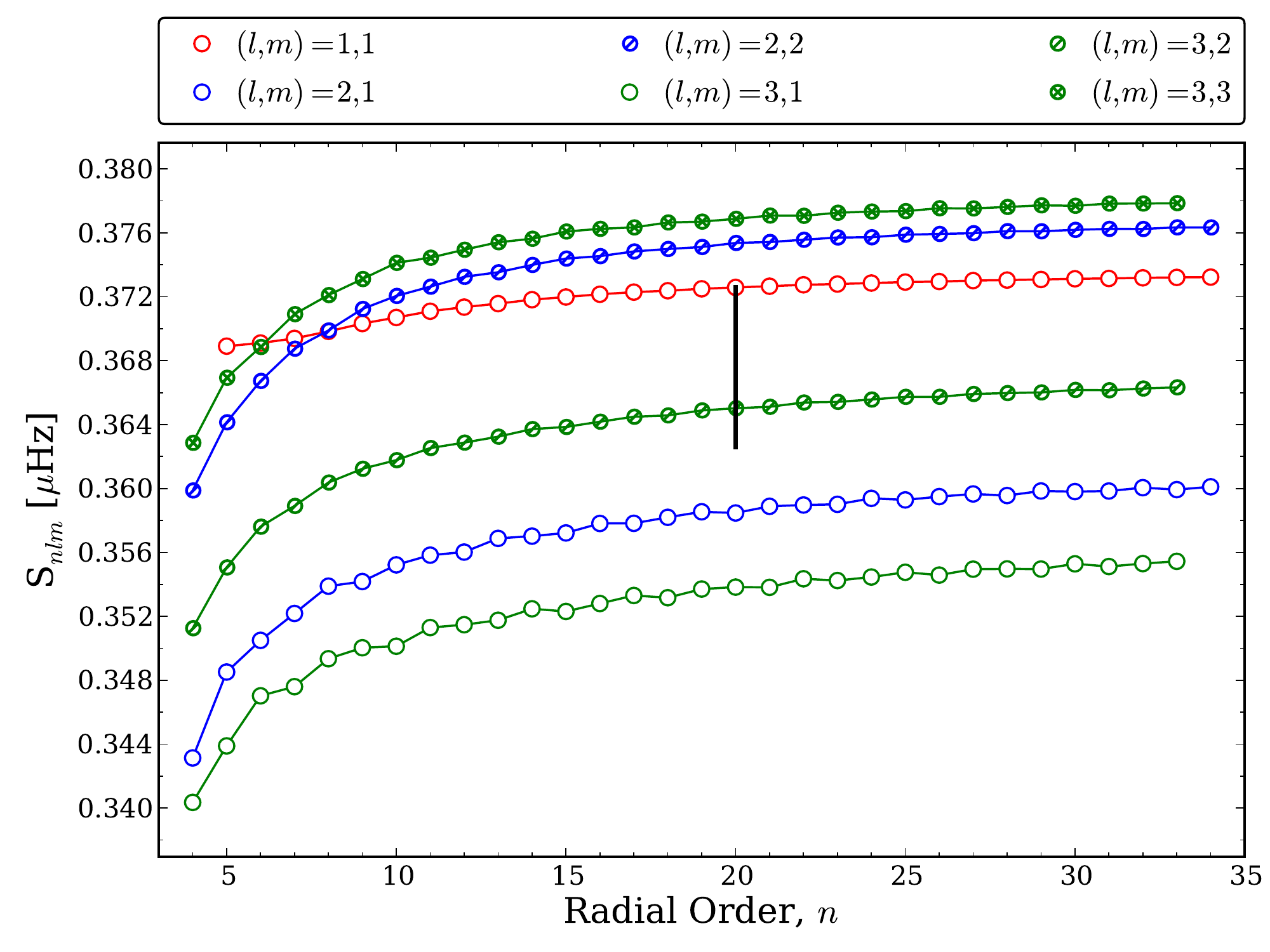}
 }
\caption{\footnotesize {\rm \textbf{Left:}} Rotation profile for a star rotating with a solar-like rotation. The small left panel gives the 2D rotation profile of the star, while the 1D rotation profiles are given in the small right panel for the subset of co-latitudes indicated on the 2D rotation profile. {\rm \textbf{Right:}} Calculated frequency splittings as a function of radial order for the solar-like rotation profile in the left panel and the rotation kernels computed from the adopted model star. The solid vertical reference bar at $n=\,$20 shows the extent of a 0.01$\rm \, \mu Hz$ frequency difference.}
\label{fig:rot_solar}
\end{figure*}

In the left panel of \fref{fig:rot_solar} the 2D profile for the solar-like rotation is shown, again with the rotation profile at specific latitudes given in the right panel. The right panel of the figure shows the resulting splittings for this rotation profile \citep[see also][]{2004SoPh..220..169G}. The first thing to notice is that now there is a clear difference in the splittings for different azimuthal orders and angular degrees.

The overall behavior of the modes corresponds to some extent to the variation seen in \fref{fig:cumkern_vol} for the amount of the kernel IP in the convective envelope. 
This can be understood if one approximates the rotation profile for a specific mode as being piecewise constant, where the parameters in \eqref{eq:pwc} are given by $\mathcal{D}$ (for $\mathcal{N}_0$) and a latitudinally weighted average of $\mathcal{A},\,\mathcal{B}$, and $\mathcal{C}$ (for $\mathcal{N}_1$) corresponding to the co-latitudes where the mode has its highest sensitivity (see \S~\ref{sec:ker}). At high radial orders the splittings keep increasing simply because there is a contribution from the inner radiative part of the star, which, as seen in \fref{fig:cumkern_vol}, increases slowly when going to the highest radial orders (equivalent to the decrease in the envelope IP).
The small wiggles seen in the splittings as a function of radial order stem from the sharp transitions in the rotation profile between radiative and convective zones. This can also be seen in \fref{fig:cumkern}. By the same reasoning the wiggles are more pronounced for, \eg, the $l=3,\, m=1$ mode than for the $l=1,\, m=1$ as the former is sensitive to higher latitudes than the latter and therefore experiences a stronger gradient between the interior and the envelope.

It is important to remember that in an actual measurement of splittings only a subset of the points in \fref{fig:rot_solar} would be obtainable with the stellar inclination angle as the main determining factor in the selection of splittings that are observable (see \S~\ref{sec:optinc}).

\subsection{Near-surface Shear Layer}

\begin{figure*}[ht]
\centering

\subfigure{
   \includegraphics[width=0.47\textwidth] {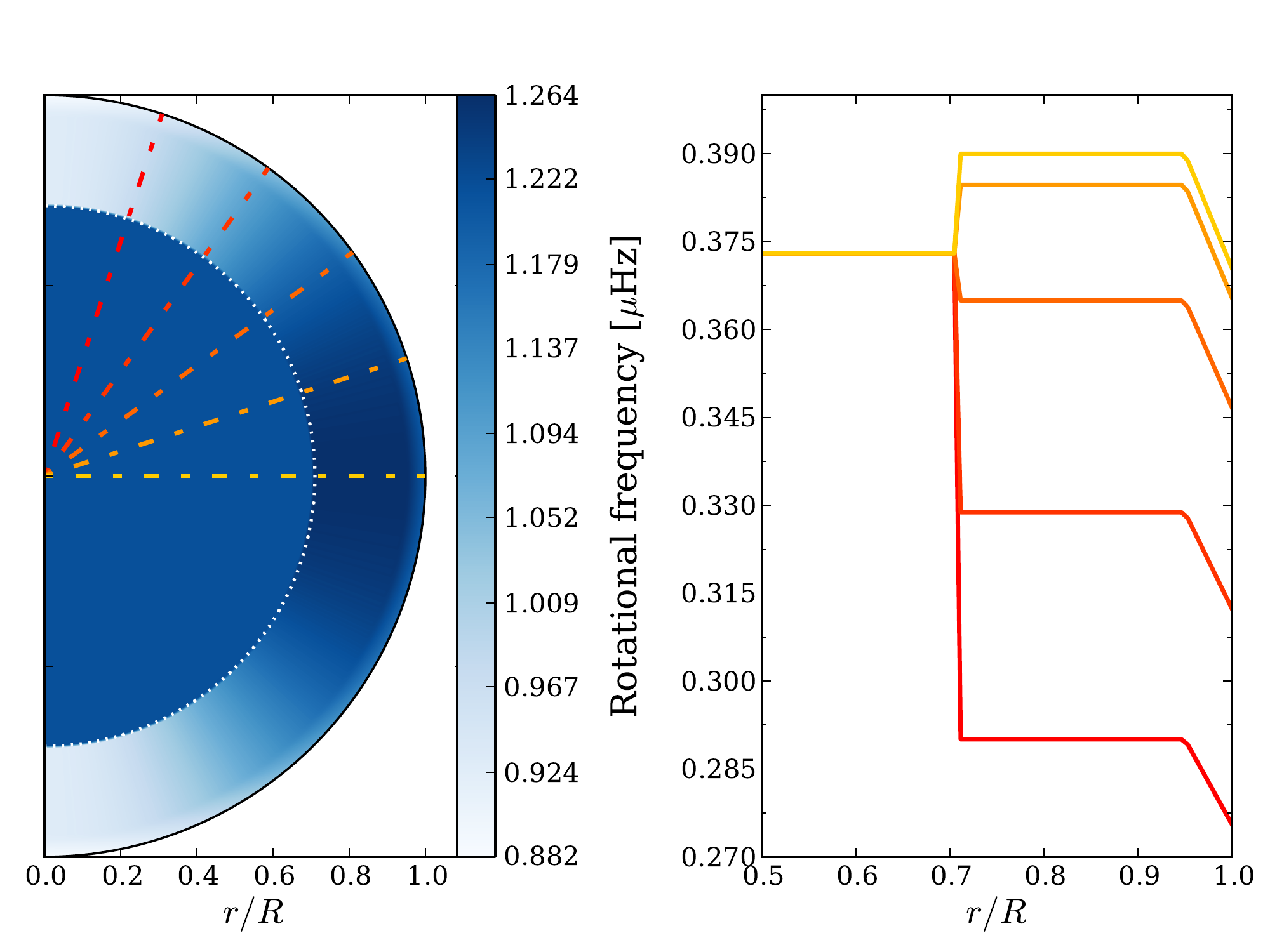}
 }\qquad
 \subfigure{
   \includegraphics[width=0.47\textwidth] {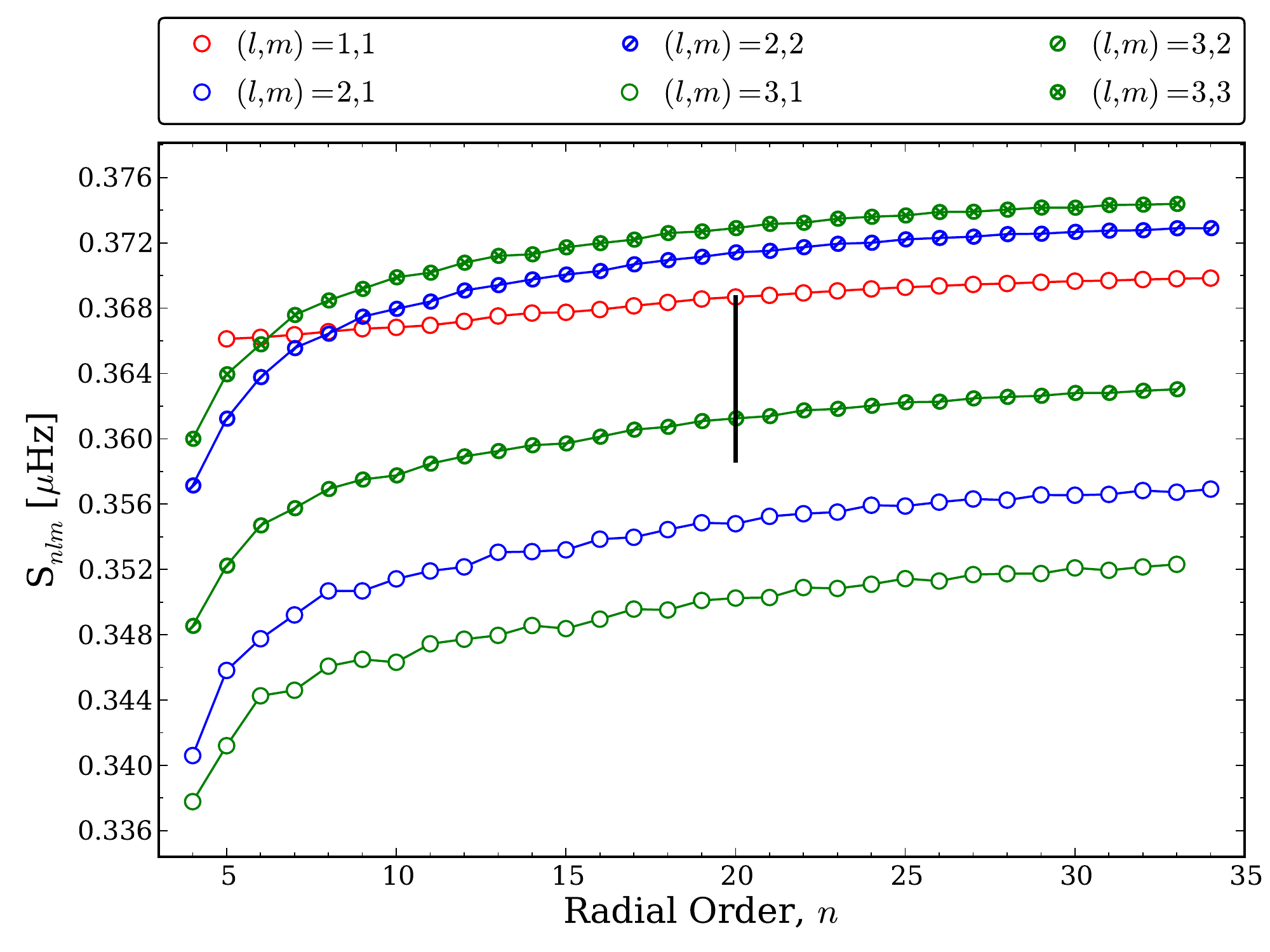}
 }
\caption{\footnotesize {\rm \textbf{Left:}} Rotation profile for a star rotating with a solar-like profile, \ie, with the poles rotating faster than the equator, but with the addition of a near-surface shear layer in the outer $5\%$ of the radius. The small left panel give the 2D rotation profile of the star, while the 1D rotation profiles are given in the small right panel for the subset of co-latitudes indicated on the 2D rotation profile. {\rm \textbf{Right:}} Calculated frequency splittings as a function of radial order for the solar-like surface shear rotation profile in the left panel and the rotation kernels computed from the adopted model star. The solid vertical reference bar at $n=\,$20 shows the extent of a 0.01$\rm \, \mu Hz$ frequency difference.}
\label{fig:rot_solar_ss}
\end{figure*}

The question now arises whether measurements of the splittings of the low-degree modes would be able to tell about some of the finer and smaller-scale details of the rotation profile. One such fine detail could, as mentioned above, be a near-surface shear layer as seen in the Sun, with a decrease in the rotation rate at all latitudes in the outer few percent of the stellar radius. Information on such a property of the rotation profile could be very important if, for instance, asteroseismic measures of the rotation were to be compared with other methods sensitive solely to the surface rotation. In the presence of a near-surface shear there would be differences in the measured rotation rates as the asteroseismic measure is sensitive not only to the surface but to the entire propagation region of the respective modes for which splittings are measured.
In the left panel of \fref{fig:rot_solar_ss} a solar-like rotation profile is shown, and it is similar to the profile of \fref{fig:rot_solar} except for the fact that a $5\%$ linear decrease in the rotation rate at each latitude is imposed in the outer $5\%$ in radius of the star. The splittings from this rotation profile are shown in the right panel of \fref{fig:rot_solar_ss}. These splittings again follow the behavior seen in \fref{fig:rot_solar}, but there are small differences. First, the splittings of all modes have been reduced, simply because they all are sensitive to the outer $5\%$ of the star. The amount by which different degrees are changed is, however, not exactly the same because the absolute change from a $5\%$ decrease in the rotation rate is largest at the equator, and thereby modes that are sensitive to the equatorial region will be affected more than the modes sensitive to higher latitudes.

\begin{figure}
\centering
\includegraphics[scale=0.4]{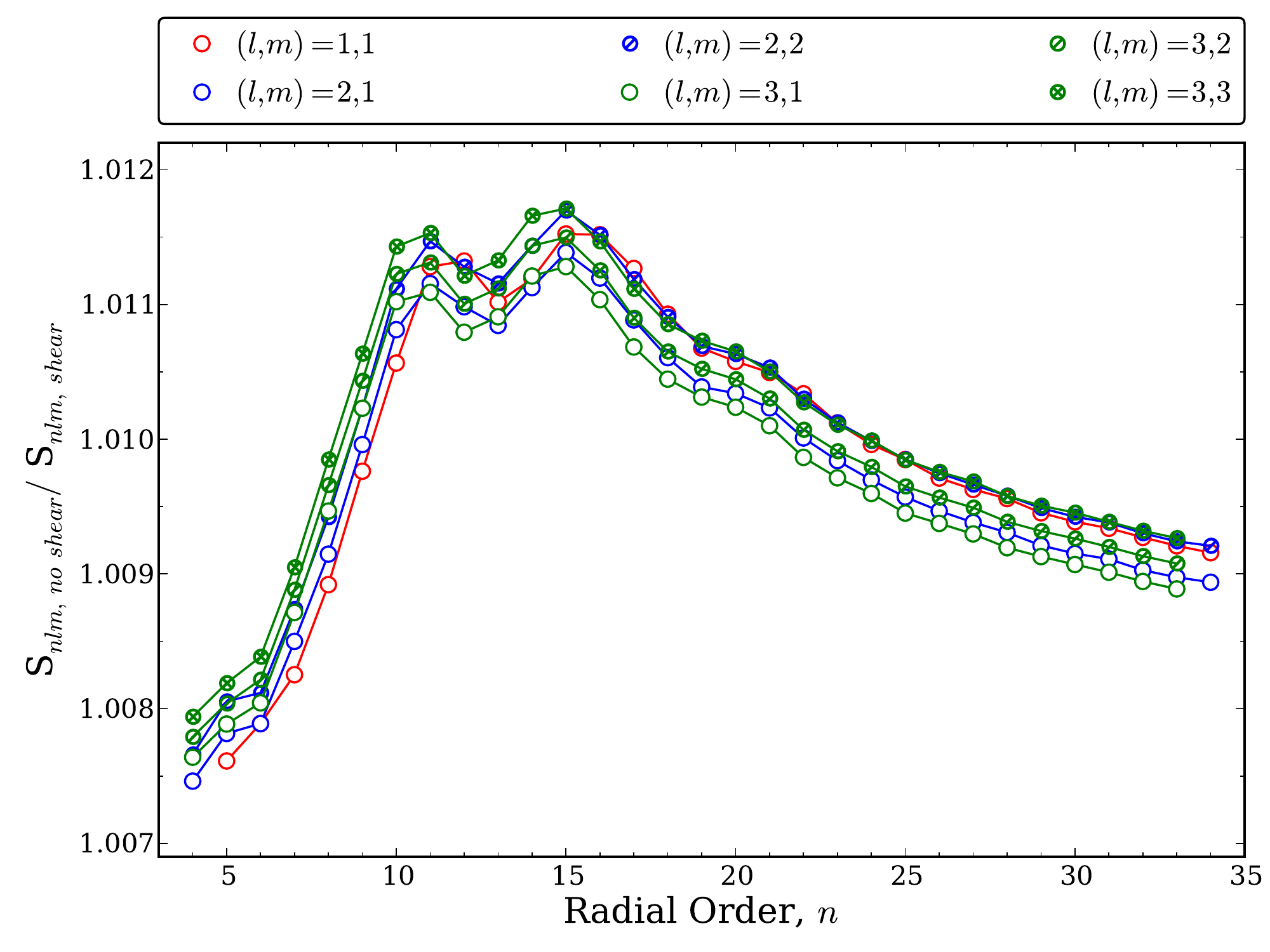}
\caption{\footnotesize Relative difference between splitting of a solar-like rotation profile (\fref{fig:rot_solar}) and a similar profile with a $5\%$ surface shear in the outer $5\%$ in radius (\fref{fig:rot_solar_ss}) as a function of radial order.}
\label{fig:solar_rot_split_over_solar_rot_shear}
\end{figure}

In \fref{fig:solar_rot_split_over_solar_rot_shear} we show the splittings from the solar rotation profile (\fref{fig:rot_solar}) divided by the ones where surface shear was included (\fref{fig:rot_solar_ss}). The variation in the curves as a function of radial order can be understood from the change in the amount of the kernel that is located in the outer convection zone (see \fref{fig:cumkern_vol}), where a maximum is seen around $n\sim 13$. As the addition of a surface shear only affects the outer $5\%$, the variation in the relative change of the splittings in \fref{fig:solar_rot_split_over_solar_rot_shear} can be seen as the change in kernel IP in the outer $5\%$ as a function of radial order. For the particular rotation rates used for the two sets of splitting curves the mean collective difference in the splittings is of the order ${\sim}1\%$. This, however, gives no information on the surface shear, only that the change in the relative difference seen in \fref{fig:solar_rot_split_over_solar_rot_shear} as a function of radial order contains potential information. This change is, on the other hand, of the order $10^{-3}$, which will certainly be beyond the realm of asteroseismology of slowly rotating solar-like stars in the foreseeable future.   

\subsection{Cylindrical Differential Rotation}
\label{sec:axial}

Cylindrical rotation profiles are generally associated with moderate ($P_{\rm rot}\sim 10$ days) to fast ($P_{\rm rot}\sim 1$ days) rotation \citep[][]{2012MNRAS.423.3344K}. 
The surface differential rotation of the \emph{CoRoT} star \emph{CoRoT-2a} was, by means of spot modeling by \citet{2009A&A...506..263F}, found to be $\Delta\Omega\sim0.11\, \rm rad\, day^{-1}$, equivalent to $\Delta\mathcal{N}\approx0.2\, \rm \mu Hz$ \citep[see also][who performed a similar analysis]{2010A&A...514A..39H}. \emph{CoRoT-2a} is a young Sun, having a mass of roughly one solar mass, but an age of only $0.5$ Gyr, and with a rotation period of $P_{\rm rot}\sim4.5$ days it can be seen as a moderate to fast rotator. For this star \citet{2011A&A...530A..48K} found by means of a mean-field model an iso-rotation profile that showed a somewhat cylindrical profile. \citet{2011A&A...530A..48K} also found a cylindrical profile when modeling a hypothetical rapidly rotating young Sun, imposing a rotation period of $P=1.33$ days. 

Another example comes with the rapidly rotating ($P_{\rm rot}\sim 0.51$ days) young dwarf star AB Doradus, observed by \citet{1997MNRAS.291....1D}. By means of mean-field modeling, \citet{2013IAUS..294..399K} were able to model the observed differential rotation for this star and found a cylindrical iso-rotation structure, once again supporting the statement of cylindrical rotation in fast rotators. The same is the case for the mean-field modeling by \citet[][]{2011MNRAS.411.1059K} of the \emph{MOST (Microvariability and Oscillations of Stars).} stars $\epsilon$ Eri \citep[][]{2006ApJ...648..607C} and $\kappa^1$ Cet \citep[][]{2007ApJ...659.1611W}.
 
\begin{figure*}[ht]
\centering

\subfigure{
   \includegraphics[width=0.47\textwidth] {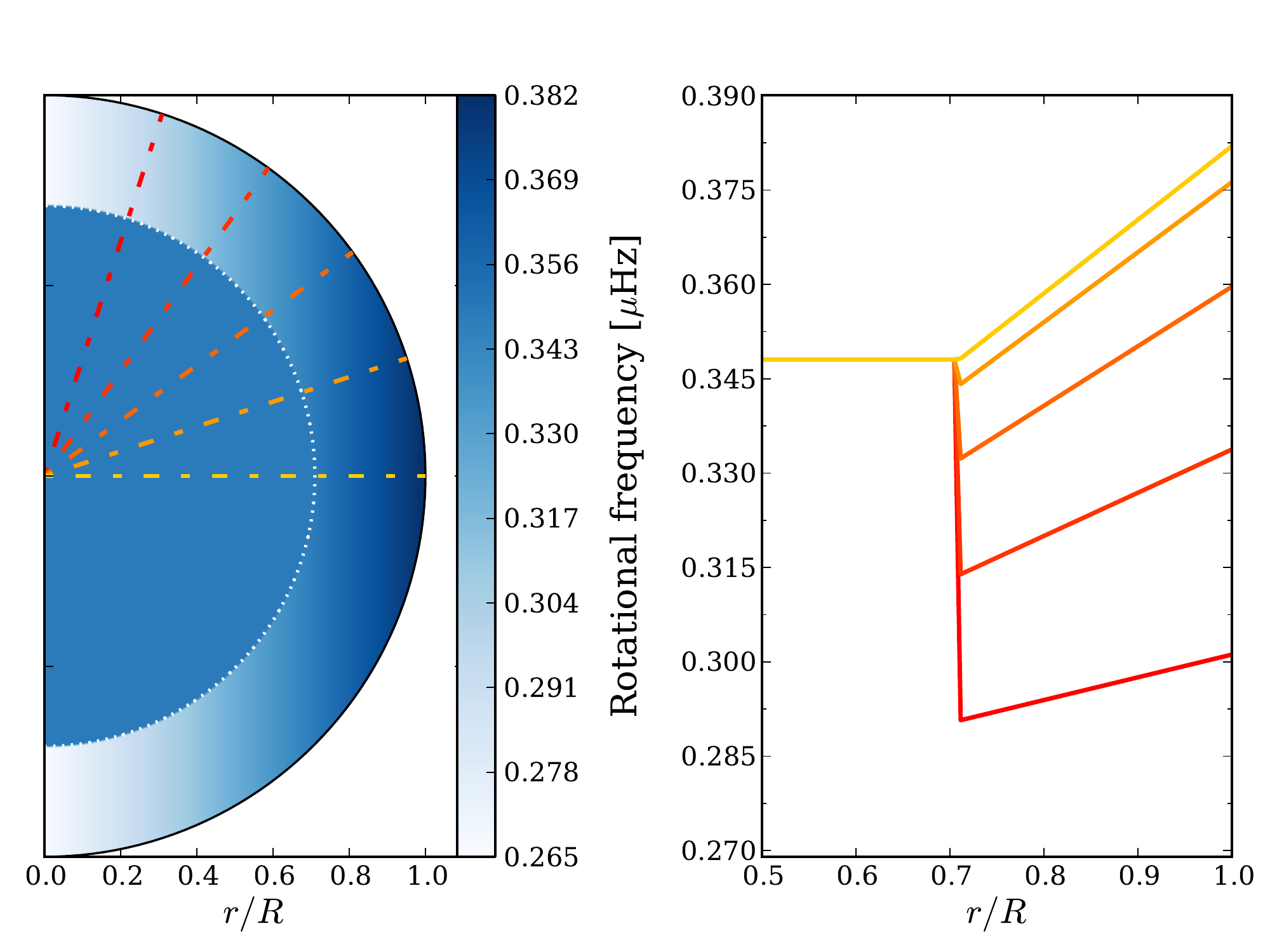}
 }\qquad
 \subfigure{
   \includegraphics[width=0.47\textwidth] {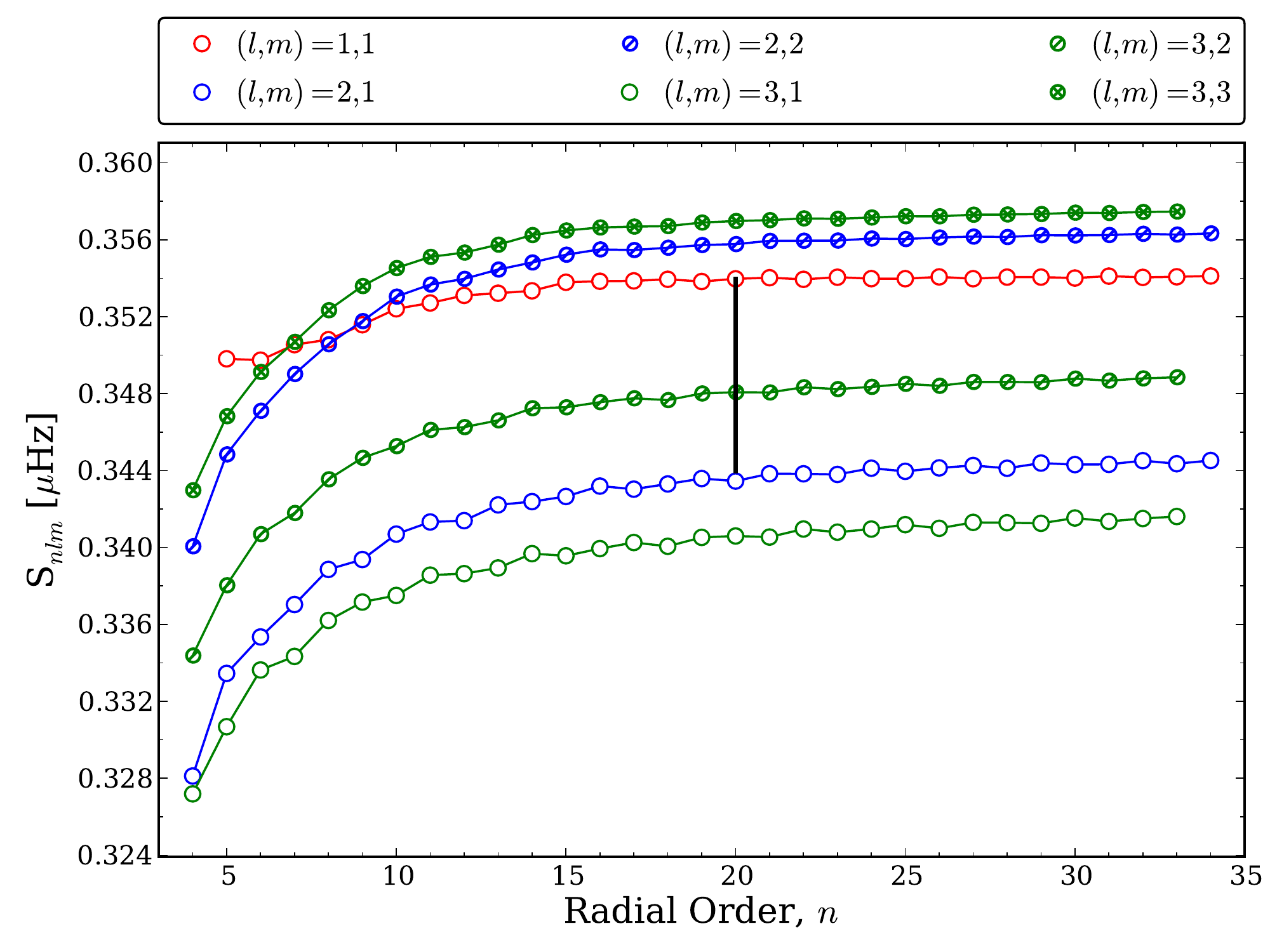}
 }
\caption{\footnotesize {\rm \textbf{Left:}} Rotation profile for a star rotating with a cylindrical profile and solar-like in the sense that the equator is rotating faster than the poles. The small left panel gives the 2D rotation profile of the star, while the 1D rotation profiles are given in the small right panel for the subset of co-latitudes indicated on the 2D rotation profile. {\rm \textbf{Right:}} Calculated frequency splittings as a function of radial order for the cylindrical rotation profile in the left panel and the rotation kernels computed from the adopted model star. The solid vertical reference bar at $n=\,$20 shows the extent of a 0.01$\rm \, \mu Hz$ frequency difference.}
\label{fig:rot_axial}
\end{figure*}

We can parameterize a cylindrical rotation profile as follows:

\begin{equation}\label{eq:axial}
\mathcal{N}(r, \theta)  =\left\{ 
\begin{array}{l l}
    \alpha + \beta\,\sin\theta\,  (r/R)   & \quad r>r_{\rm bcz}\\
    \alpha & \quad r < r_{\rm bcz}\\
  \end{array} \right.
\end{equation}
Here $\alpha$ is the rotation frequency in the radiative interior and $\theta$ is the co-latitude, so that $\sin\theta\, (r/R)$ denotes the distance from the rotation axis in units of the stellar radius. The parameter $\beta$ gives the rate of the change away from the rotation axis. 
The size of the surface differential rotation, constrained by \eqref{eq:shear}, is given by
\begin{align}\label{eq:axial_delta}
\Delta\mathcal{N} &= \mathcal{N}_{\rm eq.} - \mathcal{N}_{\rm pole}\\ \nonumber
	&= (\alpha + \beta) - \alpha\\ \nonumber
	&= \beta \, .  \nonumber
\end{align}
For a given model the parameter $\beta$ is thus given directly. The value for $\alpha$ can be found as (see Appendix~\ref{app:param})
\begin{equation}
\alpha = \langle P_{\rm rot} \rangle^{-1} - \tfrac{2}{\pi}\Delta\mathcal{N} \, .
\end{equation}
The parameters obtained for our stellar model are given in Table~\ref{tab:models3}. In the left panel of \fref{fig:rot_axial} the rotation profile for cylindrical rotation is illustrated. 

\begin{figure*}[ht]
\centering

\subfigure{
   \includegraphics[width=0.47\textwidth] {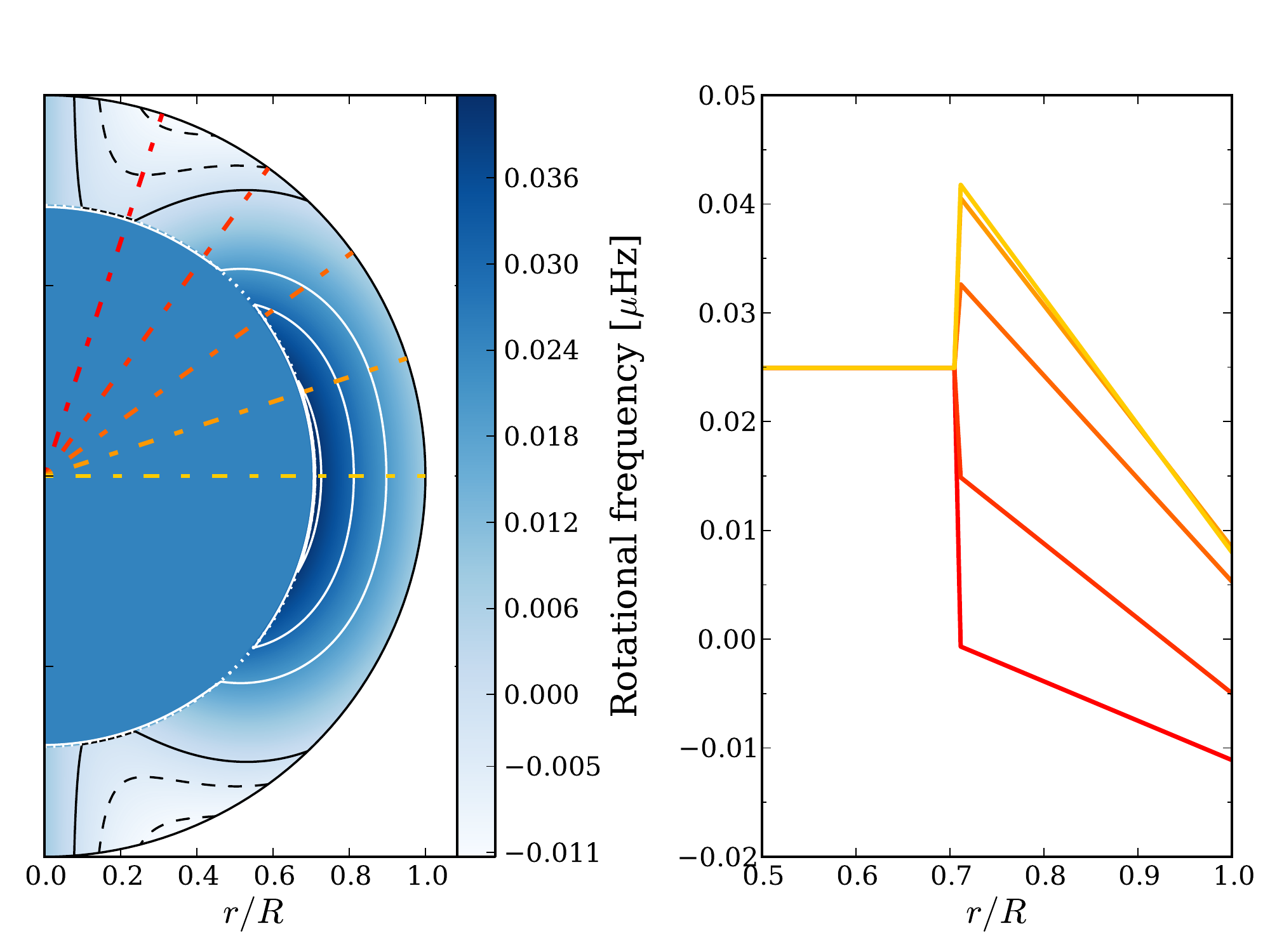}
 }\qquad
 \subfigure{
   \includegraphics[width=0.47\textwidth] {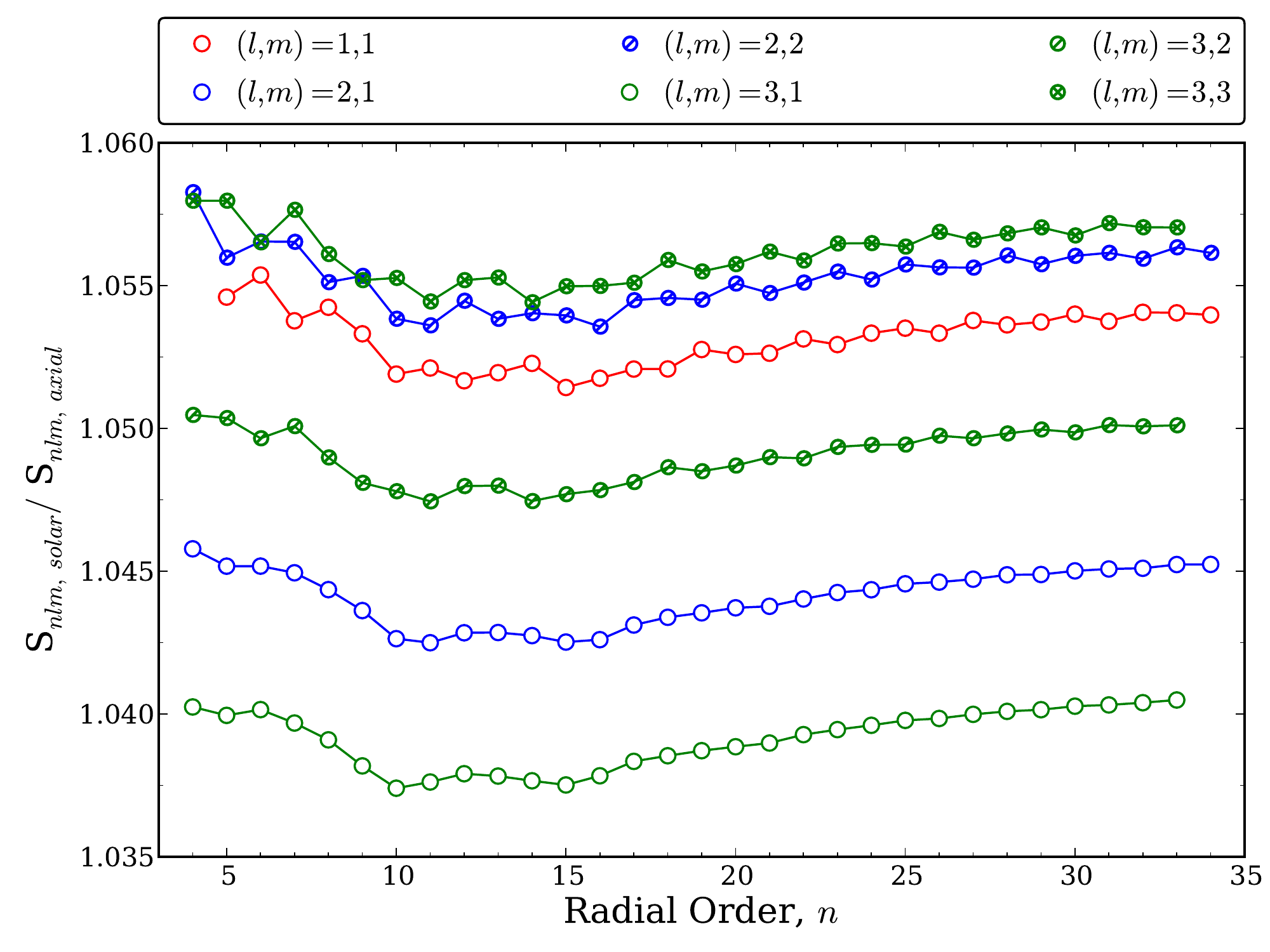}
 }
\caption{\footnotesize {\rm \textbf{Left:}} Contour plot of the difference between the solar-like rotation profile (\fref{fig:rot_solar}) and the cylindrical rotation profile (\fref{fig:rot_axial}). Rotation profiles are given in the small right panels for the subset of co-latitudes indicated on the 2D contour plot. {\rm \textbf{Right:}} Ratio between the frequency splittings as a function of radial order for the solar-like rotation profile and those for the cylindrical rotation profile.}
\label{fig:sub_axial_diff}
\end{figure*}

We present in the right panel of \fref{fig:rot_axial} the results from the computation of the splittings. The splittings for the cylindrical profile are very similar to the ones for the solar-like rotation profile (\fref{fig:rot_solar}). There are, of course, small differences, the main being that the variation with radial order is not as smooth as for the solar-like rotation. Toward high values of the radial order the splittings become nearly constant, whereas for the solar-like rotation there was a small increase in the symmetric splittings toward the highest $n$-values. The difference between the splittings is better portrayed in the right panel of \fref{fig:sub_axial_diff}, where the splittings from the solar-like rotation profile used here have been divided by the ones from the cylindrical profile. All splittings are higher for the solar-like rotation but vary for the different orders, and there is a common trend in the variation as a function of radial order for all degrees. These differences can be better understood when looking at the differences in the rotation rates for the two profiles; this is shown in the left panel of \fref{fig:sub_axial_diff}, where the rotation rates from the cylindrical profile are subtracted from those of the solar-like profile. The difference between the two profiles shows that in the bulk of the star the solar-like profile has higher rotation rates, most noticeable in the equatorial regions, and only at very high latitudes is the sign of the difference reversed. As even the modes with sensitivities at the highest latitudes only partly cover the region of "cylindrical higher than solar-like", it makes sense that all splittings are reduced for the cylindrical configuration. The variation with degree comes directly from the latitudinal sensitivities for the different modes. The variation with radial order is less trivial and must be found in the way the IPs of the kernels vary, not only with radius but also with latitude, and how the variations coincide with the regions of different signs in the difference between the two rotation profiles.

\subsection{Anti-solar Differential Rotation}

A very interesting issue in models of differential rotation is whether or not one can observe whether the differential rotation in latitude is solar or anti-solar, here in the sense that the equator is rotating faster than the pole for the solar profile. Detection of anti-solar differential rotation by spot tracking via cross-correlation of Doppler images has been reported for giant stars and binaries by a number of groups \citep[see, \eg,][]{1999ApJS..121..547V,2001A&A...374..171H,2003A&A...408.1103S,2007AN....328.1075W,2007A&A...474..165K}. See \citet[][]{2012IAUS..286..268K} for an overview. In the spectroscopic line-profile observations (Fourier method) by \citet[][]{2003A&A...398..647R} a few stars were found to have a differential rotation consistent with anti-solar; however, the authors find that the presence of polar spots presents a more likely explanation. 

\begin{figure*}[ht]
\centering

\subfigure{
   \includegraphics[width=0.47\textwidth] {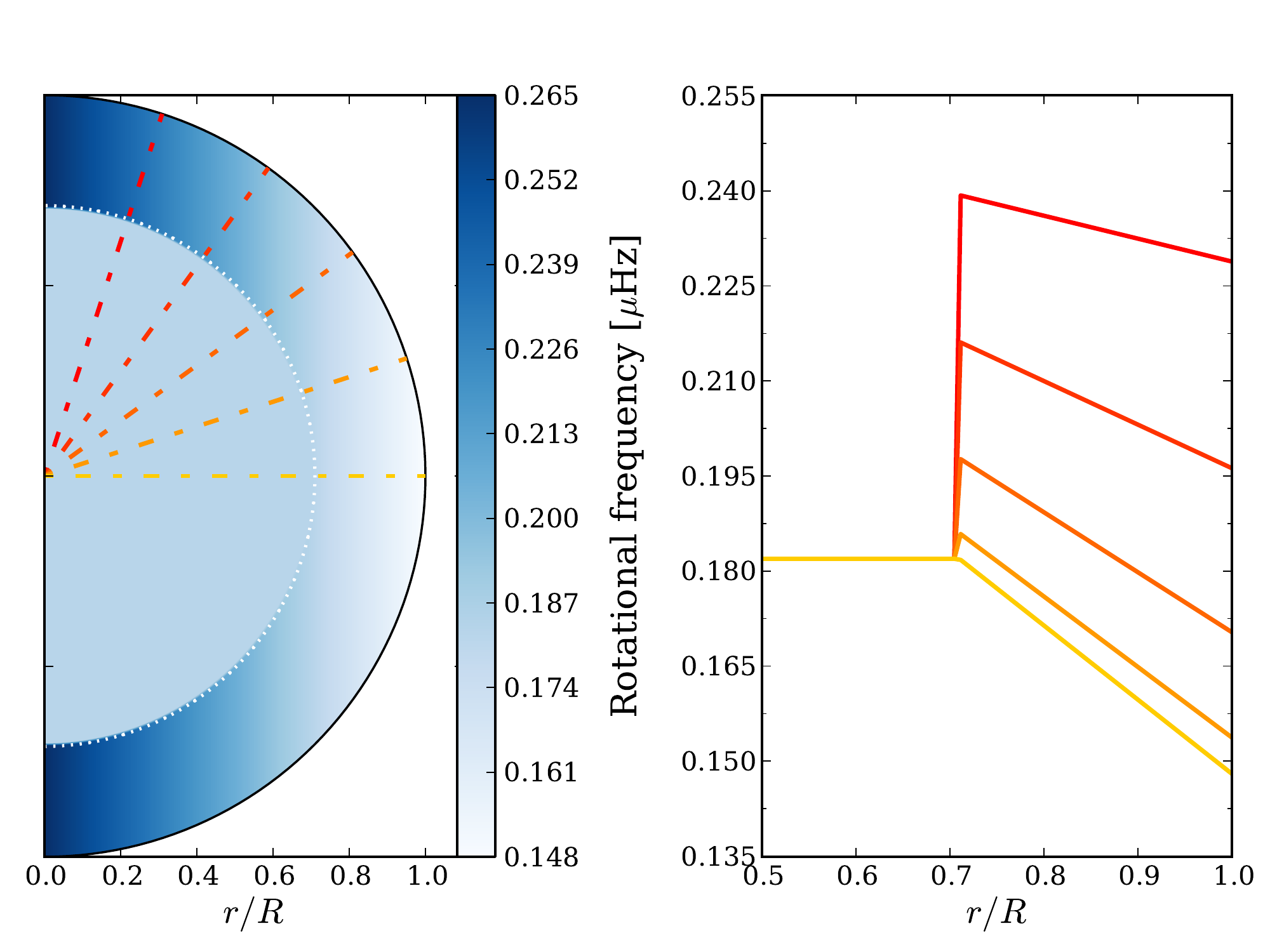}
 }\qquad
 \subfigure{
   \includegraphics[width=0.47\textwidth] {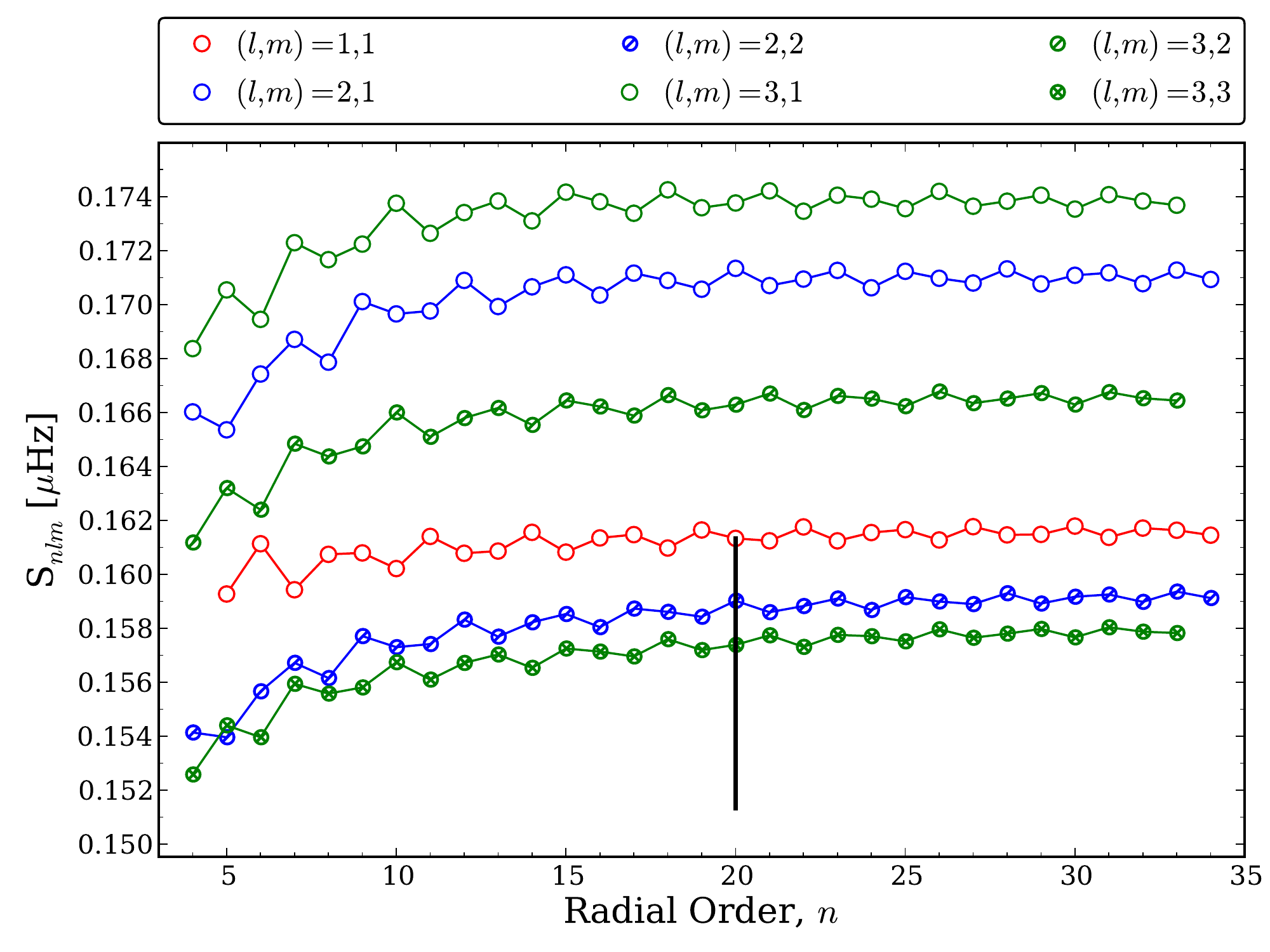}
 }
\caption{\footnotesize {\rm \textbf{Left:}} Rotation profile for a star rotating with an anti-solar cylindrical profile, \ie, with the poles rotating faster than the equator. The small left panel gives the 2D rotation profile of the star, while the 1D rotation profiles are given in the small right panel for the subset of co-latitudes indicated on the 2D rotation profile. {\rm \textbf{Right:}} Calculated frequency splittings as a function of radial order for the anti-solar rotation profile in the left panel and the rotation kernels computed from the adopted model star. The solid vertical reference bar at $n=\,$20 shows the extent of a 0.01$\rm \, \mu Hz$ frequency difference.}
\label{fig:rot_solar_anti}
\end{figure*}

In the left panel of \fref{fig:rot_solar_anti} an anti-solar cylindrical (following \eqref{eq:axial}) rotation profile is given.
The splittings for this profile are given in the right panel of \fref{fig:rot_solar_anti}. Here the relative ordering of the splittings of different degree is swapped around the position of the $l=1$ splittings. The reason is that now the modes with sensitivities in latitude away from the equator will be split more relative to the ones that sense the slowly rotating equatorial regions.
By observing the ordering of the splittings of different degrees and different azimuthal orders relative to each other, one would be able to discern whether the observed differential rotation is indeed solar or anti-solar in nature. Furthermore, only a few splittings should in principle be needed to make this distinction. 

\section{Variations on solar-like differential rotation}
\label{sec:depen}

In this section we see how the behavior of the splittings changes by varying three influential parameters individually, namely, the latitudinal gradient, the depth of the transition from differential to solid-body rotation, and the radial gradient. This approach gives a better understanding and feeling of what to expect from observations and, more importantly, how to interpret observations in a very simple manner without doing detailed inversions. All tests are performed assuming a solar-like rotation profile (see \fref{fig:rot_solar}).


\subsection{Effect of Changing the Latitudinal Gradient}
\label{sec:latgra}

\begin{figure*}
	\centering
	\includegraphics[scale=0.35]{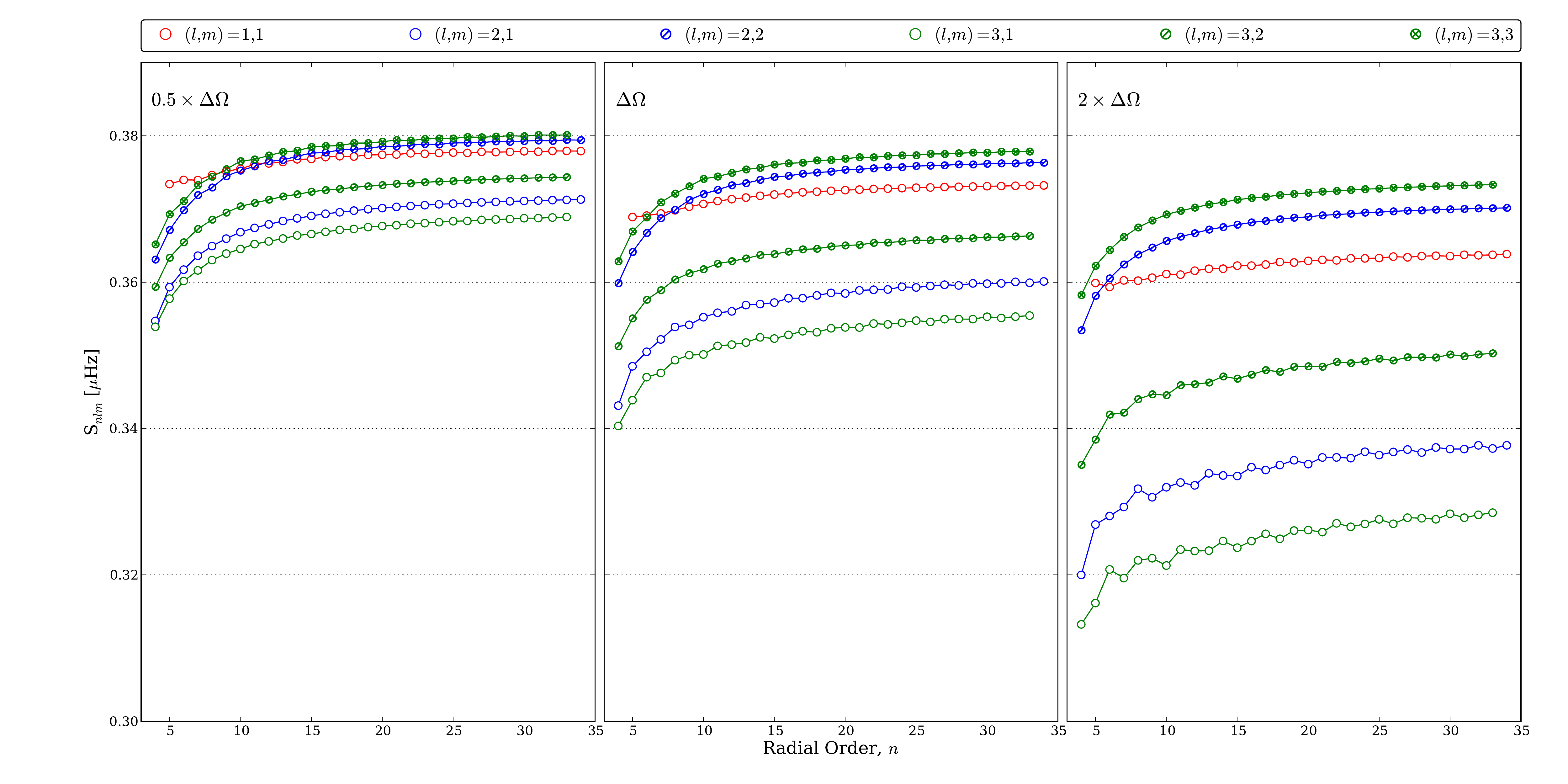}
	\caption{\footnotesize Effect in the splittings from changing the latitudinal gradient of the differential rotation for a solar-like rotation profile as a function of radial order. {\rm \textbf{Left:}} Splittings when adopting half the amount of differential rotation found for the model star, \ie, $\mathcal{B}=\,0.5\,\times\,-0.049\rm \,\mu Hz$ and $\mathcal{C}=\,0.5\,\times\,-0.068\rm \,\mu Hz$. {\rm \textbf{Middle:}} Splittings when adopting the amount of differential rotation found for the model star (see Table~\ref{tab:models3}), \ie\, $\mathcal{B}=\,-0.049\rm \,\mu Hz$ and $\mathcal{C}=\,-0.068\rm \,\mu Hz$. {\rm \textbf{Right:}} Splittings when adopting twice the amount of differential rotation found for the model star, \ie, $\mathcal{B}=\,2\,\times\,-0.049\rm \,\mu Hz$ and $\mathcal{C}=\,2\,\times\,-0.068\rm \,\mu Hz$.}
\label{fig:shear_var}
\end{figure*}

To see the effect of changing the latitudinal gradient or the surface differential rotation, the values of $\mathcal{B}$ and $\mathcal{C}$ entering \eqref{eq:solar_rot_profile} were multiplied by $0.5, 1$, and $2$, so the ratio between these two parameters stays the same. The remainder of the parameters were left unchanged. The resulting splittings are shown in \fref{fig:shear_var}. As seen, the splittings of all modes decrease with an increase in the latitudinal gradient, which makes good sense as the global mean rotation rate of the star drops with such an increase. The absolute change is greater the higher the latitudinal sensitivity of the modes, which is due to the nonlinear change in the dependence on co-latitude in \eqref{eq:solar_rot_profile}. Furthermore, the latitudinal spread in the sensitivity of the modes is important. The modes with higher latitudinal spread exhibit the largest change as they reach further into the high-latitude regimes, which again, coupled with the nonlinearity of the co-latitude dependence, makes for a larger change in splitting. The overall behavior is thus a decrease in all mode splittings but an increase in the range of splitting values. As one might have anticipated, this means that the higher the surface differential rotation is, the easier will it be from observations to separate the different splittings and in turn make inferences on the differential rotation. 

\begin{figure*}[ht]
\centering

\subfigure{
   \includegraphics[width=0.4\textwidth] {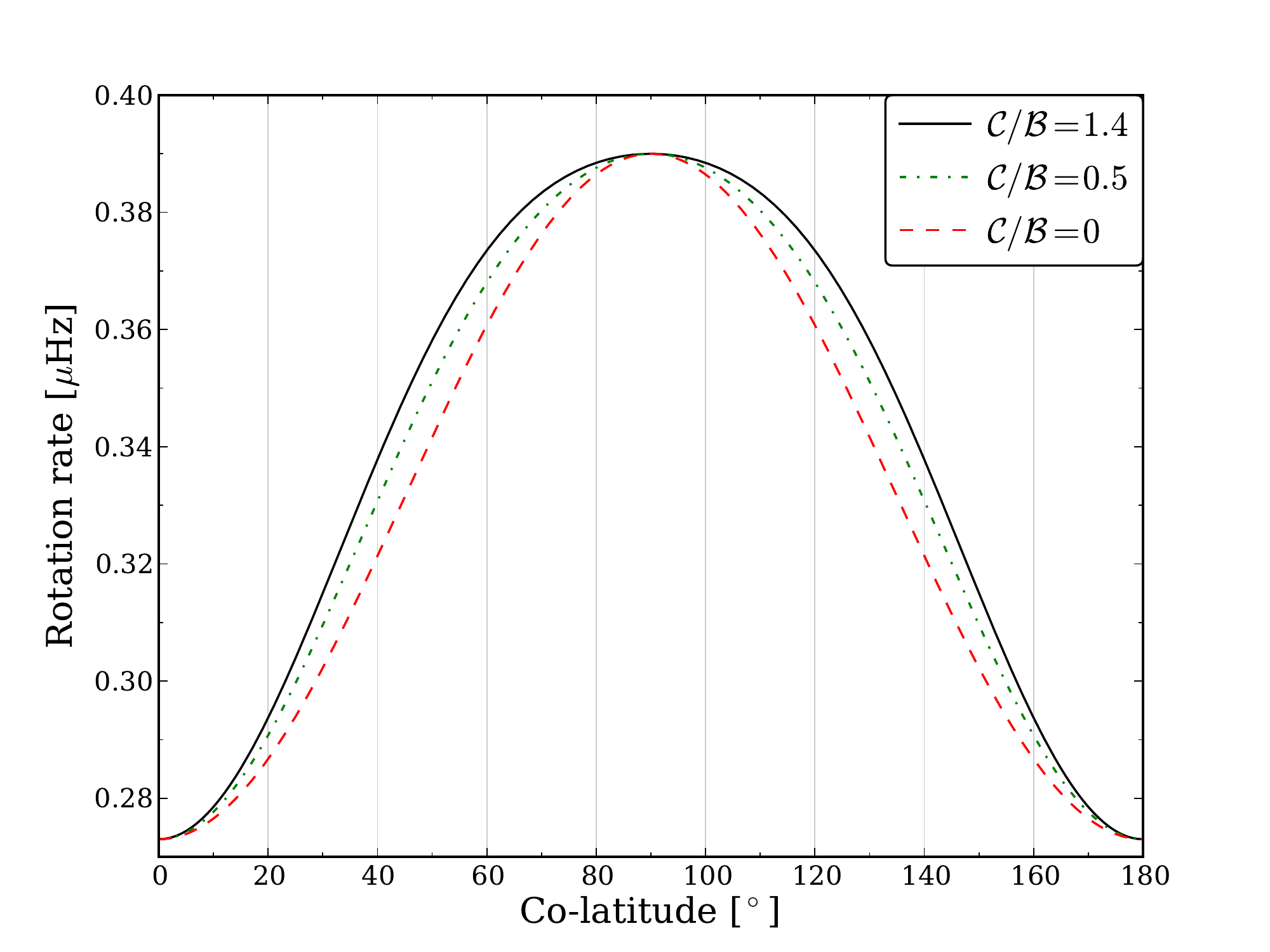}
 }\quad
 \subfigure{
   \includegraphics[width=0.5\textwidth] {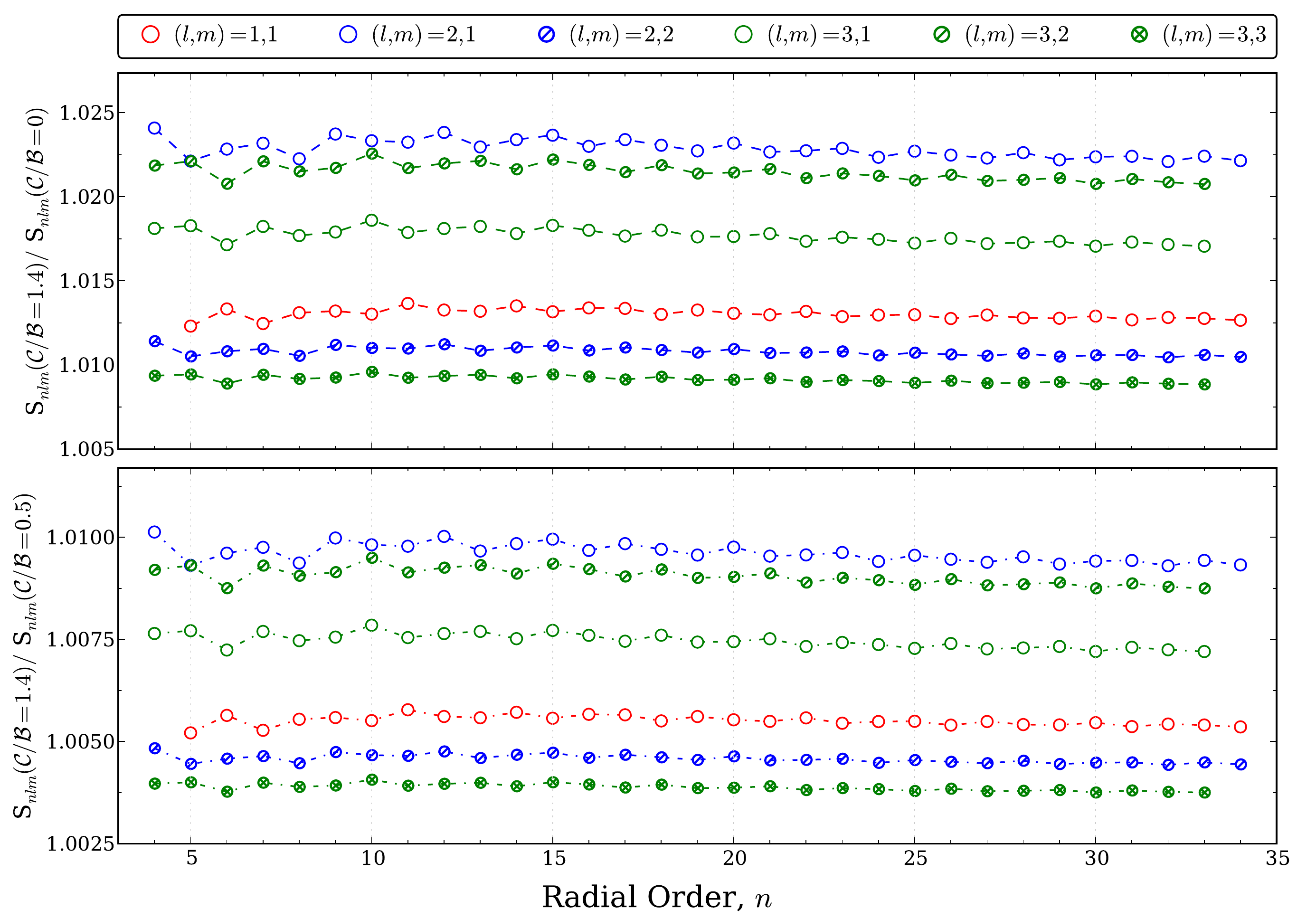}
 }
\caption{\footnotesize {\rm \textbf{Left:}} Surface rotation rates as a function of co-latitude for a solar-like differential rotation profile, with different values of the ratio $\mathcal{C}/\mathcal{B}$. The profile with $\mathcal{C}/\mathcal{B}=\,$1.4 corresponds to the one for our model star. For all profiles the total amount of surface differential rotation is the same. {\rm \textbf{Right:}} Effect on the splittings from changing the ratio of $\mathcal{C}/\mathcal{B}$ from the value applicable to our model, with ratios in splittings given as a function of radial order.}
\label{fig:CBrat}
\end{figure*}

One can, of course, also imagine that the ratio between $\mathcal{C}$ and $\mathcal{B}$ could be different from the value of ${\sim}1.4$ used for our model. By lowering the ratio, all the while keeping the sum of $\mathcal{C}$ and $\mathcal{B}$ and thereby the total amount of surface differential rotation constant, the fall-off in the rotation rate will become more steep near the equator and the rotation rates for all co-latitudes, with the exception of the equator and the poles, will be decreased (see left panel of \fref{fig:CBrat}). The opposite holds when increasing the ratio. Note that a value of $\mathcal{C}/\mathcal{B}=0$ corresponds to a profile that only includes a $\cos^2 \theta$ in \eqref{eq:solar_rot_profile} to account for the surface differential rotation. Such a description of the profile is the one most often encountered in the other methods used for detecting surface differential rotation, \eg, in spot modeling.

The fact that all co-latitudes have a lower rotation rate results in a decrease in all splitting values (see right panel of \fref{fig:CBrat}). However, as the change in rotation rate is zero at the equator, modes that are mainly sensitive to this region will not be affected as much as, \eg, $l=2,$ $m=1$ modes, which are sensitive to a co-latitude of $\theta\sim51^{\circ}$ (\fref{fig:l2m1n20}) -- more or less the co-latitude where the change in rotation rate is at its largest. Nonetheless, there is almost no change in the variation of the splittings with $n$, and therefore no way to make the distinction between a rotation profile including both $\cos^2 \theta$ and $\cos^4 \theta$ and one that only includes a $\cos^2 \theta$ term. Because of this, one should take great care when converting observed splitting values into a measure of the surface differential rotation. To make the point clear, both profiles have the same value for the surface differential rotation (between equator and pole), but they do not give exactly the same splittings. We note, however, that the differences are small and here no more than a few percent.

\subsection{Effect of Changing the Depth for the Transition from Differential to Solid-body Rotation }

\begin{figure*}[htb]
	\centering
	\includegraphics[scale=0.35]{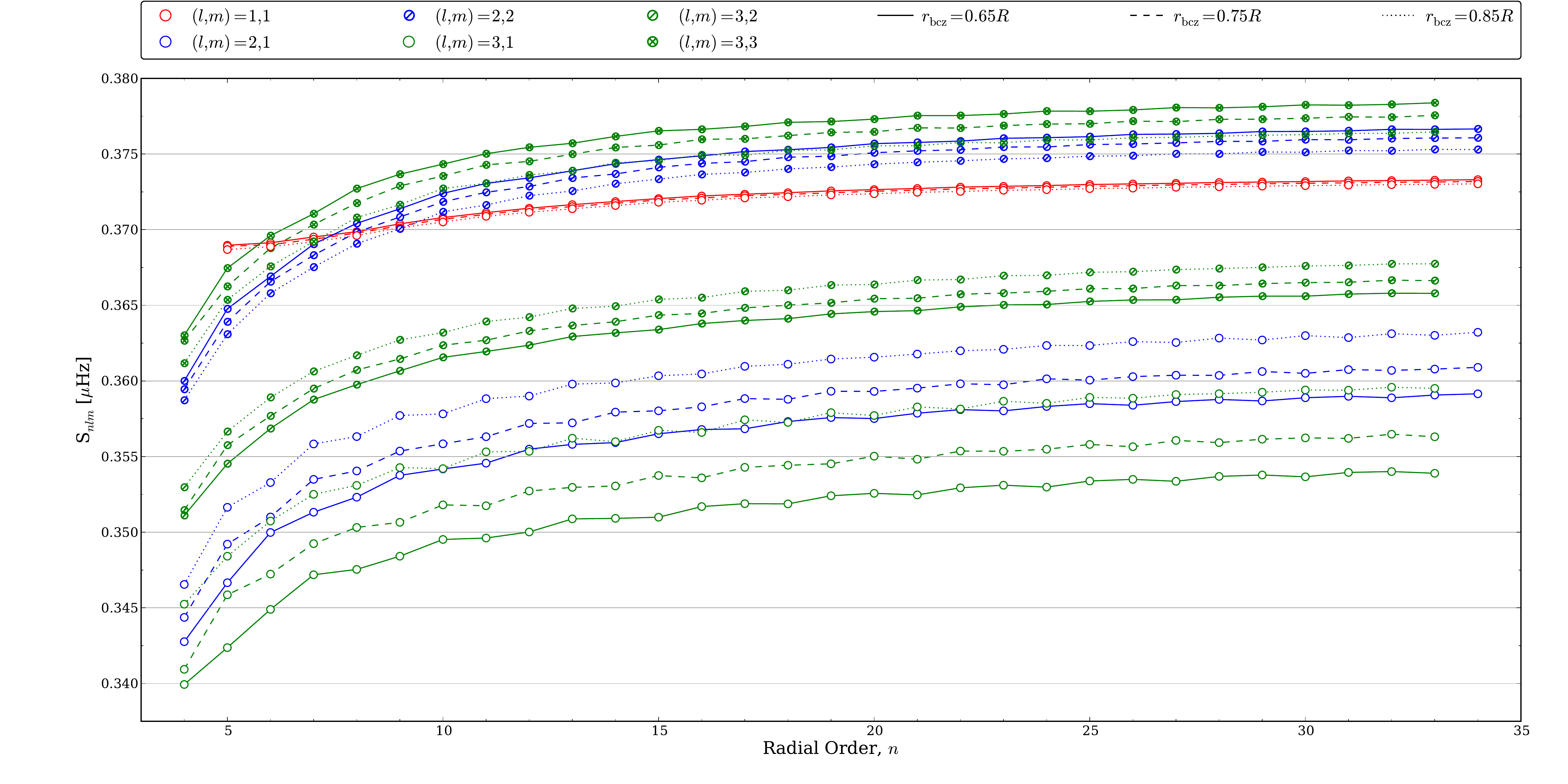}
	\caption{\footnotesize Effect in the splittings from changing the depth of the convection zone for a solar-like rotation profile as a function of radial order. The splittings for three depths are given, namely, $r_{\rm bcz}=\,$0.65$ R$, $r_{\rm bcz}=\,$0.75$ R$, and $r_{\rm bcz}=\,$0.85$ R$, and displayed with different line styles as indicated in the legend. }
\label{fig:bcz_var}
\end{figure*}

To test the dependence of the depth at which the transition from differential to solid-body rotation occurs, the value of $r_{\rm bcz}$ was changed in the rotation profile in the calculation of the splittings. \fref{fig:bcz_var} shows the resulting splittings for the values $r_{\rm bcz} = 0.65R, 0.75R, 0.85R$. The values of the profile parameters are given in Table~\ref{tab:models3}. First, it is seen that the behavior as a function of radial order stays nearly unchanged, with the small change found in the reversal of some of the wiggles. The insensitivity to radial order is to be expected as the radial gradient in rotation rate in the convection zone stays unchanged for each latitude. Second, we see a change in the splitting for the different $(l,m)$ combinations, with some increasing and some decreasing their splittings, and this with different amounts.  
Two elements come into play in describing this variation: (1) The latitude of highest sensitivity of the mode and the size of the interior-to-envelope ratio in rotation rates at that latitude; as an example, the $l=2, m=1$ mode is, according to \fref{fig:l2m1n20}, sensitive to a co-latitude of $\theta\sim 51^{\circ}$, at which the interior has a higher rotation rate than the envelope, such that the splitting decreases with an increase in the extent of the envelope and vice versa. (2) The latitudinal spread of the kernels of the modes, with a higher spread producing smaller changes. The best example of this is seen for the $l=1$ splitting, which is, to a good approximation, unaffected by the change in $r_{\rm bcz}$, as the very wide kernel of this mode samples latitudes of both signs in the radial gradient of rotation rate and the effects from the regions increasing the splittings are canceled out by the ones decreasing it. 
As with the increase in the latitudinal gradient of the surface differential rotation (\S~\ref{sec:latgra}), an increase in the depth of the convection zone will also make the detection of differential rotation easier as the splittings spread out. For this reason one might expect that it would be easier to detect differential rotation in older stars where the convective envelope has had time to grow once the star evolves off the main-sequence. On the other hand, we see from \fref{fig:gyro} that the latitudinal gradient decreases with age. The decrease is not large, but it will still act against the gain from a deeper convection zone. Furthermore, the absolute rotation rate will decrease with age (see \eqref{eq:prot}), which will have a much greater negative impact on the absolute difference between the splittings of different modes than the small gain from a deeper convection zone. 

In this test we did not assume any changes in the size of the differential rotation for a change in the depth of the convection zone. Mean-field models by \citet[][]{2011A&A...530A..48K} do, however, suggest that an increase in the depth of the convection zone will lead to a decrease in the differential rotation (latitude gradient) and thereby also make the separation of $m$-components more difficult. This is also suggested by convection simulations \citep[see][]{2012ApJ...756..169A}.
Nonetheless, the depth of the convection zone is also (in addition to the age) dependent on, for instance, the metallicity and mass of the star, with deeper convection zones for more metal-rich and less massive stars. So, on the one hand we find that a deeper convection zone for a given value of the mean rotation rate and surface shear makes it easier to detect the differential rotation from splittings. On the other hand, the older, cooler, and less massive stars having these deep convection zones will at the same time have smaller absolute values for the mean rotation rate \citep[][]{2007ApJ...669.1167B} and surface shear \citep[][]{2011A&A...530A..48K,2012MNRAS.423.3344K}, likely outweighing the potential gain from a deeper convection zone. Though it is difficult to say for sure, one might gain a bit in the detection of differential rotation in main-sequence solar-like stars by looking at the more metal-rich targets.

\subsection{Effect of Changing the Radial Gradient}

An effect from a change in the radial gradient is naturally of great interest. As mentioned in \S~\ref{sec:rates}, studies of red giants have revealed an order-of-magnitude difference between the rotation rates in the envelope and the core region of these stars. The situation is much more difficult when dealing with main-sequence solar-like stars, as we here have no mixed modes by which one could probe the core region more directly. Instead, we are faced with only a very small difference as a function of radial order between the sensitivity of the modes to the interior and the envelope (Figures~\ref{fig:cumkern} and \ref{fig:cumkern_vol}).

\begin{figure*}[htb]
	\centering
	\includegraphics[scale=0.35]{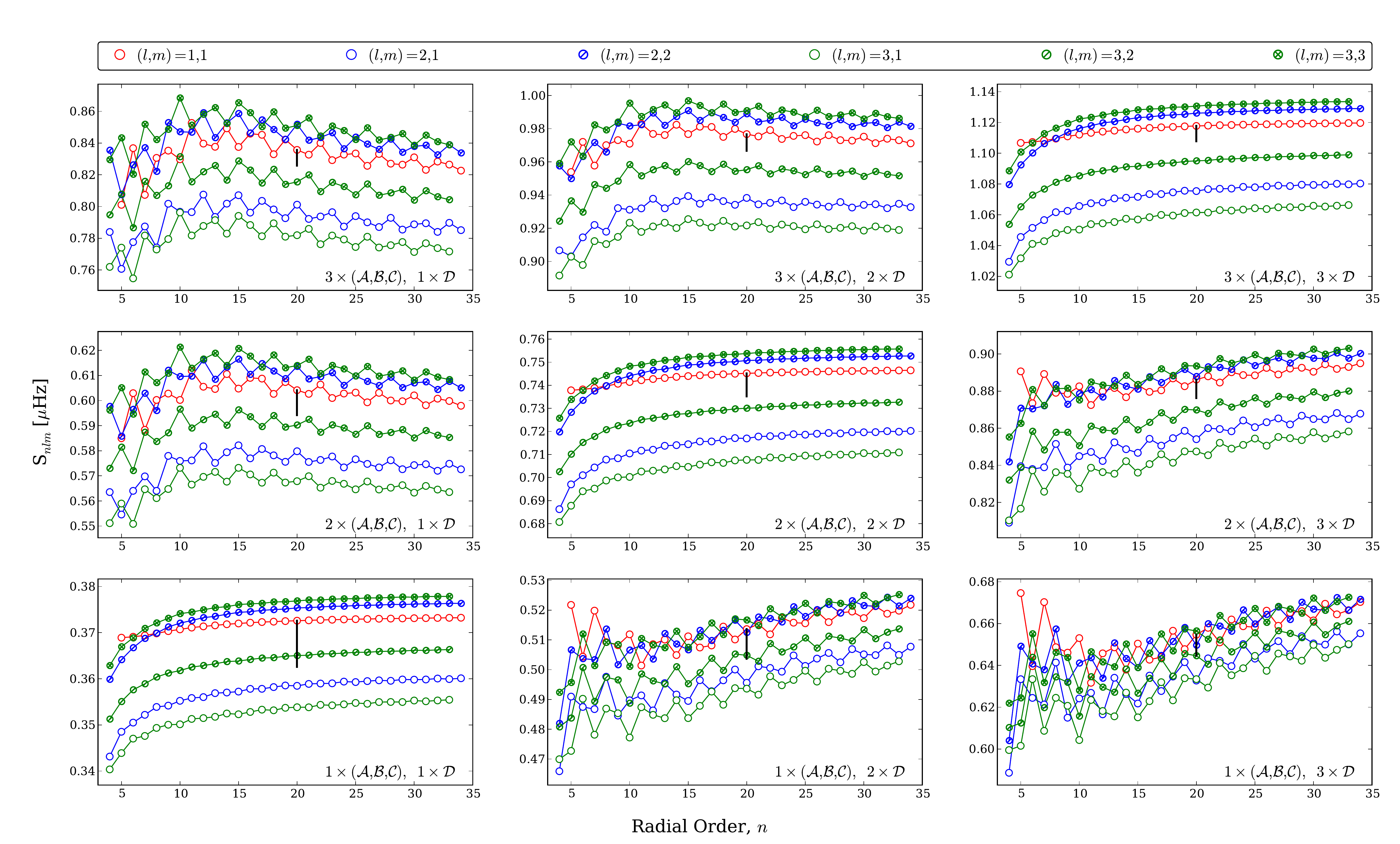}
	\caption{\footnotesize Effect in the splittings from changing the radial gradient of differential rotation for a solar-like rotation profile as a function of radial order. The splittings for nine combinations of envelope vs. interior rotation rates are given. Note the difference of the vertical scale between the panels. For a given row the rotation rate in the envelope stays constant while the interior rate increases to the right by an integer value with the column, \ie, $\mathcal{D}$ is multiplied by a factor of 1, 2, or 3 while $\mathcal{A},\,\mathcal{B}$, and $\mathcal{C}$ are left untouched. For a given column the rotation rate in the interior stays constant, while the rate in the envelope increases upward by an integer value with the row, \ie, $\mathcal{A},\,\mathcal{B}$, and $\mathcal{C}$ are multiplied by a factor of 1, 2, or 3, while $\mathcal{D}$ is left untouched. The solid vertical reference bar in each of the panels indicates the extent of a 0.01$\rm \, \mu Hz$ frequency difference. }
\label{fig:solrotgrid}
\end{figure*}

The shear layer formed at the boundary between the envelope and the interior, \ie\ the tachocline, is thought to be an important driver for the solar dynamo, and so any information on the interior-to-envelope ratio, be it just a sign for the radial change, will be of great importance to constrain models of stellar differential rotation. In \fref{fig:solrotgrid} we display the splittings obtained from different combinations of interior-to-envelope rotation rates. When all parameters entering the description of the rotation profile (\eqref{eq:solar_rot_profile}) are changed by the same amount, the behavior of the splittings for different modes relative to each other, as well as a function of radial order, does not change (diagonal panels from bottom left to top right). However, the absolute size of the splittings naturally follows the absolute increase in rotation rate; this can be seen from the $0.01\,\rm \mu Hz$ reference bar in each panel. 

We see that as the relative difference between interior and envelope rotation rates increases (maximum in upper left and lower right panels), the size of the wiggles in $n$ is enhanced. This is to be expected as the difference in splittings between two radial orders with slightly different IPs in the envelope/interior is amplified by an increase in the difference in rotation rates.  

In the extreme cases of \fref{fig:solrotgrid} where the interior/envelope rotation rate is enhanced (relative to the envelope/interior) by a factor of three (top left and bottom right panels) much of the behavior can be understood in the context of the piecewise constant profile (\S~\ref{sec:pwc}). When the envelope rotation rate is enhanced, and thereby increasingly dominates the splittings, the variation in the splittings as a function of $n$ resembles the variation seen in \fref{fig:cumkern_vol}, viz., the IP of the kernel found in the envelope -- this is similarly seen for the piecewise constant profile having a $2:1$ ratio in rotation rates between the envelope and interior. In the opposite case where the interior rotation rate is enhanced, the variation becomes more linear with $n$. However, with a continuing increase in the interior rotation rate relative to the envelope one will start to see a variation reversed compared to \fref{fig:cumkern_vol}. Here splittings of the lowest and highest radial orders will become greater than the splittings around the former peak at around $n\sim13$, which instead will become a minimum.

\section{Estimate of the feasibility of detecting differential rotation}
\label{sec:esti}

Whether or not differential rotation can be detected will depend on the frequency precision that can be achieved on the measured splittings. For an isolated $l=0$ mode, estimated as a single Lorentzian profile, the precision on the central frequency was estimated by \citet[][]{1992ApJ...387..712L} and extended to split $l=1$ modes in the solar configuration (\ie, $i\sim 90^{\circ}$) by \citet[][]{1994A&A...289..649T} as a function of the S/N of the mode, the mode line width, and the observing time. However, as we have seen, the stellar inclination has an important impact on the visibility of the modes and will consequently also affect the achievable precisions. \citet[][]{2008A&A...486..867B} took this into account and gave an expression for the splitting uncertainty including the stellar inclination. We can use this expression to obtain a rough estimate on the feasibility of detecting the differential rotation, keeping in mind that the expression assumes a uniform and constant splitting (see \eqref{eq:solid_body}) and thus will only give an estimate of error on the mean splitting. Assuming ${\sim}4$ yr of observations (max duration of \emph{Kepler} observations) of a Sun-like star with $\rm\Gamma= 1\, \mu Hz$, an inclination of $i= 55^{\circ}$, and an $\rm S/N\sim 4.5$ for $l=1$ modes ($\rm S/N\sim 1.5$ for $l=2$; here $l=3$ modes are of no use), one gets from Equation~(22) in \citet[][]{2008A&A...486..867B} uncertainties on the splitting of $\rm\sigma_{\nu_s}\sim 0.06\, \mu Hz$ for $l=1$ and $\rm\sigma_{\nu_s}\sim 0.09\, \mu Hz$ for $l=2$ when adopting a rotation rate of $\rm 3\, \mu Hz$. This equation is only applicable when the azimuthal components are well separated, such that each component in essence can be fitted as individual Lorentzian profiles. It should also be mentioned that in this calculation it is assumed that all $2l+1$ $m$-components of a given split multiplet add to the uncertainty as
\begin{equation}
\sigma_{\nu_s}^{-2} = \sum_{m=-l}^{l} m^2 \sigma_{\nu_m}^{-2} \, ,
\end{equation} 
while for the detection of differential rotation the splitting of different $m$-components is needed independently.

From the panels along the secondary diagonal (lower left to upper right) of \fref{fig:solrotgrid} it can be seen that the desired difference in $S$-values increases linearly with the overall rotation rate when all components of the differential rotation (\ie, $\mathcal{A},\,\mathcal{B}$, $\mathcal{C}$, and $\mathcal{D}$) are increased by the same amount. In the upper right panel the latitudinal-mean rotational frequency is about $\rm{\sim}1\, \mu Hz$, while the difference in the reduced splittings between $l=1$ and $l=2,\, m=1$ modes at $n=20$ is $\rm{\sim}0.043\, \mu Hz$. Scaling this to a mean rotational frequency of $\rm{\sim}3\, \mu Hz$, this difference becomes $\rm{\sim}0.13\, \mu Hz$. This is larger than the uncertainty in the reduced splittings at this $n$, which for $l=1$ would be $\sigma_{S_{l=1}} =\rm \sigma_{\nu_s}/2\sim 0.03\, \mu Hz$. Assuming that all components for the rotation profile scale in this manner, and that main-sequence Sun-like stars with excited oscillations can be found at rotation rates of $\rm{\sim}3\, \mu Hz$ (period of ${\sim}3.86$ days), it should be possible to detect latitudinal differential rotation. For $l=2$ modes it is more difficult to estimate uncertainties on reduced splittings, again from the need to isolate different $m$-components. 
How the uncertainties in $S_{nlm}$ behave for different inclinations, line widths, S/N, and rotation rates (avoiding the assumption of well-separated $m$-components) will be the subject of a future study.


\section{Discussion}
\label{sec:con}

From our analysis we have seen how different rotation profiles affect the splittings that could be observed for late-type stars using asteroseismology. It is clear that if any variation in the splittings can be significantly determined as a function of $m$, one can directly discard rotation profiles without a latitudinal dependence -- that is to say, constant, piecewise constant, and shellular profiles. Any differential rotation in latitude will be sensitive to the region between the equator and a latitude of ${\sim}40^{\circ}$ when frequencies of $l=2,$ $m=1$ modes can be measured, and up to a latitude of ${\sim}56^{\circ}$ if $l=3,$ $m=1$ modes can be measured. The detectability of the differential rotation also depends strongly on the stellar inclination angle, which should be about $i \sim 65^{\circ}$ for the optimum conditions, and the mode line width, which should be as low as possible to enable a better separation of $m$-components; this favors measurements of modes with low $n$ since $\Gamma_{nl}(n)$ is generally an increasing function (see \S~\ref{sec:ker}). The fact that low-order modes are to be preferred also holds true if any information on the radial variation in the differential rotation is to be found from \emph{p}-mode splittings. Additionally, low-$n$ modes from the Sun are known to be less sensitive to frequency shifts from stellar cycles \citep[][]{1990Natur.345..779L}. The detection of very low $n$ modes ($\lesssim9$), preferably of $l=3$, is not likely to be obtainable with photometry, even from a mission such as \emph{Kepler}, as the stellar granulation background becomes problematic toward low $n$ and furthermore the visibility for $l=3$ is very low. Such modes could possibly be observed in radial velocity data, \eg, from the planned \emph{Stellar Observations Network Group} (SONG)\footnote{At the time of writing the first node of the network is under commissioning at Observatorio del Teide on Tenerife, Spain, while the second node is being installed at the Delingha Observatory, China (F. Grundahl 2014, private communication).} network \citep[][]{2008JPhCS.118a2041G,2012AN....333.1103U} -- the problem here being the lack of temporal coverage on the same scale as possible with \emph{Kepler}. In terms of temporal coverage and photometric precision the recently selected ESA mission PLATO\footnote{PLAnetary Transits and Oscillations of stars.} \citep[see, \eg,][]{plato_ref} would offer much the same as has been possible with \emph{Kepler}, but with the advantage of a much larger sample of observed stars.

The errors on frequency measurements are likely too high for inference on differential rotation for stars comparable to our model star, which had subsolar rotation rates. However, if the measurement were made on a somewhat younger star with a lower rotation period, it should be possible to, as a first, see if there are any signs of differential rotation at all, simply by detecting a variation in splittings for different $m$-components. Whether or not it will be possible to make reliable inversions of observed splittings will be the subject of future work, although the finer details such as near-surface shear layers, etc., will surely be beyond the realm of asteroseismology for the foreseeable future. In the long term, however, a mission such as NASA's Vision Mission The Stellar Imager (SI) \citep[see, \eg,][]{2006SPIE.6268E..63C,2009ASPC..412...91C}, a UV-optical space-based interferometer with a resolution of ${\sim} 100 \, \mu \rm as$ and thereby able to resolve the surfaces of nearby stars, would revolutionize our view of stellar surface activity and rotation and could via asteroseismology of modes of degree $l\lesssim 60$ also enable detailed inference on the interior structure and rotation.

From our analysis we find that a distinction between solar-like and anti-solar-like rotation should indeed be quite possible, and this without many, or even very precise, measurements of frequency splittings. We find that a distinction between a solar-like rotation profile, as in \eqref{eq:solar_rot_profile}, and one including only a $\cos^2 \theta$ term is not readily possible from the splittings, and one might therefore opt for the simpler version. Moreover, the choice of profile will affect the interpretation made for the surface differential rotation.   

If the measurements of differential rotation from frequency splittings could be corroborated by some other measure, this would clearly be desirable. Some alternative methods for the detection of differential rotation were mentioned previously. Not many of these are, however, likely to be applicable for a main-sequence solar-like star, which in general shows little magnetic activity, thereby inhibiting methods such as Doppler imaging. Furthermore, as the rotation rate is generally quite slow for these stars, methods such as those based on the Fourier transform of spectral lines become very difficult \citep[see, \eg,][]{2001A&A...376L..13R,2002A&A...381..105R,2002A&A...384..155R}. If a star has some activity, this will also make detections of oscillations more difficult, as magnetic activity tends to suppress the amplitudes of solar-like oscillations \citep[][]{2011ApJ...732L...5C}. If the star is of type F, one might be able to use spot modulation in the time domain as these targets are more often than not somewhat active. However, for inference from asteroseismology these targets are very difficult owing to the short lifetime of their modes (and consequently large line widths), and information from splittings is not of much use. One could use the peak from the stellar rotation at very low frequencies in the power spectrum, which again relies on some spots on the stellar surface. 
As done by \citet[][]{2013A&A...557A..11R}, the difference between low-frequency peaks can be used as a measure of differential rotation, as spots occupying different latitudes will give rise to signal at slightly different frequencies in the low end of the power spectrum. If only the width of a low-frequency peak is used and interpreted as originating from differential rotation, we note that for a hypothetical star having only a very narrow active region in latitude, a similar broad-peaked signal could be caused by the decay and emergence of spots, whereby the average lifetime of spots is measured rather than the differential rotation -- this degeneracy, if present, can possibly by remedied by also including the temporal component of the signal using wavelet transforms \citep[see, \eg,][]{2013A&A...550A..32M}. Moreover, the width of the broad-peaked signal imparts no information on which latitudes are active or the sign of the surface differential rotation. 

In our analysis we have only taken the effect of rotation into account, but as mentioned in \S~\ref{sec:split}, other mechanisms can contribute to the splittings observed in the power spectrum and add to the variation in splittings as a function of $m$. One important factor in somewhat active stars could be the effect from magnetic fields. Magnetic fields are not sensitive to the sign of $m$ and would thereby only be able to produce a multiplet of $l+1$ components rather than $2l+1$ as for rotation \citep[][]{1972Ap&SS..16..386G,1989ApJ...336..403J,1991ApJ...378..326W}. In a multiplet already split into $2l+1$ components by rotation the effect from a magnetic field would therefore be to introduce a nonuniform spacing in the multiplet, and furthermore the position of the $m=0$ component would be shifted -- the symmetric splitting $S_{nlm}$ would here remain unaffected. Nonetheless, magnetic fields can become very problematic if, \eg, the symmetry axis of the field (or the pulsations) does not coincide with the axis of rotation \citep[][]{1985ApJ...296L..27D,1986MNRAS.223..557K}; in such a case up to $(2l+1)^2$ components could be seen for every $l$ \citep[][]{1985ApJ...292..238P,1990MNRAS.242...25G}.   

Further complications can be introduced to the analysis of frequency splittings, with the main contributor being that of higher-order effects from the rotation. Second-order terms in the rotational splitting will act much as the effect from a magnetic field, again being sensitive to $|m|$ and also shifting the central $m=0$ component in frequency \citep[see, \eg,][]{1992ApJ...394..670D,1998ESASP.418..385K}. The second-order terms do not affect $S_{nlm}$ (though they may complicate measurements); cubic terms would, though \citep[see, \eg,][]{1998A&A...334..911S,2011LNP...832..223G}, but for a slowly rotating solar-like star this effect would be negligible. For very rapidly rotating stars, \ie, generally speaking not main-sequence solar-like stars, the perturbative approach often followed is no longer applicable and a 2D numerical approach can be needed \citep[see, \eg,][]{2006A&A...455..607L,2006A&A...455..621R,2012A&A...547A..75O}. Other complications in the extraction and analysis of frequencies include rotationally induced mode couplings \citep[see, \eg,][]{2004ESASP.538..133G}, which changes the so-called small separation important for the modeling of the star; asymmetries in line profiles of modes in the power spectrum \citep[see, \eg,][]{1998ApJ...505L..51N,2008AN....329..440C}; effects from other large-scale velocity fields in the star, most notably the meridional circulation (effect is even in $m$) \citep[see, \eg,][]{2008SoPh..251...77R}; temporal variations in the rotation rates of the star (likely not an issue for the time spans currently available for analysis) \citep[][]{2013A&A...549A..74M}; and, lastly, effects from cycle-induced frequency shifts \citep[see][]{1990Natur.345..779L,2010Sci...329.1032G}.

Asteroseismology contains a wealth of information that has not yet been tapped, particularly concerning differential rotation in solar-like stars. The likely cause for the lack of detailed investigation of these stars, with particular emphasis on very Sun-like stars, is the great difficulty attached to the extraction and analysis of these frequency splittings. One course of action from this point on would be an analysis of the feasibility of extracting the splittings needed for any sort of inference on the differential rotation. This could, for instance, be based on the amount of \emph{Kepler} data currently available and take advantage of the forward modeling presented in this paper. Following this, the next natural step would be the detailed analysis of specific individual targets.


\section*{Acknowledgments} 
The authors are grateful to Matthias Rempel for reading and commenting on an earlier version of the manuscript.
M.N.L. would like to thank NORDITA for their hospitality during a stay where some of the work was done.
M.N.L. would furthermore like to thank Dennis Stello and his colleagues at the Sydney Institute for Astronomy (SIfA) for their hospitality during a stay where much of the presented work was done. 

Funding for the Stellar Astrophysics Centre (SAC) is provided by The Danish National Research Foundation (Grant agreement no.: DNRF106). The research is supported by the ASTERISK project (ASTERoseismic Investigations with SONG and Kepler) funded by the European Research Council (Grant agreement no.: 267864).

The National Center for Atmospheric Research is sponsored by the National Science Foundation.
This work is supported by the NASA Heliophysics Theory grant NNX11AJ36G.

The research leading to these results has received funding from the European Community's Seventh Framework Programme (FP7/2007-2013) under grant agreement No. 269194.

This research has made use of the following web resources: NASA's Astrophysics Data System Bibliographic Services (adswww.harvard.edu); arxiv.org, maintained and operated by the Cornell University Library.

\vspace{1cm}

\bibliography{MasterBIB.bib}

\begin{thebibliography}{224}
\expandafter\ifx\csname natexlab\endcsname\relax\def\natexlab#1{#1}\fi

\bibitem[{{Aerts} {et~al.}(2010){Aerts}, {Christensen-Dalsgaard}, \&
  {Kurtz}}]{2010aste.book.....A}
{Aerts}, C., {Christensen-Dalsgaard}, J., \& {Kurtz}, D.~W. 2010,
  {Asteroseismology}

\bibitem[{{Aerts} {et~al.}(2003){Aerts}, {Thoul}, {Daszy{\'n}ska}, {Scuflaire},
  {Waelkens}, {Dupret}, {Niemczura}, \& {Noels}}]{2003Sci...300.1926A}
{Aerts}, C., {Thoul}, A., {Daszy{\'n}ska}, J., {et~al.} 2003, Science, 300,
  1926

\bibitem[{{Angulo} {et~al.}(1999){Angulo}, {Arnould}, {Rayet}, {Descouvemont},
  {Baye}, {Leclercq-Willain}, {Coc}, {Barhoumi}, {Aguer}, {Rolfs}, {Kunz},
  {Hammer}, {Mayer}, {Paradellis}, {Kossionides}, {Chronidou}, {Spyrou},
  {degl'Innocenti}, {Fiorentini}, {Ricci}, {Zavatarelli}, {Providencia},
  {Wolters}, {Soares}, {Grama}, {Rahighi}, {Shotter}, \& {Lamehi
  Rachti}}]{1999NuPhA.656....3A}
{Angulo}, C., {Arnould}, M., {Rayet}, M., {et~al.} 1999, Nuclear Physics A,
  656, 3

\bibitem[{{Appourchaux} {et~al.}(2010){Appourchaux}, {Belkacem}, {Broomhall},
  {Chaplin}, {Gough}, {Houdek}, {Provost}, {Baudin}, {Boumier}, {Elsworth},
  {Garc{\'{\i}}a}, {Andersen}, {Finsterle}, {Fr{\"o}hlich}, {Gabriel}, {Grec},
  {Jim{\'e}nez}, {Kosovichev}, {Sekii}, {Toutain}, \&
  {Turck-Chi{\`e}ze}}]{2010A&ARv..18..197A}
{Appourchaux}, T., {Belkacem}, K., {Broomhall}, A.-M., {et~al.} 2010, \aapr,
  18, 197

\bibitem[{{Arentoft} {et~al.}(2008){Arentoft}, {Kjeldsen}, {Bedding}, {Bazot},
  {Christensen-Dalsgaard}, {Dall}, {Karoff}, {Carrier}, {Eggenberger},
  {Sosnowska}, {Wittenmyer}, {Endl}, {Metcalfe}, {Hekker}, {Reffert}, {Butler},
  {Bruntt}, {Kiss}, {O'Toole}, {Kambe}, {Ando}, {Izumiura}, {Sato}, {Hartmann},
  {Hatzes}, {Bouchy}, {Mosser}, {Appourchaux}, {Barban}, {Berthomieu},
  {Garcia}, {Michel}, {Provost}, {Turck-Chi{\`e}ze}, {Marti{\'c}}, {Lebrun},
  {Schmitt}, {Bertaux}, {Bonanno}, {Benatti}, {Claudi}, {Cosentino}, {Leccia},
  {Frandsen}, {Brogaard}, {Glowienka}, {Grundahl}, \&
  {Stempels}}]{2008ApJ...687.1180A}
{Arentoft}, T., {Kjeldsen}, H., {Bedding}, T.~R., {et~al.} 2008, \apj, 687,
  1180

\bibitem[{{Asplund} {et~al.}(2009){Asplund}, {Grevesse}, {Sauval}, \&
  {Scott}}]{2009ARA&A..47..481A}
{Asplund}, M., {Grevesse}, N., {Sauval}, A.~J., \& {Scott}, P. 2009, \araa, 47,
  481

\bibitem[{{Augustson} {et~al.}(2012){Augustson}, {Brown}, {Brun}, {Miesch}, \&
  {Toomre}}]{2012ApJ...756..169A}
{Augustson}, K.~C., {Brown}, B.~P., {Brun}, A.~S., {Miesch}, M.~S., \&
  {Toomre}, J. 2012, \apj, 756, 169

\bibitem[{{Auvergne} {et~al.}(2009){Auvergne}, {Bodin}, {Boisnard}, {Buey},
  {Chaintreuil}, {Epstein}, {Jouret}, {Lam-Trong}, {Levacher}, {Magnan},
  {Perez}, {Plasson}, {Plesseria}, {Peter}, {Steller}, {Tiph{\`e}ne}, {Baglin},
  {Agogu{\'e}}, {Appourchaux}, {Barbet}, {Beaufort}, {Bellenger}, {Berlin},
  {Bernardi}, {Blouin}, {Boumier}, {Bonneau}, {Briet}, {Butler}, {Cautain},
  {Chiavassa}, {Costes}, {Cuvilho}, {Cunha-Parro}, {de Oliveira Fialho},
  {Decaudin}, {Defise}, {Djalal}, {Docclo}, {Drummond}, {Dupuis}, {Exil},
  {Faur{\'e}}, {Gaboriaud}, {Gamet}, {Gavalda}, {Grolleau}, {Gueguen},
  {Guivarc'h}, {Guterman}, {Hasiba}, {Huntzinger}, {Hustaix}, {Imbert},
  {Jeanville}, {Johlander}, {Jorda}, {Journoud}, {Karioty}, {Kerjean},
  {Lafond}, {Lapeyrere}, {Landiech}, {Larqu{\'e}}, {Laudet}, {Le Merrer},
  {Leporati}, {Leruyet}, {Levieuge}, {Llebaria}, {Martin}, {Mazy}, {Mesnager},
  {Michel}, {Moalic}, {Monjoin}, {Naudet}, {Neukirchner}, {Nguyen-Kim},
  {Ollivier}, {Orcesi}, {Ottacher}, {Oulali}, {Parisot}, {Perruchot},
  {Piacentino}, {Pinheiro da Silva}, {Platzer}, {Pontet}, {Pradines},
  {Quentin}, {Rohbeck}, {Rolland}, {Rollenhagen}, {Romagnan}, {Russ}, {Samadi},
  {Schmidt}, {Schwartz}, {Sebbag}, {Smit}, {Sunter}, {Tello}, {Toulouse},
  {Ulmer}, {Vandermarcq}, {Vergnault}, {Wallner}, {Waultier}, \&
  {Zanatta}}]{2009A&A...506..411A}
{Auvergne}, M., {Bodin}, P., {Boisnard}, L., {et~al.} 2009, \aap, 506, 411

\bibitem[{{Ballot}(2010)}]{2010AN....331..933B}
{Ballot}, J. 2010, Astronomische Nachrichten, 331, 933

\bibitem[{{Ballot} {et~al.}(2008){Ballot}, {Appourchaux}, {Toutain}, \&
  {Guittet}}]{2008A&A...486..867B}
{Ballot}, J., {Appourchaux}, T., {Toutain}, T., \& {Guittet}, M. 2008, \aap,
  486, 867

\bibitem[{{Balmforth}(1992)}]{1992MNRAS.255..603B}
{Balmforth}, N.~J. 1992, \mnras, 255, 603

\bibitem[{{Barnes} {et~al.}(2005){Barnes}, {Collier Cameron}, {Donati},
  {James}, {Marsden}, \& {Petit}}]{2005MNRAS.357L...1B}
{Barnes}, J.~R., {Collier Cameron}, A., {Donati}, J.-F., {et~al.} 2005, \mnras,
  357, L1

\bibitem[{{Barnes}(2003)}]{2003ApJ...586..464B}
{Barnes}, S.~A. 2003, \apj, 586, 464

\bibitem[{{Barnes}(2007)}]{2007ApJ...669.1167B}
---. 2007, \apj, 669, 1167

\bibitem[{{Basu} \& {Antia}(2003)}]{2003ApJ...585..553B}
{Basu}, S., \& {Antia}, H.~M. 2003, \apj, 585, 553

\bibitem[{{Beck}(2000)}]{2000SoPh..191...47B}
{Beck}, J.~G. 2000, \solphys, 191, 47

\bibitem[{{Beck} {et~al.}(2012){Beck}, {Montalban}, {Kallinger}, {De Ridder},
  {Aerts}, {Garc{\'{\i}}a}, {Hekker}, {Dupret}, {Mosser}, {Eggenberger},
  {Stello}, {Elsworth}, {Frandsen}, {Carrier}, {Hillen}, {Gruberbauer},
  {Christensen-Dalsgaard}, {Miglio}, {Valentini}, {Bedding}, {Kjeldsen},
  {Girouard}, {Hall}, \& {Ibrahim}}]{2012Natur.481...55B}
{Beck}, P.~G., {Montalban}, J., {Kallinger}, T., {et~al.} 2012, \nat, 481, 55

\bibitem[{{Belkacem} {et~al.}(2011){Belkacem}, {Goupil}, {Dupret}, {Samadi},
  {Baudin}, {Noels}, \& {Mosser}}]{2011A&A...530A.142B}
{Belkacem}, K., {Goupil}, M.~J., {Dupret}, M.~A., {et~al.} 2011, \aap, 530,
  A142

\bibitem[{{Bessolaz} \& {Brun}(2011)}]{2011ApJ...728..115B}
{Bessolaz}, N., \& {Brun}, A.~S. 2011, \apj, 728, 115

\bibitem[{{Birch} \& {Kosovichev}(1998)}]{1998ApJ...503L.187B}
{Birch}, A.~C., \& {Kosovichev}, A.~G. 1998, \apjl, 503, L187

\bibitem[{{B{\"o}hm-Vitense}(1958)}]{1958ZA.....46..108B}
{B{\"o}hm-Vitense}, E. 1958, \zap, 46, 108

\bibitem[{{Bonomo} \& {Lanza}(2012)}]{2012A&A...547A..37B}
{Bonomo}, A.~S., \& {Lanza}, A.~F. 2012, \aap, 547, A37

\bibitem[{{Bouvier} {et~al.}(1997){Bouvier}, {Forestini}, \&
  {Allain}}]{1997A&A...326.1023B}
{Bouvier}, J., {Forestini}, M., \& {Allain}, S. 1997, \aap, 326, 1023

\bibitem[{{Brown} {et~al.}(2008){Brown}, {Browning}, {Brun}, {Miesch}, \&
  {Toomre}}]{2008ApJ...689.1354B}
{Brown}, B.~P., {Browning}, M.~K., {Brun}, A.~S., {Miesch}, M.~S., \& {Toomre},
  J. 2008, \apj, 689, 1354

\bibitem[{{Brown} {et~al.}(1989){Brown}, {Christensen-Dalsgaard},
  {Dziembowski}, {Goode}, {Gough}, \& {Morrow}}]{1989ApJ...343..526B}
{Brown}, T.~M., {Christensen-Dalsgaard}, J., {Dziembowski}, W.~A., {et~al.}
  1989, \apj, 343, 526

\bibitem[{{Browning}(2008)}]{2008ApJ...676.1262B}
{Browning}, M.~K. 2008, \apj, 676, 1262

\bibitem[{{Brun}(2011)}]{2011EAS....44...81B}
{Brun}, A.~S. 2011, in EAS Publications Series, Vol.~44, EAS Publications
  Series, ed. H.~{Wozniak} \& G.~{Hensler}, 81--95

\bibitem[{{Brun} {et~al.}(2004){Brun}, {Miesch}, \&
  {Toomre}}]{2004ApJ...614.1073B}
{Brun}, A.~S., {Miesch}, M.~S., \& {Toomre}, J. 2004, \apj, 614, 1073

\bibitem[{{Brun} \& {Toomre}(2002)}]{2002ApJ...570..865B}
{Brun}, A.~S., \& {Toomre}, J. 2002, \apj, 570, 865

\bibitem[{{Busse}(1970)}]{1970JFM....44..441B}
{Busse}, F.~H. 1970, Journal of Fluid Mechanics, 44, 441

\bibitem[{{Cantiello} {et~al.}(2014){Cantiello}, {Mankovich}, {Bildsten},
  {Christensen-Dalsgaard}, \& {Paxton}}]{2014ApJ...788...93C}
{Cantiello}, M., {Mankovich}, C., {Bildsten}, L., {Christensen-Dalsgaard}, J.,
  \& {Paxton}, B. 2014, \apj, 788, 93

\bibitem[{{Carpenter} {et~al.}(2006){Carpenter}, {Schrijver}, \&
  {Karovska}}]{2006SPIE.6268E..63C}
{Carpenter}, K.~G., {Schrijver}, C.~J., \& {Karovska}, M. 2006, in Society of
  Photo-Optical Instrumentation Engineers (SPIE) Conference Series, Vol. 6268,
  Society of Photo-Optical Instrumentation Engineers (SPIE) Conference Series

\bibitem[{{Carpenter} {et~al.}(2009){Carpenter}, {Schrijver}, {Karovska}, \&
  {Si Vision Mission Team}}]{2009ASPC..412...91C}
{Carpenter}, K.~G., {Schrijver}, C.~J., {Karovska}, M., \& {Si Vision Mission
  Team}. 2009, in Astronomical Society of the Pacific Conference Series, Vol.
  412, The Biggest, Baddest, Coolest Stars, ed. D.~G. {Luttermoser}, B.~J.
  {Smith}, \& R.~E. {Stencel}, 91

\bibitem[{{Chaplin} {et~al.}(2001){Chaplin}, {Elsworth}, {Isaak}, {Marchenkov},
  {Miller}, \& {New}}]{2001MNRAS.327.1127C}
{Chaplin}, W.~J., {Elsworth}, Y., {Isaak}, G.~R., {et~al.} 2001, \mnras, 327,
  1127

\bibitem[{{Chaplin} {et~al.}(2008){Chaplin}, {Elsworth}, \&
  {Toutain}}]{2008AN....329..440C}
{Chaplin}, W.~J., {Elsworth}, Y., \& {Toutain}, T. 2008, Astronomische
  Nachrichten, 329, 440

\bibitem[{{Chaplin} \& {Miglio}(2013)}]{2013ARA&A..51..353C}
{Chaplin}, W.~J., \& {Miglio}, A. 2013, \araa, 51, 353

\bibitem[{{Chaplin} {et~al.}(2004){Chaplin}, {Sekii}, {Elsworth}, \&
  {Gough}}]{2004MNRAS.355..535C}
{Chaplin}, W.~J., {Sekii}, T., {Elsworth}, Y., \& {Gough}, D.~O. 2004, \mnras,
  355, 535

\bibitem[{{Chaplin} {et~al.}(1999){Chaplin}, {Christensen-Dalsgaard},
  {Elsworth}, {Howe}, {Isaak}, {Larsen}, {New}, {Schou}, {Thompson}, \&
  {Tomczyk}}]{1999MNRAS.308..405C}
{Chaplin}, W.~J., {Christensen-Dalsgaard}, J., {Elsworth}, Y., {et~al.} 1999,
  \mnras, 308, 405

\bibitem[{{Chaplin} {et~al.}(2011){Chaplin}, {Bedding}, {Bonanno}, {Broomhall},
  {Garc{\'{\i}}a}, {Hekker}, {Huber}, {Verner}, {Basu}, {Elsworth}, {Houdek},
  {Mathur}, {Mosser}, {New}, {Stevens}, {Appourchaux}, {Karoff}, {Metcalfe},
  {Molenda-{\.Z}akowicz}, {Monteiro}, {Thompson}, {Christensen-Dalsgaard},
  {Gilliland}, {Kawaler}, {Kjeldsen}, {Ballot}, {Benomar}, {Corsaro},
  {Campante}, {Gaulme}, {Hale}, {Handberg}, {Jarvis}, {R{\'e}gulo}, {Roxburgh},
  {Salabert}, {Stello}, {Mullally}, {Li}, \& {Wohler}}]{2011ApJ...732L...5C}
{Chaplin}, W.~J., {Bedding}, T.~R., {Bonanno}, A., {et~al.} 2011, \apjl, 732,
  L5

\bibitem[{{Chaplin} {et~al.}(2013){Chaplin}, {Sanchis-Ojeda}, {Campante},
  {Handberg}, {Stello}, {Winn}, {Basu}, {Christensen-Dalsgaard}, {Davies},
  {Metcalfe}, {Buchhave}, {Fischer}, {Bedding}, {Cochran}, {Elsworth},
  {Gilliland}, {Hekker}, {Huber}, {Isaacson}, {Karoff}, {Kawaler}, {Kjeldsen},
  {Latham}, {Lund}, {Lundkvist}, {Marcy}, {Miglio}, {Barclay}, \&
  {Lissauer}}]{2013ApJ...766..101C}
{Chaplin}, W.~J., {Sanchis-Ojeda}, R., {Campante}, T.~L., {et~al.} 2013, \apj,
  766, 101

\bibitem[{{Charbonneau} \& {MacGregor}(1993)}]{1993ApJ...417..762C}
{Charbonneau}, P., \& {MacGregor}, K.~B. 1993, \apj, 417, 762

\bibitem[{{Charbonneau} {et~al.}(1998){Charbonneau}, {Tomczyk}, {Schou}, \&
  {Thompson}}]{1998ApJ...496.1015C}
{Charbonneau}, P., {Tomczyk}, S., {Schou}, J., \& {Thompson}, M.~J. 1998, \apj,
  496, 1015

\bibitem[{{Christensen-Dalsgaard}(2008{\natexlab{a}})}]{2008Ap&SS.316...13C}
{Christensen-Dalsgaard}, J. 2008{\natexlab{a}}, \apss, 316, 13

\bibitem[{{Christensen-Dalsgaard}(2008{\natexlab{b}})}]{2008Ap&SS.316..113C}
---. 2008{\natexlab{b}}, \apss, 316, 113

\bibitem[{{Christensen-Dalsgaard} \& {Berthomieu}(1991)}]{1991sia..book..401C}
{Christensen-Dalsgaard}, J., \& {Berthomieu}, G. 1991, {Theory of solar
  oscillations}, ed. A.~N. {Cox}, W.~C. {Livingston}, \& M.~S. {Matthews},
  401--478

\bibitem[{{Christensen-Dalsgaard} {et~al.}(1991){Christensen-Dalsgaard},
  {Gough}, \& {Thompson}}]{1991ApJ...378..413C}
{Christensen-Dalsgaard}, J., {Gough}, D.~O., \& {Thompson}, M.~J. 1991, \apj,
  378, 413

\bibitem[{{Corbard} \& {Thompson}(2002)}]{2002SoPh..205..211C}
{Corbard}, T., \& {Thompson}, M.~J. 2002, \solphys, 205, 211

\bibitem[{{Croll} {et~al.}(2006){Croll}, {Walker}, {Kuschnig}, {Matthews},
  {Rowe}, {Walker}, {Rucinski}, {Hatzes}, {Cochran}, {Robb}, {Guenther},
  {Moffat}, {Sasselov}, \& {Weiss}}]{2006ApJ...648..607C}
{Croll}, B., {Walker}, G.~A.~H., {Kuschnig}, R., {et~al.} 2006, \apj, 648, 607

\bibitem[{{Cuypers}(1980)}]{1980A&A....89..207C}
{Cuypers}, J. 1980, \aap, 89, 207

\bibitem[{{Deheuvels} {et~al.}(2012){Deheuvels}, {Garc{\'{\i}}a}, {Chaplin},
  {Basu}, {Antia}, {Appourchaux}, {Benomar}, {Davies}, {Elsworth}, {Gizon},
  {Goupil}, {Reese}, {Regulo}, {Schou}, {Stahn}, {Casagrande},
  {Christensen-Dalsgaard}, {Fischer}, {Hekker}, {Kjeldsen}, {Mathur}, {Mosser},
  {Pinsonneault}, {Valenti}, {Christiansen}, {Kinemuchi}, \&
  {Mullally}}]{2012ApJ...756...19D}
{Deheuvels}, S., {Garc{\'{\i}}a}, R.~A., {Chaplin}, W.~J., {et~al.} 2012, \apj,
  756, 19

\bibitem[{{Deheuvels} {et~al.}(2014){Deheuvels}, {Do{\u g}an}, {Goupil},
  {Appourchaux}, {Benomar}, {Bruntt}, {Campante}, {Casagrande}, {Ceillier},
  {Davies}, {De Cat}, {Fu}, {Garc{\'{\i}}a}, {Lobel}, {Mosser}, {Reese},
  {Regulo}, {Schou}, {Stahn}, {Thygesen}, {Yang}, {Chaplin},
  {Christensen-Dalsgaard}, {Eggenberger}, {Gizon}, {Mathis},
  {Molenda-{\.Z}akowicz}, \& {Pinsonneault}}]{2014A&A...564A..27D}
{Deheuvels}, S., {Do{\u g}an}, G., {Goupil}, M.~J., {et~al.} 2014, \aap, 564,
  A27

\bibitem[{{Denissenkov} {et~al.}(2010){Denissenkov}, {Pinsonneault},
  {Terndrup}, \& {Newsham}}]{2010ApJ...716.1269D}
{Denissenkov}, P.~A., {Pinsonneault}, M., {Terndrup}, D.~M., \& {Newsham}, G.
  2010, \apj, 716, 1269

\bibitem[{{Donahue} {et~al.}(1996){Donahue}, {Saar}, \&
  {Baliunas}}]{1996ApJ...466..384D}
{Donahue}, R.~A., {Saar}, S.~H., \& {Baliunas}, S.~L. 1996, \apj, 466, 384

\bibitem[{{Donati} \& {Collier Cameron}(1997)}]{1997MNRAS.291....1D}
{Donati}, J.-F., \& {Collier Cameron}, A. 1997, \mnras, 291, 1

\bibitem[{{Dupret} {et~al.}(2004){Dupret}, {Thoul}, {Scuflaire},
  {Daszy{\'n}ska-Daszkiewicz}, {Aerts}, {Bourge}, {Waelkens}, \&
  {Noels}}]{2004A&A...415..251D}
{Dupret}, M.-A., {Thoul}, A., {Scuflaire}, R., {et~al.} 2004, \aap, 415, 251

\bibitem[{Dziembowski(1977)}]{1977AcA....27..203D}
Dziembowski, W. 1977, \actaa, 27, 203

\bibitem[{{Dziembowski} \& {Goode}(1985)}]{1985ApJ...296L..27D}
{Dziembowski}, W., \& {Goode}, P.~R. 1985, \apjl, 296, L27

\bibitem[{{Dziembowski} \& {Goode}(1992)}]{1992ApJ...394..670D}
{Dziembowski}, W.~A., \& {Goode}, P.~R. 1992, \apj, 394, 670

\bibitem[{{Eff-Darwich} \& {Korzennik}(1998)}]{1998ESASP.418..685E}
{Eff-Darwich}, A., \& {Korzennik}, S.~G. 1998, in ESA Special Publication, Vol.
  418, Structure and Dynamics of the Interior of the Sun and Sun-like Stars,
  ed. S.~{Korzennik}, 685

\bibitem[{{Eff-Darwich} \& {Korzennik}(2013)}]{2013SoPh..287...43E}
{Eff-Darwich}, A., \& {Korzennik}, S.~G. 2013, \solphys, 287, 43

\bibitem[{{Eggenberger} {et~al.}(2005){Eggenberger}, {Maeder}, \&
  {Meynet}}]{2005A&A...440L...9E}
{Eggenberger}, P., {Maeder}, A., \& {Meynet}, G. 2005, \aap, 440, L9

\bibitem[{{Eggenberger} {et~al.}(2012){Eggenberger}, {Montalb{\'a}n}, \&
  {Miglio}}]{2012A&A...544L...4E}
{Eggenberger}, P., {Montalb{\'a}n}, J., \& {Miglio}, A. 2012, \aap, 544, L4

\bibitem[{{Ekstr{\"o}m} {et~al.}(2012){Ekstr{\"o}m}, {Georgy}, {Eggenberger},
  {Meynet}, {Mowlavi}, {Wyttenbach}, {Granada}, {Decressin}, {Hirschi},
  {Frischknecht}, {Charbonnel}, \& {Maeder}}]{2012A&A...537A.146E}
{Ekstr{\"o}m}, S., {Georgy}, C., {Eggenberger}, P., {et~al.} 2012, \aap, 537,
  A146

\bibitem[{{Elsworth} {et~al.}(1995){Elsworth}, {Howe}, {Isaak}, {McLeod},
  {Miller}, {New}, {Wheeler}, \& {Gough}}]{1995Natur.376..669E}
{Elsworth}, Y., {Howe}, R., {Isaak}, G.~R., {et~al.} 1995, \nat, 376, 669

\bibitem[{{Epstein} \& {Pinsonneault}(2014)}]{2014ApJ...780..159E}
{Epstein}, C.~R., \& {Pinsonneault}, M.~H. 2014, \apj, 780, 159

\bibitem[{{Ferguson} {et~al.}(2005){Ferguson}, {Alexander}, {Allard}, {Barman},
  {Bodnarik}, {Hauschildt}, {Heffner-Wong}, \& {Tamanai}}]{2005ApJ...623..585F}
{Ferguson}, J.~W., {Alexander}, D.~R., {Allard}, F., {et~al.} 2005, \apj, 623,
  585

\bibitem[{{Foukal}(1972)}]{1972ApJ...173..439F}
{Foukal}, P. 1972, \apj, 173, 439

\bibitem[{{Foukal} \& {Jokipii}(1975)}]{1975ApJ...199L..71F}
{Foukal}, P., \& {Jokipii}, J.~R. 1975, \apjl, 199, L71

\bibitem[{{Fr{\"o}hlich} {et~al.}(2012){Fr{\"o}hlich}, {Frasca}, {Catanzaro},
  {Bonanno}, {Corsaro}, {Molenda-{\.Z}akowicz}, {Klutsch}, \&
  {Montes}}]{2012A&A...543A.146F}
{Fr{\"o}hlich}, H.-E., {Frasca}, A., {Catanzaro}, G., {et~al.} 2012, \aap, 543,
  A146

\bibitem[{{Fr{\"o}hlich} {et~al.}(2009){Fr{\"o}hlich}, {K{\"u}ker}, {Hatzes},
  \& {Strassmeier}}]{2009A&A...506..263F}
{Fr{\"o}hlich}, H.-E., {K{\"u}ker}, M., {Hatzes}, A.~P., \& {Strassmeier},
  K.~G. 2009, \aap, 506, 263

\bibitem[{{Garc{\'{\i}}a} {et~al.}(2013){Garc{\'{\i}}a}, {Ceillier}, {Mathur},
  \& {Salabert}}]{2013arXiv1307.4163G}
{Garc{\'{\i}}a}, R.~A., {Ceillier}, T., {Mathur}, S., \& {Salabert}, D. 2013,
  ArXiv e-prints 1307.4163

\bibitem[{{Garc{\'{\i}}a} {et~al.}(2010{\natexlab{a}}){Garc{\'{\i}}a},
  {Mathur}, {Salabert}, {Ballot}, {R{\'e}gulo}, {Metcalfe}, \&
  {Baglin}}]{2010Sci...329.1032G}
{Garc{\'{\i}}a}, R.~A., {Mathur}, S., {Salabert}, D., {et~al.}
  2010{\natexlab{a}}, Science, 329, 1032

\bibitem[{{Garc{\'{\i}}a} {et~al.}(2007){Garc{\'{\i}}a}, {Turck-Chi{\`e}ze},
  {Jim{\'e}nez-Reyes}, {Ballot}, {Pall{\'e}}, {Eff-Darwich}, {Mathur}, \&
  {Provost}}]{2007Sci...316.1591G}
{Garc{\'{\i}}a}, R.~A., {Turck-Chi{\`e}ze}, S., {Jim{\'e}nez-Reyes}, S.~J.,
  {et~al.} 2007, Science, 316, 1591

\bibitem[{{Garc{\'{\i}}a} {et~al.}(2010{\natexlab{b}}){Garc{\'{\i}}a},
  {Ballot}, {Eff-Darwich}, {Garrido}, {Jimenez}, {Mathis}, {Mathur}, {Moya},
  {Palle}, {Regulo}, {Salabert}, {Suarez}, \&
  {Turck-Chieze}}]{2010arXiv1007.4445G}
{Garc{\'{\i}}a}, R.~A., {Ballot}, J., {Eff-Darwich}, A., {et~al.}
  2010{\natexlab{b}}, ArXiv e-prints 1007.4445

\bibitem[{{Gastine} {et~al.}(2013){Gastine}, {Wicht}, \&
  {Aurnou}}]{2013Icar..225..156G}
{Gastine}, T., {Wicht}, J., \& {Aurnou}, J.~M. 2013, \icarus, 225, 156

\bibitem[{{Gastine} {et~al.}(2014){Gastine}, {Yadav}, {Morin}, {Reiners}, \&
  {Wicht}}]{2014MNRAS.438L..76G}
{Gastine}, T., {Yadav}, R.~K., {Morin}, J., {Reiners}, A., \& {Wicht}, J. 2014,
  \mnras, 438, L76

\bibitem[{{Gilliland} {et~al.}(2010){Gilliland}, {Brown},
  {Christensen-Dalsgaard}, {Kjeldsen}, {Aerts}, {Appourchaux}, {Basu},
  {Bedding}, {Chaplin}, {Cunha}, {De Cat}, {De Ridder}, {Guzik}, {Handler},
  {Kawaler}, {Kiss}, {Kolenberg}, {Kurtz}, {Metcalfe}, {Monteiro}, {Szab{\'o}},
  {Arentoft}, {Balona}, {Debosscher}, {Elsworth}, {Quirion}, {Stello},
  {Su{\'a}rez}, {Borucki}, {Jenkins}, {Koch}, {Kondo}, {Latham}, {Rowe}, \&
  {Steffen}}]{2010PASP..122..131G}
{Gilliland}, R.~L., {Brown}, T.~M., {Christensen-Dalsgaard}, J., {et~al.} 2010,
  \pasp, 122, 131

\bibitem[{{Gilman}(1977)}]{1977GApFD...8...93G}
{Gilman}, P.~A. 1977, Geophysical and Astrophysical Fluid Dynamics, 8, 93

\bibitem[{{Gilman}(1983)}]{1983ApJS...53..243G}
---. 1983, \apjs, 53, 243

\bibitem[{{Gilman} \& {Howe}(2003)}]{2003ESASP.517..283G}
{Gilman}, P.~A., \& {Howe}, R. 2003, in ESA Special Publication, Vol. 517,
  GONG+ 2002. Local and Global Helioseismology: the Present and Future, ed.
  H.~{Sawaya-Lacoste}, 283--285

\bibitem[{{Gizon} \& {Solanki}(2003)}]{2003ApJ...589.1009G}
{Gizon}, L., \& {Solanki}, S.~K. 2003, \apj, 589, 1009

\bibitem[{{Gizon} \& {Solanki}(2004)}]{2004SoPh..220..169G}
---. 2004, \solphys, 220, 169

\bibitem[{{Gizon} {et~al.}(2013){Gizon}, {Ballot}, {Michel}, {Stahn},
  {Vauclair}, {Bruntt}, {Quirion}, {Benomar}, {Vauclair}, {Appourchaux},
  {Auvergne}, {Baglin}, {Barban}, {Baudin}, {Bazot}, {Campante}, {Catala},
  {Chaplin}, {Creevey}, {Deheuvels}, {Dolez}, {Elsworth}, {Garcia}, {Gaulme},
  {Mathis}, {Mathur}, {Mosser}, {Regulo}, {Roxburgh}, {Salabert}, {Samadi},
  {Sato}, {Verner}, {Hanasoge}, \& {Sreenivasan}}]{2013PNAS..11013267G}
{Gizon}, L., {Ballot}, J., {Michel}, E., {et~al.} 2013, Proceedings of the
  National Academy of Science, 110, 13267

\bibitem[{{Glatzmaier}(1984)}]{1984JCoPh..55..461G}
{Glatzmaier}, G.~A. 1984, Journal of Computational Physics, 55, 461

\bibitem[{{Goldreich} {et~al.}(1994){Goldreich}, {Murray}, \&
  {Kumar}}]{1994ApJ...424..466G}
{Goldreich}, P., {Murray}, N., \& {Kumar}, P. 1994, \apj, 424, 466

\bibitem[{{Goode} {et~al.}(1991){Goode}, {Dziembowski}, {Korzennik}, \&
  {Rhodes}}]{1991ApJ...367..649G}
{Goode}, P.~R., {Dziembowski}, W.~A., {Korzennik}, S.~G., \& {Rhodes}, Jr.,
  E.~J. 1991, \apj, 367, 649

\bibitem[{{Goossens}(1972)}]{1972Ap&SS..16..386G}
{Goossens}, M. 1972, \apss, 16, 386

\bibitem[{{Gough}(1981)}]{1981MNRAS.196..731G}
{Gough}, D.~O. 1981, \mnras, 196, 731

\bibitem[{{Gough} \& {McIntyre}(1998)}]{1998Natur.394..755G}
{Gough}, D.~O., \& {McIntyre}, M.~E. 1998, \nat, 394, 755

\bibitem[{{Gough} \& {Thompson}(1990)}]{1990MNRAS.242...25G}
{Gough}, D.~O., \& {Thompson}, M.~J. 1990, \mnras, 242, 25

\bibitem[{{Goupil}(2011)}]{2011LNP...832..223G}
{Goupil}, M. 2011, in Lecture Notes in Physics, Berlin Springer Verlag, Vol.
  832, Lecture Notes in Physics, Berlin Springer Verlag, ed. J.-P. {Rozelot} \&
  C.~{Neiner}, 223

\bibitem[{{Goupil} {et~al.}(1996){Goupil}, {Dziembowski}, {Goode}, \&
  {Michel}}]{1996A&A...305..487G}
{Goupil}, M.-J., {Dziembowski}, W.~A., {Goode}, P.~R., \& {Michel}, E. 1996,
  \aap, 305, 487

\bibitem[{{Goupil} {et~al.}(2013){Goupil}, {Mosser}, {Marques}, {Ouazzani},
  {Belkacem}, {Lebreton}, \& {Samadi}}]{2013A&A...549A..75G}
{Goupil}, M.~J., {Mosser}, B., {Marques}, J.~P., {et~al.} 2013, \aap, 549, A75

\bibitem[{{Goupil} {et~al.}(2004){Goupil}, {Samadi}, {Lochard}, {Dziembowski},
  \& {Pamyatnykh}}]{2004ESASP.538..133G}
{Goupil}, M.~J., {Samadi}, R., {Lochard}, J., {Dziembowski}, W.~A., \&
  {Pamyatnykh}, A. 2004, in ESA Special Publication, Vol. 538, Stellar
  Structure and Habitable Planet Finding, ed. F.~{Favata}, S.~{Aigrain}, \&
  A.~{Wilson}, 133--140

\bibitem[{{Gray}(1976)}]{1976oasp.book.....G}
{Gray}, D.~F. 1976, {The observation and analysis of stellar photospheres}

\bibitem[{{Gray}(2005)}]{2005oasp.book.....G}
---. 2005, {The Observation and Analysis of Stellar Photospheres}

\bibitem[{{Grundahl} {et~al.}(2008){Grundahl}, {Arentoft},
  {Christensen-Dalsgaard}, {Frandsen}, {Kjeldsen}, \&
  {Rasmussen}}]{2008JPhCS.118a2041G}
{Grundahl}, F., {Arentoft}, T., {Christensen-Dalsgaard}, J., {et~al.} 2008,
  Journal of Physics Conference Series, 118, 012041

\bibitem[{{Guerrero} {et~al.}(2013){Guerrero}, {Smolarkiewicz}, {Kosovichev},
  \& {Mansour}}]{2013IAUS..294..417G}
{Guerrero}, G., {Smolarkiewicz}, P.~K., {Kosovichev}, A., \& {Mansour}, N.
  2013, in IAU Symposium, Vol. 294, IAU Symposium, ed. A.~G. {Kosovichev},
  E.~{de Gouveia Dal Pino}, \& Y.~{Yan}, 417--425

\bibitem[{{Hackman} {et~al.}(2001){Hackman}, {Jetsu}, \&
  {Tuominen}}]{2001A&A...374..171H}
{Hackman}, T., {Jetsu}, L., \& {Tuominen}, I. 2001, \aap, 374, 171

\bibitem[{{Hansen} {et~al.}(1977){Hansen}, {Cox}, \& {van
  Horn}}]{1977ApJ...217..151H}
{Hansen}, C.~J., {Cox}, J.~P., \& {van Horn}, H.~M. 1977, \apj, 217, 151

\bibitem[{{Harvey}(1985)}]{1985ESASP.235..199H}
{Harvey}, J. 1985, in ESA Special Publication, Vol. 235, Future Missions in
  Solar, Heliospheric \& Space Plasma Physics, ed. E.~{Rolfe} \& B.~{Battrick},
  199--208

\bibitem[{{Henry} {et~al.}(1995){Henry}, {Eaton}, {Hamer}, \&
  {Hall}}]{1995ApJS...97..513H}
{Henry}, G.~W., {Eaton}, J.~A., {Hamer}, J., \& {Hall}, D.~S. 1995, \apjs, 97,
  513

\bibitem[{{Hotta} \& {Yokoyama}(2011)}]{2011ApJ...740...12H}
{Hotta}, H., \& {Yokoyama}, T. 2011, \apj, 740, 12

\bibitem[{{Houdek} {et~al.}(1999){Houdek}, {Balmforth},
  {Christensen-Dalsgaard}, \& {Gough}}]{1999A&A...351..582H}
{Houdek}, G., {Balmforth}, N.~J., {Christensen-Dalsgaard}, J., \& {Gough},
  D.~O. 1999, \aap, 351, 582

\bibitem[{{Howard} {et~al.}(1984){Howard}, {Gilman}, \&
  {Gilman}}]{1984ApJ...283..373H}
{Howard}, R., {Gilman}, P.~I., \& {Gilman}, P.~A. 1984, \apj, 283, 373

\bibitem[{{Howe}(2009)}]{2009LRSP....6....1H}
{Howe}, R. 2009, Living Reviews in Solar Physics, 6, 1

\bibitem[{{Howe} {et~al.}(2005){Howe}, {Christensen-Dalsgaard}, {Hill}, {Komm},
  {Schou}, \& {Thompson}}]{2005ApJ...634.1405H}
{Howe}, R., {Christensen-Dalsgaard}, J., {Hill}, F., {et~al.} 2005, \apj, 634,
  1405

\bibitem[{{Howe} {et~al.}(2006){Howe}, {Rempel}, {Christensen-Dalsgaard},
  {Hill}, {Komm}, {Larsen}, {Schou}, \& {Thompson}}]{2006ApJ...649.1155H}
{Howe}, R., {Rempel}, M., {Christensen-Dalsgaard}, J., {et~al.} 2006, \apj,
  649, 1155

\bibitem[{{Huber} {et~al.}(2010){Huber}, {Czesla}, {Wolter}, \&
  {Schmitt}}]{2010A&A...514A..39H}
{Huber}, K.~F., {Czesla}, S., {Wolter}, U., \& {Schmitt}, J.~H.~M.~M. 2010,
  \aap, 514, A39

\bibitem[{{Iglesias} \& {Rogers}(1996)}]{1996ApJ...464..943I}
{Iglesias}, C.~A., \& {Rogers}, F.~J. 1996, \apj, 464, 943

\bibitem[{Johansson {et~al.}(2010)}]{mpmath}
Johansson, F., {et~al.} 2010, mpmath: a {P}ython library for
  arbitrary-precision floating-point arithmetic (version 0.14), {\tt
  http://code.google.com/p/mpmath/}

\bibitem[{{Jones} {et~al.}(2009){Jones}, {Kuzanyan}, \&
  {Mitchell}}]{2009JFM...634..291J}
{Jones}, C.~A., {Kuzanyan}, K.~M., \& {Mitchell}, R.~H. 2009, Journal of Fluid
  Mechanics, 634, 291

\bibitem[{{Jones} {et~al.}(1989){Jones}, {Hansen}, {Pesnell}, \&
  {Kawaler}}]{1989ApJ...336..403J}
{Jones}, P.~W., {Hansen}, C.~J., {Pesnell}, W.~D., \& {Kawaler}, S.~D. 1989,
  \apj, 336, 403

\bibitem[{{K{\"a}pyl{\"a}} {et~al.}(2012){K{\"a}pyl{\"a}}, {Mantere}, \&
  {Brandenburg}}]{2012ApJ...755L..22K}
{K{\"a}pyl{\"a}}, P.~J., {Mantere}, M.~J., \& {Brandenburg}, A. 2012, \apjl,
  755, L22

\bibitem[{{K{\"a}pyl{\"a}} {et~al.}(2011){K{\"a}pyl{\"a}}, {Mantere},
  {Guerrero}, {Brandenburg}, \& {Chatterjee}}]{2011A&A...531A.162K}
{K{\"a}pyl{\"a}}, P.~J., {Mantere}, M.~J., {Guerrero}, G., {Brandenburg}, A.,
  \& {Chatterjee}, P. 2011, \aap, 531, A162

\bibitem[{{Karoff}(2012)}]{2012MNRAS.421.3170K}
{Karoff}, C. 2012, \mnras, 421, 3170

\bibitem[{{Kawaler}(1988)}]{1988ApJ...333..236K}
{Kawaler}, S.~D. 1988, \apj, 333, 236

\bibitem[{{Kawaler}(2009)}]{2009ASPC..416..385K}
{Kawaler}, S.~D. 2009, in Astronomical Society of the Pacific Conference
  Series, Vol. 416, Solar-Stellar Dynamos as Revealed by Helio- and
  Asteroseismology: GONG 2008/SOHO 21, ed. M.~{Dikpati}, T.~{Arentoft},
  I.~{Gonz{\'a}lez Hern{\'a}ndez}, C.~{Lindsey}, \& F.~{Hill}, 385

\bibitem[{{Kawaler} \& {Hostler}(2005)}]{2005ApJ...621..432K}
{Kawaler}, S.~D., \& {Hostler}, S.~R. 2005, \apj, 621, 432

\bibitem[{{Kawaler} {et~al.}(1999){Kawaler}, {Sekii}, \&
  {Gough}}]{1999ApJ...516..349K}
{Kawaler}, S.~D., {Sekii}, T., \& {Gough}, D. 1999, \apj, 516, 349

\bibitem[{{Kitchatinov}(2013)}]{2013IAUS..294..399K}
{Kitchatinov}, L.~L. 2013, in IAU Symposium, Vol. 294, IAU Symposium, ed. A.~G.
  {Kosovichev}, E.~{de Gouveia Dal Pino}, \& Y.~{Yan}, 399--410

\bibitem[{{Kitchatinov} \& {Olemskoy}(2011)}]{2011MNRAS.411.1059K}
{Kitchatinov}, L.~L., \& {Olemskoy}, S.~V. 2011, \mnras, 411, 1059

\bibitem[{{Kitchatinov} \& {Olemskoy}(2012)}]{2012MNRAS.423.3344K}
---. 2012, \mnras, 423, 3344

\bibitem[{{Kitchatinov} \& {R{\"u}diger}(1999)}]{1999A&A...344..911K}
{Kitchatinov}, L.~L., \& {R{\"u}diger}, G. 1999, \aap, 344, 911

\bibitem[{{Kitchatinov} \& {R{\"u}diger}(2004)}]{2004AN....325..496K}
---. 2004, Astronomische Nachrichten, 325, 496

\bibitem[{{Kitchatinov} \& {Ruediger}(1995)}]{1995A&A...299..446K}
{Kitchatinov}, L.~L., \& {Ruediger}, G. 1995, \aap, 299, 446

\bibitem[{{Kjeldsen} {et~al.}(1998){Kjeldsen}, {Arentoft}, {Bedding},
  {Christensen-Dalsgaard}, {Frandsen}, \& {Thompson}}]{1998ESASP.418..385K}
{Kjeldsen}, H., {Arentoft}, T., {Bedding}, T., {et~al.} 1998, in ESA Special
  Publication, Vol. 418, Structure and Dynamics of the Interior of the Sun and
  Sun-like Stars, ed. S.~{Korzennik}, 385

\bibitem[{{Kjeldsen} \& {Bedding}(1995)}]{1995A&A...293...87K}
{Kjeldsen}, H., \& {Bedding}, T.~R. 1995, \aap, 293, 87

\bibitem[{{Korhonen}(2012)}]{2012IAUS..286..268K}
{Korhonen}, H. 2012, in IAU Symposium, Vol. 286, IAU Symposium, ed. C.~H.
  {Mandrini} \& D.~F. {Webb}, 268--278

\bibitem[{{Korhonen} {et~al.}(2000){Korhonen}, {Berdyugina}, {Hackman},
  {Strassmeier}, \& {Tuominen}}]{2000A&A...360.1067K}
{Korhonen}, H., {Berdyugina}, S.~V., {Hackman}, T., {Strassmeier}, K.~G., \&
  {Tuominen}, I. 2000, \aap, 360, 1067

\bibitem[{{Korhonen} {et~al.}(2013){Korhonen}, {Gonz{\'a}lez}, {Briquet},
  {Flores Soriano}, {Hubrig}, {Savanov}, {Hackman}, {Ilyin}, {Eulaers}, \&
  {Pessemier}}]{2013A&A...553A..27K}
{Korhonen}, H., {Gonz{\'a}lez}, J.~F., {Briquet}, M., {et~al.} 2013, \aap, 553,
  A27

\bibitem[{{Kosovichev} {et~al.}(1997){Kosovichev}, {Schou}, {Scherrer},
  {Bogart}, {Bush}, {Hoeksema}, {Aloise}, {Bacon}, {Burnette}, {de Forest},
  {Giles}, {Leibrand}, {Nigam}, {Rubin}, {Scott}, {Williams}, {Basu},
  {Christensen-Dalsgaard}, {Dappen}, {Rhodes}, {Duvall}, {Howe}, {Thompson},
  {Gough}, {Sekii}, {Toomre}, {Tarbell}, {Title}, {Mathur}, {Morrison}, {Saba},
  {Wolfson}, {Zayer}, \& {Milford}}]{1997SoPh..170...43K}
{Kosovichev}, A.~G., {Schou}, J., {Scherrer}, P.~H., {et~al.} 1997, \solphys,
  170, 43

\bibitem[{{Kov{\'a}ri} {et~al.}(2007){Kov{\'a}ri}, {Bartus}, {Strassmeier},
  {Vida}, {{\v S}vanda}, \& {Ol{\'a}h}}]{2007A&A...474..165K}
{Kov{\'a}ri}, Z., {Bartus}, J., {Strassmeier}, K.~G., {et~al.} 2007, \aap, 474,
  165

\bibitem[{{Kov{\'a}ri} {et~al.}(2004){Kov{\'a}ri}, {Strassmeier}, {Granzer},
  {Weber}, {Ol{\'a}h}, \& {Rice}}]{2004A&A...417.1047K}
{Kov{\'a}ri}, Z., {Strassmeier}, K.~G., {Granzer}, T., {et~al.} 2004, \aap,
  417, 1047

\bibitem[{{K{\"u}ker} \& {R{\"u}diger}(2005)}]{2005AN....326..265K}
{K{\"u}ker}, M., \& {R{\"u}diger}, G. 2005, Astronomische Nachrichten, 326, 265

\bibitem[{{K{\"u}ker} {et~al.}(2011){K{\"u}ker}, {R{\"u}diger}, \&
  {Kitchatinov}}]{2011A&A...530A..48K}
{K{\"u}ker}, M., {R{\"u}diger}, G., \& {Kitchatinov}, L.~L. 2011, \aap, 530,
  A48

\bibitem[{{Kumar} \& {Quataert}(1997)}]{1997ApJ...475L.143K}
{Kumar}, P., \& {Quataert}, E.~J. 1997, \apjl, 475, L143

\bibitem[{{Kurtz} \& {Shibahashi}(1986)}]{1986MNRAS.223..557K}
{Kurtz}, D.~W., \& {Shibahashi}, H. 1986, \mnras, 223, 557

\bibitem[{{Lanza} {et~al.}(2014){Lanza}, {Das Chagas}, \& {De
  Medeiros}}]{2014A&A...564A..50L}
{Lanza}, A.~F., {Das Chagas}, M.~L., \& {De Medeiros}, J.~R. 2014, \aap, 564,
  A50

\bibitem[{{Ledoux}(1951)}]{1951ApJ...114..373L}
{Ledoux}, P. 1951, \apj, 114, 373

\bibitem[{{Libbrecht}(1988)}]{1988ApJ...334..510L}
{Libbrecht}, K.~G. 1988, \apj, 334, 510

\bibitem[{{Libbrecht}(1992)}]{1992ApJ...387..712L}
---. 1992, \apj, 387, 712

\bibitem[{{Libbrecht} \& {Woodard}(1990)}]{1990Natur.345..779L}
{Libbrecht}, K.~G., \& {Woodard}, M.~F. 1990, \nat, 345, 779

\bibitem[{{Ligni{\`e}res} {et~al.}(2006){Ligni{\`e}res}, {Rieutord}, \&
  {Reese}}]{2006A&A...455..607L}
{Ligni{\`e}res}, F., {Rieutord}, M., \& {Reese}, D. 2006, \aap, 455, 607

\bibitem[{{Maeder} \& {Meynet}(2000)}]{2000ARA&A..38..143M}
{Maeder}, A., \& {Meynet}, G. 2000, \araa, 38, 143

\bibitem[{{Maeder} \& {Meynet}(2004)}]{2004A&A...422..225M}
---. 2004, \aap, 422, 225

\bibitem[{{Maeder} \& {Zahn}(1998)}]{1998A&A...334.1000M}
{Maeder}, A., \& {Zahn}, J.-P. 1998, \aap, 334, 1000

\bibitem[{{Marques} {et~al.}(2013){Marques}, {Goupil}, {Lebreton}, {Talon},
  {Palacios}, {Belkacem}, {Ouazzani}, {Mosser}, {Moya}, {Morel}, {Pichon},
  {Mathis}, {Zahn}, {Turck-Chi{\`e}ze}, \& {Nghiem}}]{2013A&A...549A..74M}
{Marques}, J.~P., {Goupil}, M.~J., {Lebreton}, Y., {et~al.} 2013, \aap, 549,
  A74

\bibitem[{{Mathis}(2009)}]{2009A&A...506..811M}
{Mathis}, S. 2009, \aap, 506, 811

\bibitem[{{Mathur} {et~al.}(2013){Mathur}, {Garc{\'{\i}}a}, {Morgenthaler},
  {Salabert}, {Petit}, {Ballot}, {R{\'e}gulo}, \&
  {Catala}}]{2013A&A...550A..32M}
{Mathur}, S., {Garc{\'{\i}}a}, R.~A., {Morgenthaler}, A., {et~al.} 2013, \aap,
  550, A32

\bibitem[{{Mathur} {et~al.}(2011){Mathur}, {Hekker}, {Trampedach}, {Ballot},
  {Kallinger}, {Buzasi}, {Garc{\'{\i}}a}, {Huber}, {Jim{\'e}nez}, {Mosser},
  {Bedding}, {Elsworth}, {R{\'e}gulo}, {Stello}, {Chaplin}, {De Ridder},
  {Hale}, {Kinemuchi}, {Kjeldsen}, {Mullally}, \&
  {Thompson}}]{2011ApJ...741..119M}
{Mathur}, S., {Hekker}, S., {Trampedach}, R., {et~al.} 2011, \apj, 741, 119

\bibitem[{{McQuillan} {et~al.}(2013){McQuillan}, {Aigrain}, \&
  {Mazeh}}]{2013MNRAS.432.1203M}
{McQuillan}, A., {Aigrain}, S., \& {Mazeh}, T. 2013, \mnras, 432, 1203

\bibitem[{{McQuillan} {et~al.}(2014){McQuillan}, {Mazeh}, \&
  {Aigrain}}]{2014ApJS..211...24M}
{McQuillan}, A., {Mazeh}, T., \& {Aigrain}, S. 2014, \apjs, 211, 24

\bibitem[{{Meibom} {et~al.}(2009){Meibom}, {Mathieu}, \&
  {Stassun}}]{2009ApJ...695..679M}
{Meibom}, S., {Mathieu}, R.~D., \& {Stassun}, K.~G. 2009, \apj, 695, 679

\bibitem[{{Metcalfe} {et~al.}(2012){Metcalfe}, {Chaplin}, {Appourchaux},
  {Garc{\'{\i}}a}, {Basu}, {Brand{\~a}o}, {Creevey}, {Deheuvels}, {Do{\v g}an},
  {Eggenberger}, {Karoff}, {Miglio}, {Stello}, {Y{\i}ld{\i}z}, {{\c C}elik},
  {Antia}, {Benomar}, {Howe}, {R{\'e}gulo}, {Salabert}, {Stahn}, {Bedding},
  {Davies}, {Elsworth}, {Gizon}, {Hekker}, {Mathur}, {Mosser}, {Bryson},
  {Still}, {Christensen-Dalsgaard}, {Gilliland}, {Kawaler}, {Kjeldsen},
  {Ibrahim}, {Klaus}, \& {Li}}]{2012ApJ...748L..10M}
{Metcalfe}, T.~S., {Chaplin}, W.~J., {Appourchaux}, T., {et~al.} 2012, \apjl,
  748, L10

\bibitem[{{Michel} {et~al.}(2009){Michel}, {Samadi}, {Baudin}, {Barban},
  {Appourchaux}, \& {Auvergne}}]{2009A&A...495..979M}
{Michel}, E., {Samadi}, R., {Baudin}, F., {et~al.} 2009, \aap, 495, 979

\bibitem[{{Miesch}(2005)}]{2005LRSP....2....1M}
{Miesch}, M.~S. 2005, Living Reviews in Solar Physics, 2, 1

\bibitem[{{Miesch} {et~al.}(2008){Miesch}, {Brun}, {De Rosa}, \&
  {Toomre}}]{2008ApJ...673..557M}
{Miesch}, M.~S., {Brun}, A.~S., {De Rosa}, M.~L., \& {Toomre}, J. 2008, \apj,
  673, 557

\bibitem[{{Miesch} {et~al.}(2006){Miesch}, {Brun}, \&
  {Toomre}}]{2006ApJ...641..618M}
{Miesch}, M.~S., {Brun}, A.~S., \& {Toomre}, J. 2006, \apj, 641, 618

\bibitem[{{Miesch} \& {Hindman}(2011)}]{2011ApJ...743...79M}
{Miesch}, M.~S., \& {Hindman}, B.~W. 2011, \apj, 743, 79

\bibitem[{{Miesch} \& {Toomre}(2009)}]{2009AnRFM..41..317M}
{Miesch}, M.~S., \& {Toomre}, J. 2009, Annual Review of Fluid Mechanics, 41,
  317

\bibitem[{{Mosser} {et~al.}(2012{\natexlab{a}}){Mosser}, {Goupil}, {Belkacem},
  {Michel}, {Stello}, {Marques}, {Elsworth}, {Barban}, {Beck}, {Bedding}, {De
  Ridder}, {Garc{\'{\i}}a}, {Hekker}, {Kallinger}, {Samadi}, {Stumpe},
  {Barclay}, \& {Burke}}]{2012A&A...540A.143M}
{Mosser}, B., {Goupil}, M.~J., {Belkacem}, K., {et~al.} 2012{\natexlab{a}},
  \aap, 540, A143

\bibitem[{{Mosser} {et~al.}(2012{\natexlab{b}}){Mosser}, {Goupil}, {Belkacem},
  {Marques}, {Beck}, {Bloemen}, {De Ridder}, {Barban}, {Deheuvels}, {Elsworth},
  {Hekker}, {Kallinger}, {Ouazzani}, {Pinsonneault}, {Samadi}, {Stello},
  {Garc{\'{\i}}a}, {Klaus}, {Li}, {Mathur}, \& {Morris}}]{2012A&A...548A..10M}
---. 2012{\natexlab{b}}, \aap, 548, A10

\bibitem[{{Nielsen} {et~al.}(2013){Nielsen}, {Gizon}, {Schunker}, \&
  {Karoff}}]{2013A&A...557L..10N}
{Nielsen}, M.~B., {Gizon}, L., {Schunker}, H., \& {Karoff}, C. 2013, \aap, 557,
  L10

\bibitem[{{Nigam} \& {Kosovichev}(1998)}]{1998ApJ...505L..51N}
{Nigam}, R., \& {Kosovichev}, A.~G. 1998, \apjl, 505, L51

\bibitem[{{Osaki}(1990)}]{1990LNP...367...75O}
{Osaki}, Y. 1990, in Lecture Notes in Physics, Berlin Springer Verlag, Vol.
  367, Progress of Seismology of the Sun and Stars, ed. Y.~{Osaki} \&
  H.~{Shibahashi}, 75

\bibitem[{{Ouazzani} {et~al.}(2012){Ouazzani}, {Dupret}, \&
  {Reese}}]{2012A&A...547A..75O}
{Ouazzani}, R.-M., {Dupret}, M.-A., \& {Reese}, D.~R. 2012, \aap, 547, A75

\bibitem[{{Ouazzani} \& {Goupil}(2012)}]{2012A&A...542A..99O}
{Ouazzani}, R.-M., \& {Goupil}, M.-J. 2012, \aap, 542, A99

\bibitem[{{Pamyatnykh} {et~al.}(2004){Pamyatnykh}, {Handler}, \&
  {Dziembowski}}]{2004MNRAS.350.1022P}
{Pamyatnykh}, A.~A., {Handler}, G., \& {Dziembowski}, W.~A. 2004, \mnras, 350,
  1022

\bibitem[{{Pesnell}(1985)}]{1985ApJ...292..238P}
{Pesnell}, W.~D. 1985, \apj, 292, 238

\bibitem[{{Pinsonneault} {et~al.}(1989){Pinsonneault}, {Kawaler}, {Sofia}, \&
  {Demarque}}]{1989ApJ...338..424P}
{Pinsonneault}, M.~H., {Kawaler}, S.~D., {Sofia}, S., \& {Demarque}, P. 1989,
  \apj, 338, 424

\bibitem[{{Plumb} \& {McEwan}(1978)}]{Plumb:1978}
{Plumb}, R.~A., \& {McEwan}, A.~D. 1978, J. Atmos. Sci., 35, 1827

\bibitem[{{Rauer} {et~al.}(2013){Rauer}, {Catala}, {Aerts}, {Appourchaux}, {Benz}}]{plato_ref}
{Rauer}, H., {Catala}, C., {Aerts}, C., {et~al.} 2013, ArXiv e-prints 1310.0696

\bibitem[{{Reese} {et~al.}(2006){Reese}, {Ligni{\`e}res}, \&
  {Rieutord}}]{2006A&A...455..621R}
{Reese}, D., {Ligni{\`e}res}, F., \& {Rieutord}, M. 2006, \aap, 455, 621

\bibitem[{{Reiners}(2006)}]{2006A&A...446..267R}
{Reiners}, A. 2006, \aap, 446, 267

\bibitem[{{Reiners} \& {Mohanty}(2012)}]{2012ApJ...746...43R}
{Reiners}, A., \& {Mohanty}, S. 2012, \apj, 746, 43

\bibitem[{{Reiners} \& {Schmitt}(2002)}]{2002A&A...384..155R}
{Reiners}, A., \& {Schmitt}, J.~H.~M.~M. 2002, \aap, 384, 155

\bibitem[{{Reiners} \& {Schmitt}(2003)}]{2003A&A...398..647R}
---. 2003, \aap, 398, 647

\bibitem[{{Reiners} {et~al.}(2001){Reiners}, {Schmitt}, \&
  {K{\"u}rster}}]{2001A&A...376L..13R}
{Reiners}, A., {Schmitt}, J.~H.~M.~M., \& {K{\"u}rster}, M. 2001, \aap, 376,
  L13

\bibitem[{{Reinhold} \& {Reiners}(2013)}]{2013A&A...557A..11R}
{Reinhold}, T., \& {Reiners}, A. 2013, \aap, 557, A11

\bibitem[{{Reinhold} {et~al.}(2013){Reinhold}, {Reiners}, \&
  {Basri}}]{2013A&A...560A...4R}
{Reinhold}, T., {Reiners}, A., \& {Basri}, G. 2013, \aap, 560, A4

\bibitem[{{Rempel}(2005)}]{2005ApJ...622.1320R}
{Rempel}, M. 2005, \apj, 622, 1320

\bibitem[{{Robinson} \& {Chan}(2001)}]{2001MNRAS.321..723R}
{Robinson}, F.~J., \& {Chan}, K.~L. 2001, \mnras, 321, 723

\bibitem[{{Rogers} {et~al.}(1996){Rogers}, {Swenson}, \&
  {Iglesias}}]{1996ApJ...456..902R}
{Rogers}, F.~J., {Swenson}, F.~J., \& {Iglesias}, C.~A. 1996, \apj, 456, 902

\bibitem[{{Rogers} {et~al.}(2013){Rogers}, {Lin}, {McElwaine}, \&
  {Lau}}]{2013ApJ...772...21R}
{Rogers}, T.~M., {Lin}, D.~N.~C., {McElwaine}, J.~N., \& {Lau}, H.~H.~B. 2013,
  \apj, 772, 21

\bibitem[{{Roth} \& {Stix}(2008)}]{2008SoPh..251...77R}
{Roth}, M., \& {Stix}, M. 2008, \solphys, 251, 77

\bibitem[{{Royer} {et~al.}(2002){Royer}, {Gerbaldi}, {Faraggiana}, \&
  {G{\'o}mez}}]{2002A&A...381..105R}
{Royer}, F., {Gerbaldi}, M., {Faraggiana}, R., \& {G{\'o}mez}, A.~E. 2002,
  \aap, 381, 105

\bibitem[{R\"{u}diger(1989)}]{rudig89}
R\"{u}diger, G. 1989, Differential Rotation and Stellar Convection (New York:
  Gordon and Breach)

\bibitem[{{Schatzman}(1993)}]{1993A&A...279..431S}
{Schatzman}, E. 1993, \aap, 279, 431

\bibitem[{{Schou} {et~al.}(1994){Schou}, {Christensen-Dalsgaard}, \&
  {Thompson}}]{1994ApJ...433..389S}
{Schou}, J., {Christensen-Dalsgaard}, J., \& {Thompson}, M.~J. 1994, \apj, 433,
  389

\bibitem[{{Schou} {et~al.}(1998){Schou}, {Antia}, {Basu}, {Bogart}, {Bush},
  {Chitre}, {Christensen-Dalsgaard}, {di Mauro}, {Dziembowski}, {Eff-Darwich},
  {Gough}, {Haber}, {Hoeksema}, {Howe}, {Korzennik}, {Kosovichev}, {Larsen},
  {Pijpers}, {Scherrer}, {Sekii}, {Tarbell}, {Title}, {Thompson}, \&
  {Toomre}}]{1998ApJ...505..390S}
{Schou}, J., {Antia}, H.~M., {Basu}, S., {et~al.} 1998, \apj, 505, 390

\bibitem[{{Sekiguchi} \& {Fukugita}(2000)}]{2000AJ....120.1072S}
{Sekiguchi}, M., \& {Fukugita}, M. 2000, \aj, 120, 1072

\bibitem[{{Shu} {et~al.}(1994){Shu}, {Najita}, {Ostriker}, {Wilkin}, {Ruden},
  \& {Lizano}}]{1994ApJ...429..781S}
{Shu}, F., {Najita}, J., {Ostriker}, E., {et~al.} 1994, \apj, 429, 781

\bibitem[{{Sills} \& {Pinsonneault}(2000)}]{2000ApJ...540..489S}
{Sills}, A., \& {Pinsonneault}, M.~H. 2000, \apj, 540, 489

\bibitem[{{Skumanich}(1972)}]{1972ApJ...171..565S}
{Skumanich}, A. 1972, \apj, 171, 565

\bibitem[{{Snodgrass}(1983)}]{1983ApJ...270..288S}
{Snodgrass}, H.~B. 1983, \apj, 270, 288

\bibitem[{{Snodgrass} {et~al.}(1984){Snodgrass}, {Howard}, \&
  {Webster}}]{1984SoPh...90..199S}
{Snodgrass}, H.~B., {Howard}, R., \& {Webster}, L. 1984, \solphys, 90, 199

\bibitem[{{Soufi} {et~al.}(1998){Soufi}, {Goupil}, \&
  {Dziembowski}}]{1998A&A...334..911S}
{Soufi}, F., {Goupil}, M.~J., \& {Dziembowski}, W.~A. 1998, \aap, 334, 911

\bibitem[{{Spiegel} \& {Zahn}(1992)}]{1992A&A...265..106S}
{Spiegel}, E.~A., \& {Zahn}, J.-P. 1992, \aap, 265, 106

\bibitem[{{Spruit}(1999)}]{1999A&A...349..189S}
{Spruit}, H.~C. 1999, \aap, 349, 189

\bibitem[{{Spruit}(2002)}]{2002A&A...381..923S}
---. 2002, \aap, 381, 923

\bibitem[{{Strassmeier} {et~al.}(2003){Strassmeier}, {Kratzwald}, \&
  {Weber}}]{2003A&A...408.1103S}
{Strassmeier}, K.~G., {Kratzwald}, L., \& {Weber}, M. 2003, \aap, 408, 1103

\bibitem[{{Su{\'a}rez} {et~al.}(2010){Su{\'a}rez}, {Andrade}, {Goupil}, \&
  {Janot-Pacheco}}]{2010AN....331.1073S}
{Su{\'a}rez}, J.~C., {Andrade}, L., {Goupil}, M.~J., \& {Janot-Pacheco}, E.
  2010, Astronomische Nachrichten, 331, 1073

\bibitem[{{Su{\'a}rez} {et~al.}(2006){Su{\'a}rez}, {Goupil}, \&
  {Morel}}]{2006A&A...449..673S}
{Su{\'a}rez}, J.~C., {Goupil}, M.~J., \& {Morel}, P. 2006, \aap, 449, 673

\bibitem[{{Talon} \& {Charbonnel}(2003)}]{2003A&A...405.1025T}
{Talon}, S., \& {Charbonnel}, C. 2003, \aap, 405, 1025

\bibitem[{{Talon} \& {Charbonnel}(2005)}]{2005A&A...440..981T}
---. 2005, \aap, 440, 981

\bibitem[{{Talon} \& {Charbonnel}(2008)}]{2008A&A...482..597T}
---. 2008, \aap, 482, 597

\bibitem[{{Talon} {et~al.}(2002){Talon}, {Kumar}, \&
  {Zahn}}]{2002ApJ...574L.175T}
{Talon}, S., {Kumar}, P., \& {Zahn}, J.-P. 2002, \apjl, 574, L175

\bibitem[{{Tayler}(1973)}]{1973MNRAS.161..365T}
{Tayler}, R.~J. 1973, \mnras, 161, 365

\bibitem[{{Thompson} {et~al.}(2003){Thompson}, {Christensen-Dalsgaard},
  {Miesch}, \& {Toomre}}]{2003ARA&A..41..599T}
{Thompson}, M.~J., {Christensen-Dalsgaard}, J., {Miesch}, M.~S., \& {Toomre},
  J. 2003, \araa, 41, 599

\bibitem[{{Thompson} {et~al.}(1996){Thompson}, {Toomre}, {Anderson}, {Antia},
  {Berthomieu}, {Burtonclay}, {Chitre}, {Christensen-Dalsgaard}, {Corbard}, {De
  Rosa}, {Genovese}, {Gough}, {Haber}, {Harvey}, {Hill}, {Howe}, {Korzennik},
  {Kosovichev}, {Leibacher}, {Pijpers}, {Provost}, {Rhodes}, {Schou}, {Sekii},
  {Stark}, \& {Wilson}}]{1996Sci...272.1300T}
{Thompson}, M.~J., {Toomre}, J., {Anderson}, E.~R., {et~al.} 1996, Science,
  272, 1300

\bibitem[{{Tomczyk} {et~al.}(1995){Tomczyk}, {Schou}, \&
  {Thompson}}]{1995ApJ...448L..57T}
{Tomczyk}, S., {Schou}, J., \& {Thompson}, M.~J. 1995, \apjl, 448, L57

\bibitem[{{Toutain} \& {Appourchaux}(1994)}]{1994A&A...289..649T}
{Toutain}, T., \& {Appourchaux}, T. 1994, \aap, 289, 649

\bibitem[{{Uytterhoeven} {et~al.}(2012){Uytterhoeven}, {Pall{\'e}}, {Grundahl},
  {Frandsen}, {Christensen-Dalsgaard}, {Trivi{\~n}o Hage}, \& {SONG
  Team}}]{2012AN....333.1103U}
{Uytterhoeven}, K., {Pall{\'e}}, P.~L., {Grundahl}, F., {et~al.} 2012,
  Astronomische Nachrichten, 333, 1103

\bibitem[{{Van Eylen} {et~al.}(2014){Van Eylen}, {Lund}, {Silva Aguirre},
  {Arentoft}, {Kjeldsen}, {Albrecht}, {Chaplin}, {Isaacson}, {Pedersen},
  {Jessen-Hansen}, {Tingley}, {Christensen-Dalsgaard}, {Aerts}, {Campante}, \&
  {Bryson}}]{2014ApJ...782...14V}
{Van Eylen}, V., {Lund}, M.~N., {Silva Aguirre}, V., {et~al.} 2014, \apj, 782,
  14

\bibitem[{{van Saders} \& {Pinsonneault}(2013)}]{2013ApJ...776...67V}
{van Saders}, J.~L., \& {Pinsonneault}, M.~H. 2013, \apj, 776, 67

\bibitem[{{Vogt} {et~al.}(1999){Vogt}, {Hatzes}, {Misch}, \&
  {K{\"u}rster}}]{1999ApJS..121..547V}
{Vogt}, S.~S., {Hatzes}, A.~P., {Misch}, A.~A., \& {K{\"u}rster}, M. 1999,
  \apjs, 121, 547

\bibitem[{{Vorontsov} {et~al.}(2002){Vorontsov}, {Christensen-Dalsgaard},
  {Schou}, {Strakhov}, \& {Thompson}}]{2002Sci...296..101V}
{Vorontsov}, S.~V., {Christensen-Dalsgaard}, J., {Schou}, J., {Strakhov},
  V.~N., \& {Thompson}, M.~J. 2002, Science, 296, 101

\bibitem[{{Walker} {et~al.}(2007){Walker}, {Croll}, {Kuschnig}, {Walker},
  {Rucinski}, {Matthews}, {Guenther}, {Moffat}, {Sasselov}, \&
  {Weiss}}]{2007ApJ...659.1611W}
{Walker}, G.~A.~H., {Croll}, B., {Kuschnig}, R., {et~al.} 2007, \apj, 659, 1611

\bibitem[{{Weber}(2007)}]{2007AN....328.1075W}
{Weber}, M. 2007, Astronomische Nachrichten, 328, 1075

\bibitem[{{Wilcox} \& {Howard}(1970)}]{1970SoPh...13..251W}
{Wilcox}, J.~M., \& {Howard}, R. 1970, \solphys, 13, 251

\bibitem[{{Winget} {et~al.}(1991){Winget}, {Nather}, {Clemens}, {Provencal},
  {Kleinman}, {Bradley}, {Wood}, {Claver}, {Frueh}, {Grauer}, {Hine}, {Hansen},
  {Fontaine}, {Achilleos}, {Wickramasinghe}, {Marar}, {Seetha}, {Ashoka},
  {O'Donoghue}, {Warner}, {Kurtz}, {Buckley}, {Brickhill}, {Vauclair}, {Dolez},
  {Chevreton}, {Barstow}, {Solheim}, {Kanaan}, {Kepler}, {Henry}, \&
  {Kawaler}}]{1991ApJ...378..326W}
{Winget}, D.~E., {Nather}, R.~E., {Clemens}, J.~C., {et~al.} 1991, \apj, 378,
  326

\bibitem[{{Zahn}(1992)}]{1992A&A...265..115Z}
{Zahn}, J.-P. 1992, \aap, 265, 115

\bibitem[{{Zahn} {et~al.}(1997){Zahn}, {Talon}, \&
  {Matias}}]{1997A&A...322..320Z}
{Zahn}, J.-P., {Talon}, S., \& {Matias}, J. 1997, \aap, 322, 320

\end{thebibliography}

\clearpage

\appendix

\section{\normalsize Appendix A: Kernels}
\label{app:ker}

In the following we will go through our calculation of the rotation kernel entering the description of the splitting as in \eqref{eq:split} \citep[][]{1977ApJ...217..151H, 1981MNRAS.196..731G}:
\begin{equation}
\delta\nu_{nlm} = m \int_0^{\pi}\int_0^R \mathcal{K}_{nlm}(r, \theta)\mathcal{N}(r, \theta) r dr d\theta.
\end{equation} 
We define $L^2 = l(l+1)$, and we use the normalized associated Legendre function $\Theta_{l}^{m}(x)$ given by
\begin{equation}
\Theta_{l}^{m}(x) = \left[\frac{(2l +1)(l-|m|)!}{2(l+|m|)!} \right]^{1/2} P_{l}^{m}(x),
\end{equation}
where $P_{l}^m(x)$ is the normal associated Legendre polynomial. $\Theta_{l}^m(x)$ has the property
\begin{equation}
\int_{-1}^{1} \Theta_{l}^m(x)^2 dx = 1.
\end{equation}

Following the prescription by \citet[][]{1994ApJ...433..389S} and \citet[][]{2010aste.book.....A} we write the rotation kernel as
\small{
\begin{equation}\label{eq:kernel_ref}
\mathcal{K}_{nlm}(r, \theta) = \mathcal{I}_{nlm}^{-1} \left\lbrace  \xi_r(r)\left[\xi_r(r) - 2\xi_h(r)   \right] \Theta_{l}^m(\cos\theta)^2 + \xi_h^2\left[\left(\frac{d\Theta_{l}^m}{d\theta} \right)^2  - 2 \Theta_{l}^m(\cos\theta)  \frac{d\Theta_{l}^m}{d\theta} \frac{\cos\theta}{\sin\theta} + \frac{m^2}{\sin^2\theta} \Theta_{l}^m(\cos\theta)^2 \right]  \right\rbrace \rho(r) r \sin\theta\, .
\end{equation}
}%
Here $\rho(r)$ is the density as a function of radius, $\xi_r(r)$ and $\xi_h(r)$\footnote{Sometimes written as $L^{-1}\eta_{nl}$.} are the radial and horizontal displacement eigenfunctions, respectively \citep[][]{1991sia..book..401C}, and $\mathcal{I}_{nlm}$ is the mode inertia (see below).
We make the variable change $x = \cos\theta$ ($\sin\theta=\sqrt{1-x^2}$) and use
\small{
\begin{equation}
\qquad \frac{d}{d\theta} = \frac{d}{d\theta} (\cos\theta)\frac{d}{d(\cos\theta)}  =-\sin\theta \frac{d}{d(\cos\theta)} = -\sqrt{1-x^2} \frac{d}{dx}\, ,
\end{equation}
}%
to get\footnote{Note that there seems to be a sign error in \citet{1994ApJ...433..389S} on the $2 \Theta_{l}^m(x)  \frac{d\Theta_{l}^m}{dx} x$ term. }
\small{
\begin{equation}
\mathcal{K}_{nlm}(r, \theta)= \mathcal{I}_{nlm}^{-1} \left\lbrace  \xi_r(r)\left[\xi_r(r) - 2\xi_h(r)   \right] \Theta_{l}^m(x)^2 + \xi_h^2\left[\left(\frac{d\Theta_{l}^m}{dx} \right)^2 \left(1-x^2\right) + 2 \Theta_{l}^m(x)  \frac{d\Theta_{l}^m}{dx} x + \frac{m^2}{1-x^2} \Theta_{l}^m(x)^2 \right]  \right\rbrace \rho(r) r \sin\theta\, .
\end{equation}
}%
From ADIPLS we do not directly get the displacement eigenfunctions; rather, we get the following density-weighted versions, $Z_1(r)$ and $Z_2(r)$, from which we can get $\xi_r(r)$ and $\xi_h(r)$:
\begin{equation}
\setlength\arraycolsep{0.3em}
  \begin{array}{llll}
\displaystyle Z_1(r) &= \sqrt{\frac{4\pi r^3 \rho(r)}{M}} \frac{\xi_r(r)}{R}, &\text{\ie,} &\xi_r(r) = Z_1(r) \sqrt{\frac{R^2M}{4\pi r^3 \rho(r)}}\\[0.5em]
\displaystyle Z_2(r) &= \sqrt{\frac{4\pi r^3 \rho(r)}{M}}\frac{\xi_h(r)L}{R}, &\text{\ie,} &\xi_h(r) = Z_2(r)\frac{1}{L} \sqrt{\frac{R^2M}{4\pi r^3 \rho(r)}}\, .
\end{array}
 \end{equation} 
To shorten our notation, we define
\small{
\begin{equation}
\Lambda = \left[\left(\frac{d\Theta_{l}^m}{dx} \right)^2 \left(1-x^2\right) + 2 \Theta_{l}^m(x)  \frac{d\Theta_{l}^m}{dx} x + \frac{m^2}{1-x^2} \Theta_{l}^m(x)^2 \right]\, .
\end{equation}
}%

Using the above, and changing to the fractional radius $u$, such that $r = Ru,\,dr = Rdu$, we get the following expression for the kernel:
\small{
\begin{align}
\mathcal{K}_{nl m}(r, \theta) &= \mathcal{I}_{nlm}^{-1}\left\lbrace \left(\frac{R^2M}{4\pi r^3 \rho}\right) \left[\left[Z_1(u)^2 - 2Z_1(u)\frac{Z_2(u)}{L} \right]\Theta_{l}^m(x)^2 +  \frac{Z_2(u)^2}{L^2} \Lambda  \right]\right\rbrace \rho(r) r \sin\theta\\
&=\mathcal{I}_{nl m}^{-1} \left(\frac{R^2M}{4\pi}\right) \frac{\sqrt{1-x^2}}{(Ru)^2} \left\lbrace  \left[Z_1(u)^2 - 2Z_1(u)\frac{Z_2(u)}{L} \right]\Theta_{l}^m(x)^2 +  \frac{Z_2(u)^2}{L^2} \Lambda  \right\rbrace\, , \label{eq:kernel_fin}
\end{align}
}%
with the mode inertia, $\mathcal{I}_{nlm}$, given as
\small{
\begin{align}\label{eq:III}
\mathcal{I}_{nlm} &= \int_0^{\pi} \int_0^R \left\lbrace |\xi_r(r)|^2\Theta_{l}^m(\cos\theta)^2 + |\xi_h(r)|^2 \left[\left(\frac{d\Theta_{l}^m}{d\theta} \right)^2 + \frac{m^2}{\sin^2\theta}\Theta_{l}^m(\cos\theta)^2 \right]\right\rbrace \rho(r)r^2 \sin\theta  dr d\theta \\
&= \int_0^R \left[\xi_r(r)^2 + L^2\xi_h(r)^2 \right]\rho(r)r^2 dr  \notag\\ 
&=\int_0^R \left( \frac{R^2M}{4\pi r^3 \rho(r)} \right) \left[Z_1(u)^2 + \frac{L^2Z_2(u)^2}{L^2}  \right]\rho(r)r^2 dr \notag\\
&= \left(\frac{R^2M}{4\pi}\right) \int_{0}^{1} \frac{Z_1(u)^2 + Z_2(u)^2}{u} du  \notag\\
&= \left(\frac{R^2M}{4\pi}\right)\,\,  \mathcal{F}\, . \notag
\end{align}
}%
We therefore get for the kernel
\small{
\begin{equation}
\mathcal{K}_{nl m}(u, \theta) =  \frac{\sqrt{1-x^2}}{(Ru)^2\mathcal{F}} \left\lbrace \, \left[\, \left[Z_1(u)^2 - 2Z_1(u)\frac{Z_2(u)}{L} \right]\Theta_{l}^m(x)^2 +  \frac{Z_2(u)^2}{L^2} \Lambda  \right]\, \right\rbrace\, ,
\end{equation}
}%
and finally we get for the splitting
\small{
\begin{align}\label{eq:split_final}
\delta\nu_{nl m} &= m \int_0^{\pi}\int_0^R \mathcal{K}_{nl m}(r, \theta)\mathcal{N}(r, \theta) r dr d\theta\\
&= \frac{m}{\mathcal{F}}\int_0^{\pi}\int_0^1 \frac{\sqrt{1-x^2}}{u} \left\lbrace \, \left[\, \left[Z_1(u)^2 - 2Z_1(u)\frac{Z_2(u)}{L} \right]\Theta_{l}^m(x)^2 +  \frac{Z_2(u)^2}{L^2} \Lambda  \right]\, \right\rbrace \mathcal{N}(u, \theta) du d\theta \, . \notag
\end{align}
}%
For the integration of \eqref{eq:split_final} we use a tanh-sinh quadrature numerical scheme as implemented in the \texttt{Python} module \texttt{mpmath} \citep[][]{mpmath}.\\

In the case in which the rotation profile, $\mathcal{N}(r, \theta)$, can be parameterised in a certain way, the evaluation of the integral in \eqref{eq:split_final} becomes much simpler. If the rotation profile can be expanded as

\begin{equation}
\mathcal{N}(r, \theta) = \sum_{s=0}^{s_{\rm max}} \mathcal{N}_{s}(r) x^{2s} \, ,
\end{equation}
as is the case with our solar rotation profile in \eqref{eq:solar_rot_profile}, the integrals over $\theta$ ($x=\cos\theta$) can be evaluated analytically with the following expression for the splitting \citep[][]{1980A&A....89..207C}:

\begin{equation}
\delta\nu_{nl m} = \sum_{s=0}^{s_{max}} \int_0^R \mathcal{K}_{nlms}(r)\mathcal{N}_{s}(r) dr\, .
\end{equation}
The new kernel $\mathcal{K}_{nlms}(r)$ is given by \citep[][]{1989ApJ...343..526B}

\begin{equation}
\mathcal{K}_{nlms}(r) = \rho(r) r^2 \mathcal{I}_{nlm}^{-1} \left\lbrace \left[ (\xi_r(r) - \xi_h(r))^2 + (L^2 -2s^2 -3s-2)\xi_h^2(r) \right]Q_{lms} + s(2s-1)\xi_h^2(r) Q_{lms-1} \right\rbrace \, ,
\end{equation}
with
\begin{equation}
Q_{lms} = \int_{-1}^{1} x^{2s}\Theta_{l}^m(x)^2 dx
\end{equation}
and $\mathcal{I}_{nlm}$ as given in \eqref{eq:III}. For an expansion up to $s=2$ (\ie, up to $\cos^4\theta$) the $Q_{lms}$ factors are given as
\small{
\begin{align}
Q_{lm0} &= 1\\
Q_{lm1} &= \frac{2L^2 - 2m^2 - 1}{4L^2 -3}\\
Q_{lm2} &= R_{l+1}^m (R_{l+2}^m+R_{l+1}^m+R_{l}^m) + R_{l}^m (R_{l+1}^m +R_{l}^m +R_{l-1}^m), \quad \text{with} \,\,\, R_{l}^m = \frac{l^2 - m^2}{4l^2 -1}.
\end{align}
}%

With this one can give the splitting for the solar-like differential rotation profile in \eqref{eq:solar_rot_profile} as
\small{
\begin{equation}
\delta\nu_{nl m} =  \mathcal{A}\int^R_{r_{bcz}} \mathcal{K}_{nlm0}(r) dr + \mathcal{B}\int^R_{r_{bcz}} \mathcal{K}_{nlm1}(r) dr + \mathcal{C}\int^R_{r_{bcz}} \mathcal{K}_{nlm2}(r) dr + \mathcal{D}\int_0^{r_{bcz}} \mathcal{K}_{nlm0}(r) dr\, ,
\end{equation}
}%
with the kernels given by
\small{
\begin{align}
\mathcal{K}_{nlm0}(r) &= \rho r^2 \mathcal{I}_{nlm}^{-1} \left\lbrace \left[ (\xi_r(r) - \xi_h(r))^2 + (L^2 -2)\xi_h^2(r) \right] \right\rbrace\\
\mathcal{K}_{nlm1}(r) &= \rho r^2 \mathcal{I}_{nlm}^{-1} \left\lbrace \left[ (\xi_r(r) - \xi_h(r))^2 + (L^2 -7)\xi_h^2(r) \right]Q_{lm1} + \xi_h^2(r) \right\rbrace\\
\mathcal{K}_{nlm2}(r) &= \rho r^2 \mathcal{I}_{nlm}^{-1} \left\lbrace \left[ (\xi_r(r) - \xi_h(r))^2 + (L^2 -16)\xi_h^2(r) \right]Q_{lm2} + 6\xi_h^2(r) Q_{lm1} \right\rbrace\, ,
\end{align}
}%
or in terms of $Z_1(u)$ and $Z_2(u)$ as
\small{
\begin{align}
\mathcal{K}_{nlm0}(u) &= (uR)^{-1}\mathcal{F}^{-1} \left[Z_1^2(u) +Z_2^2(u) - Z_2^2(u)L^{-2} - 2Z_1(u)Z_2(u)L^{-1} \right]\\
\mathcal{K}_{nlm1}(u) &= (uR)^{-1}\mathcal{F}^{-1} \left\lbrace \left[ Z_1^2(u) +Z_2^2(u) - 6Z_2^2(u)L^{-2} - 2Z_1(u)Z_2(u)L^{-1} \right]Q_{lm1} + Z^2_2(u)L^{-2} \right\rbrace\\
\mathcal{K}_{nlm2}(u) &= (uR)^{-1}\mathcal{F}^{-1} \left\lbrace \left[ Z_1^2(u) +Z_2^2(u) - 15Z_2^2(u)L^{-2} - 2Z_1(u)Z_2(u)L^{-1}\right]Q_{lm2} + 6Z^2_2(u)L^{-2} Q_{lm1} \right\rbrace
\end{align}
}%


\section{\normalsize  Appendix B: Parameters for the rotation profiles}
\label{app:param}

\subsection{Solar-like}

First, the co-latitude, $\theta_{\rm bind}$, where the rotation rate in the convection zone equals the rate of the interior is found. In order to have the parameters in the rotation profile relate to each other as they do in the Sun, $\theta_{\rm bind}$ is found as
\small{
\begin{align}\notag
\mathcal{D}_{\odot} &= \mathcal{A}_{\odot}+\mathcal{B}_{\odot} \cos^2\theta_{\rm bind}   +\mathcal{C}_{\odot} \cos^4\theta_{\rm bind} \, ,\notag
\end{align}
}%
with the solution
\small{
\begin{align}\notag
\theta_{\rm bind} &= \cos^{-1}\left(\pm\sqrt{\frac{-\mathcal{B}_{\odot} \pm \sqrt{\mathcal{B}_{\odot}^2 - 4\mathcal{C}_{\odot}(\mathcal{A}_{\odot}-\mathcal{D}_{\odot})}}{2\mathcal{C}_{\odot}}} \right)\, .
\end{align}
}%
The solar parameters used are as given in \citet{2004SoPh..220..169G}, viz., $\mathcal{A}_{\odot}=454\, \rm  nHz$, $\mathcal{B}_{\odot}=-55\, \rm  nHz$, $\mathcal{C}_{\odot}=-76\, \rm  nHz$, and $\mathcal{D}_{\sun}=435\, \rm  nHz$.    

Secondly, we note that the differential rotation $\Delta\Omega$ is related to the parameters of the rotation profile as
\small{
\begin{equation}
\Delta\mathcal{N} \equiv \mathcal{N}_{\rm eq.} - \mathcal{N}_{\rm pole} = \mathcal{A} - (\mathcal{A}+\mathcal{B}+\mathcal{C})  = -(\mathcal{B}+\mathcal{C})\, .
\end{equation}
}%
With this we can set the sum of $\mathcal{B}$ and $\mathcal{C}$ for a given assumed value of $\Delta\mathcal{N}$. To set the individual parameters, we use the ratio between these in the Sun, denoted by $\chi_{\odot} = \mathcal{C}_{\odot}/\mathcal{B}_{\odot}$, to get
\small{
\begin{align}
\Delta\mathcal{N} &= -\mathcal{B}(1 + \chi_{\odot})\Rightarrow\\ \label{eq:BandC}
\mathcal{B} &= \frac{-\Delta\mathcal{N}}{1+\chi_{\odot}} \quad \text{and} \quad \mathcal{C} = -\left(\frac{\varrho_{\odot}}{1+\chi_{\odot}} \right)\Delta\mathcal{N}\, .
\end{align}
}%

To get $\mathcal{A}$, we need to relate the mean surface rotation period $P_{\rm rot}$ as given by, \eg, \eqref{eq:prot} to the rotation profile parameters. We define the inverse of the mean rotation period to equal the latitudinally averaged rotation rate at the surface to get
\small{
\begin{align}\label{eq:meanrotangle}
\langle \mathcal{N}(R) \rangle_{\theta} &= \frac{1}{\pi-0}\int_0^{\pi} \mathcal{N}(R, \theta) d\theta\\ \nonumber
&= \frac{1}{\pi}\int_0^{\pi} \left(\mathcal{A}+\mathcal{B} \cos^2\theta   +\mathcal{C} \cos^4\theta \right)\, d\theta\\ \nonumber
& =  \mathcal{A} + \tfrac{1}{2}\mathcal{B} + \tfrac{3}{8}\mathcal{C}\\ \nonumber
& = \langle P_{\rm rot} \rangle^{-1}\, .
\end{align}
}%

Using this in combination with \eqref{eq:BandC}, we get $\mathcal{A}$ as
\small{
\begin{align}
\langle P_{\rm rot} \rangle^{-1} &= \mathcal{A} + \frac{-\Delta\mathcal{N}}{2(1+\varrho_{\odot})} -  \frac{3}{8}\left(\frac{\varrho_{\odot}}{1+\varrho_{\odot}} \right)\Delta\mathcal{N}\\\nonumber
&= \mathcal{A} - \frac{\Delta\mathcal{N}}{1+\varrho_{\odot}}\left(\frac{1}{2} + \frac{3}{8}\varrho_{\odot} \right) \, ,
\end{align}
}%
or
\small{
\begin{align}
\mathcal{A} &= \langle P_{\rm rot} \rangle^{-1} + \frac{\Delta\mathcal{N}}{1+\varrho_{\odot}}\left(\frac{1}{2} + \frac{3}{8}\varrho_{\odot} \right)\, .
\end{align}
}%

Finally, we use the found values for $\mathcal{A}$, $\mathcal{B}$, $\mathcal{C}$, and $\theta_{\rm bind}$ to get $\mathcal{D}$ as
\small{
\begin{equation}
\mathcal{D} = \mathcal{A} +\mathcal{B} \cos^2\theta_{\rm bind}   +\mathcal{C}  \cos^4\theta_{\rm bind}\, . 
\end{equation}
}

\subsection{Cylindrical}
To get the value of $\alpha$ entering the description of the cylindrical rotation profile (\eqref{eq:axial}), we first define the inverse of the mean rotation period as the latitudinally averaged rotation rate at the surface:
\begin{align}\label{eq:axial_param1}
\langle \mathcal{N}(R) \rangle_{\theta} &= \frac{1}{\pi-0}\int_0^{\pi} \mathcal{N}(R, \theta) d\theta\\ \nonumber
&= \frac{1}{\pi}\int_0^{\pi} \left( \alpha +\beta \sin\theta \right) d\theta\\ \nonumber
& = \frac{\pi \alpha + 2\beta}{\pi}\\ \nonumber
& =  \alpha + \tfrac{2}{\pi}\beta\\ \nonumber
& = \langle P_{\rm rot} \rangle^{-1}\, . \nonumber
\end{align}

From \eqref{eq:axial_delta} it is given that $\beta=\Delta\mathcal{N}$, whereby $\alpha$ is obtained as

\begin{equation}
\alpha = \langle P_{\rm rot} \rangle^{-1} - \tfrac{2}{\pi}\Delta\mathcal{N} \, .
\end{equation}


\end{document}